\documentclass[rmp,aps,amssymb,twocolumn,nofootinbib,floatfix,showpacs]{revtex4}

\usepackage{amsmath,amssymb,bm}
\usepackage{graphics}
\usepackage{adjustbox}
\usepackage{graphicx}
\usepackage{multirow}
\usepackage{booktabs}
\usepackage{array}
\usepackage{tikz}
\usepackage{rotating}
\usepackage{hhline}
\usepackage[caption=false]{subfig}

\usepackage[colorlinks,linkcolor=blue,citecolor=blue,breaklinks=true]{hyperref}
\usepackage{breakurl}

\begin{document}

\title{Community detection in networks: A user guide}

\author{Santo Fortunato}
\email{santo@indiana.edu}
\affiliation{Center for Complex Networks and Systems Research, School
  of Informatics and Computing and Indiana University Network Science
  Institute (IUNI), Indiana University, Bloomington, USA}
\affiliation{Department of Computer Science, Aalto University School of Science, P.O. Box 15400, FI-00076}
\author{Darko Hric}
\affiliation{Department of Computer Science, Aalto University School
  of Science, P.O. Box 15400, FI-00076}

\date{\today}

\begin{abstract}

Community detection in networks is one of the most popular topics of modern network science. Communities, or clusters, are usually groups of vertices 
having higher probability of being connected to each other than to members of other groups, though other patterns are possible.
Identifying communities is an ill-defined problem. There are no universal protocols on the fundamental ingredients, like 
the definition of community itself, nor on other crucial issues, like the validation of algorithms and the comparison of their performances. 
This has generated a number of confusions and misconceptions, which undermine the progress in the field. 
We offer a guided tour through the main aspects of the problem. We also point out strengths and weaknesses of popular methods, and give directions to their use.

\end{abstract}


\maketitle
\tableofcontents

\begin{sloppypar}

\section{Introduction}
\label{sec-intro}

\begin{figure}[h!]
\includegraphics[width=\columnwidth]{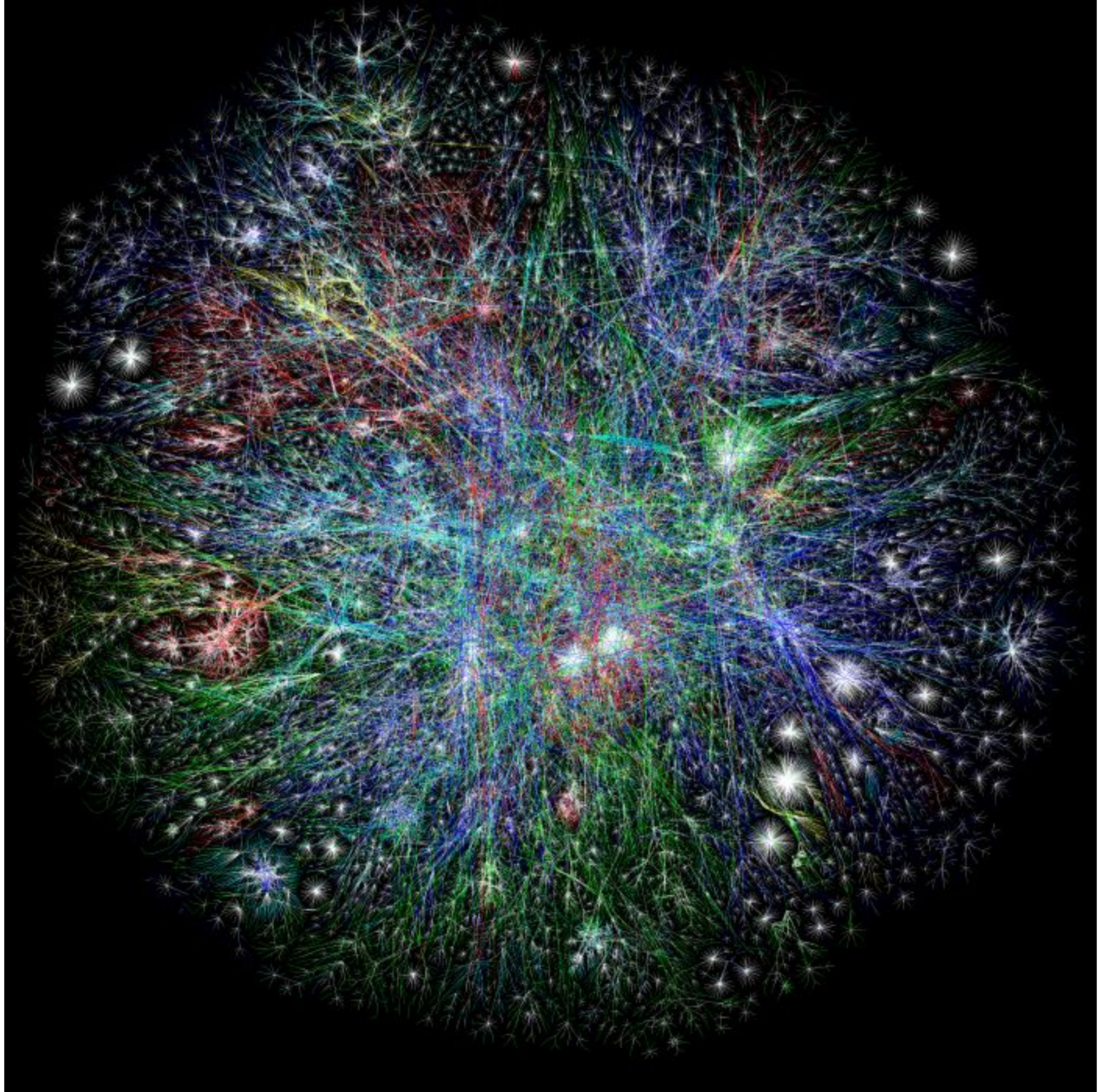}
\caption{Internet network. 
Reprinted figure with permission from www.opte.org.}
\label{figgennet}
\end{figure}

The science of networks is a modern discipline spanning the natural, social and computer sciences, 
as well as engineering~\cite{caldarelli07,barrat08,cohen10,newman10,estrada11b,dorogovtsev13,estrada15}.
Networks, or graphs, consist of {\it vertices} and {\it edges}. An edge typically connects a pair of vertices\footnote{There may be connections between three vertices or more. In this case 
one speaks of {\it hyperedges} and the network is a {\it hypergraph}.}. Networks
occur in an huge variety of contexts. Facebook, for instance, is 
a large social network, where more than one billion people are connected via virtual acquaintanceships.
Another famous example is the Internet, the physical network of computers, routers and modems which are linked via
cables or wireless signals (Fig.~\ref{figgennet}).
Many other examples come from biology, physics, economics, engineering, computer science, ecology, marketing, social and political sciences, etc..

\begin{figure}[h!]
\begin{center}
\includegraphics[width=\columnwidth]{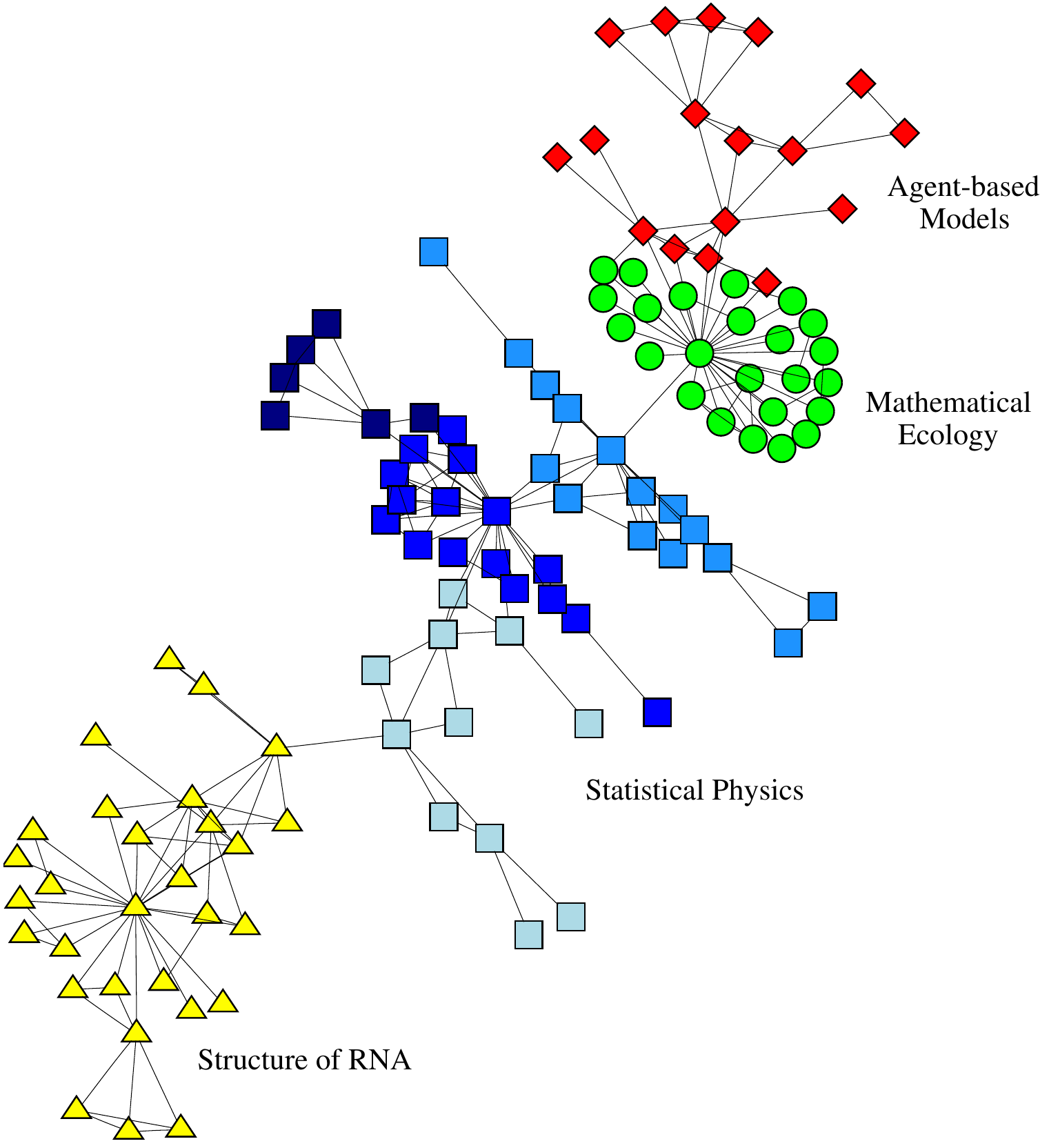}
\end{center}
\caption{Collaboration network of scientists working at the Santa Fe Institute (SFI). Edges connect scientists that have coauthored at least one paper.
Symbols indicate the research areas of the scientists.
Naturally, there are more edges between scholars working on the same area than between scholars working in different areas. 
Reprinted figure with permission from~\cite{girvan02}. 
\copyright\,2002, by the National Academy of Sciences, USA.}
\label{figSFI}
\end{figure}

Most networks of interest display {\it community structure}, i. e., their vertices are organised into groups, called {\it communities}, {\it clusters} or {\it modules}.
In Fig.~\ref{figSFI} we show a collaboration network of scientists working at the Santa Fe Institute (SFI) in Santa Fe, New Mexico. Vertices are scientists, edges join coauthors. 
Edges are concentrated within groups of vertices representing scientists working on the same research topic, where collaborations are more natural.
Likewise, communities could represent proteins with similar function in protein-protein interaction networks, groups of friends in social networks,
websites on similar topics on the Web graph, and so on.

Identifying communities may offer insight on how the network is organised. It allows us to focus on regions having some degree of autonomy within the graph.
It helps to classify the vertices, based on their role with respect to the communities they belong to. For instance we can distinguish vertices totally embedded within their clusters
from vertices at the boundary of the clusters, which may act as brokers between the modules and, in that case, could play a major role both in holding 
the modules together and in the dynamics of spreading processes across the network.

Community detection in networks, also called {\it graph} or {\it network clustering}, is an ill-defined problem though.
There is no universal definition of the objects that one should be looking for. Consequently, there are no clear-cut guidelines on how
to assess the performance of different algorithms and how to compare them with each other.
On the one hand, such ambiguity leaves a lot of freedom to propose diverse approaches to the problem, which often depend on the specific research question and (or) the particular system at study.
On the other hand, it 
has introduced a lot of noise into the field, slowing down progress. In particular, it has favoured the diffusion of questionable concepts and convictions, on which a large number of 
methods are based.

This work presents a critical analysis of the problem of community detection, intended to practitioners but 
accessible to readers with basic notions of network science.  
It is not meant to be an exhaustive survey.
The focus is on the general aspects of the problem, especially in the light of recent findings. Also, we discuss some popular classes of algorithms and give 
advice on their usage. More info on network clustering can be found in 
several review articles~\cite{schaeffer07,porter09,fortunato10,coscia11,parthasarathy11,newman12,malliaros13,xie13,chakraborty16}.

The contents are organised in three main sections. Section~\ref{sec-comm} deals with 
the concept of community, describing its evolution from the classic subgraph-based notions to the modern statistical interpretation.
Next we discuss the critical issue of validation (Section~\ref{sec-valid}), emphasising the role of artificial benchmarks, the importance of the choice of partition similarity scores,
the conditions under which clusters are detectable, the usefulness of metadata and the structural peculiarities of communities in real networks. Section~\ref{sec-meth}  
hosts a critical discussion of some popular clustering approaches. It also tackles important general methodological aspects, such as the determination of the number of 
clusters, which is a necessary input for several techniques, the possibility to generate robust solutions by combining multiple partitions, the main approaches to discover 
dynamic communities, as well as the assessment of the significance of 
clusterings. In Section~\ref{sec-soft} we indicate where to find useful software. The concluding remarks of Section~\ref{sec-OL} close the work.

\section{What are communities?}
\label{sec-comm}

\subsection{Variables}
\label{sec-var}

We start with a subgraph ${C}$ of a graph ${G}$.
The number of vertices and edges are $n$, $m$ for ${G}$ and $n_{C}$, $m_{C}$ for ${C}$, respectively. The adjacency matrix of $G$ is $A$, its element $A_{ij}$ equals $1$ if 
vertices $i$ and $j$ are neighbours, otherwise it equals $0$.
We assume that the subgraph is connected because communities usually are\footnote{The variables defined in this section hold for any subgraph, connected or not.}. 
Other types of group structures do not require connectedness (Section~\ref{sec-MV}).

\begin{figure}[h!]
\begin{center}
\includegraphics[width=\columnwidth]{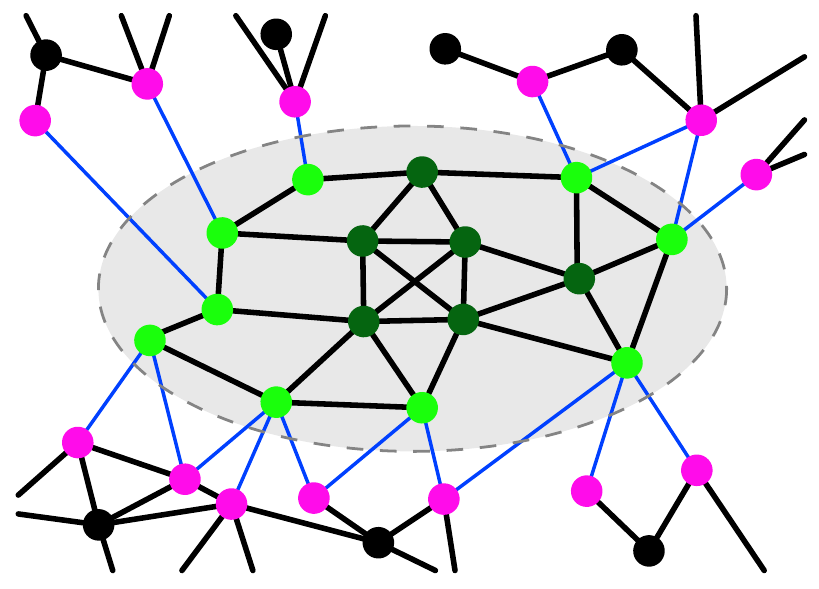}
\end{center}
\caption{Schematic picture of a connected subgraph.}
\label{figstruct}
\end{figure}

The subgraph is schematically illustrated in Fig.~\ref{figstruct}. 
Its vertices are enclosed by the dashed contour.
The magenta dots are the external vertices connected to the subgraph, while the black ones are the remaining vertices of the network. 
The blue lines indicate the edges connecting the subgraph to the rest of the network.

The {\it internal} and {\it external} 
degree $k_{i}^{int}$ and $k_{i}^{ext}$ of a vertex $i$ of the network with respect to subgraph ${C}$ are 
the number of edges connecting $i$ to vertices of ${C}$ and to the rest of the graph, respectively.  
Both definitions can be expressed in compact form via the adjacency matrix $A$:
$k^{int}_i=\sum_{j \in C}A_{ij}$ and $k^{ext}_i=\sum_{j \notin C}A_{ij}$, where the sums run over all vertices $j$ inside and outside $C$, respectively.
Naturally, 
the degree $k_i$ of $i$ is the sum of $k_{i}^{int}$ and $k_{i}^{ext}$: $k_i=\sum_{j}A_{ij}$.
If $k_{i}^{ext}=0$ and $k_{i}^{int}>0$ $i$ has neighbours only within ${C}$ and is an {\it internal vertex} of ${C}$ (dark green dots in the figure).
If $k_{i}^{ext}>0$ and $k_{i}^{int}>0$ $i$ has neighbours outside ${C}$ and is a {\it boundary vertex} of ${C}$ (bright green dots in the figure).
If $k_{i}^{int}=0$, instead, the vertex is disjoint from ${C}$. The {\it embeddedness} $\xi_{i}$ is the ratio
between the internal degree and the degree of vertex $i$: $\xi_i=k_i^{int}/k_i$. The larger $\xi_i$, the stronger the 
relationship between the vertex and its community. The {\it mixing parameter} $\mu_{i}$ is the ratio
between the external degree and the degree of vertex $i$: $\mu_i=k_i^{ext}/k_i$. By definition, $\mu_i=1-\xi_i$.

Now we present a number of variables related to the subgraph as a whole. We distinguish them in three classes.

The first class comprises measures based on internal connectedness, i. e., on how cohesive the subgraph is.
The main variables are:
\begin{itemize}
\item {\it Internal degree} $k^{int}_{C}$. The sum of the internal degrees of the vertices of ${C}$. It equals twice the number $m_{C}$ 
of internal edges, as each edge contributes two units of degree. In matrix form, $k^{int}_C=\sum_{i,j \in C}A_{ij}$.
\item {\it Average internal degree} $k^{avg\mbox{-}int}_{C}$. Average degree of vertices of ${C}$, considering only internal edges: 
$k^{avg-int}_{C}=k^{int}_{C}/n_{C}$.
\item {\it Internal edge density} $\delta^{int}_{C}$. The
ratio between the number of internal edges of ${C}$ and the number of all possible internal edges:
\begin{equation} 
\delta^{int}_{C}=\frac{k^{int}_{C}}{n_{C}(n_{C}-1)}.
\label{eq1}
\end{equation}
We remark that $n_{C}(n_{C}-1)/2$ is the maximum number of internal edges that a simple graph with $n_{C}$ vertices may have\footnote{A simple graph 
has at most one edge running between any pair of vertices and no self-loops, i. e., no edges connecting a vertex to itself.}.
\end{itemize}

The second class includes measures based on external connectedness, i. e., on how embedded the subgraph is in the network or, equivalently, how separated 
the subgraph is from it.
The main variables are:
\begin{itemize}
\item {\it External degree}, or {\it cut}, $k^{ext}_{C}$. The sum of the external degrees of the vertices of ${C}$. It gives the number of external edges of the subgraph (blue lines
in Fig.~\ref{figstruct}). In matrix form, $k^{ext}_C=\sum_{i \in C, j \notin C}A_{ij}$.
\item {\it Average external degree}, or {\it expansion}, $k^{avg\mbox{-}ext}_{C}$. Average degree of vertices of ${C}$, considering only external edges: 
$k^{avg-ext}_{C}=k^{ext}_{C}/n_{C}$.
\item {\it External edge density}, or {\it cut ratio}, $\delta^{ext}_{C}$. The
ratio between the number of external edges of ${C}$ and the number of all possible external edges:
\begin{equation} 
\delta^{ext}_{C}=\frac{k^{ext}_{C}}{n_{C}(n-n_{C})}.
\label{eq2}
\end{equation}
\end{itemize}

\begin{sidewaystable}
\centering
{\renewcommand*{\arraystretch}{2.0}
\begin{tabular}{|c|c|c||c|c|c|}
\multicolumn{3}{c}{{\bf Unweighted networks}} & \multicolumn{3}{c}{{\bf Weighted networks}} \\
\hline
  Name & Symbol & Definition & Name & Symbol & Definition \\ \hline

 Internal degree & $k^{int}_i$ & $\sum_{j \in C}A_{ij}$ & Internal strength & $w^{int}_i$ & $\sum_{j \in C}W_{ij}$ \\ \hline

 External degree & $k^{ext}_i$ & $\sum_{j \notin C}A_{ij}$ & External strength & $w^{ext}_i$ & $\sum_{j\notin C}W_{ij}$ \\ \hline

Degree & $k_i$ & $\sum_{j}A_{ij}$ & Strength & $w_i$ & $\sum_{j}W_{ij}$ \\ \hline

Embeddedness &$\xi_i$ & $\frac{k^{int}_i}{k_i}$& Weighted embeddedness &$\xi^{w}_i$ & $\frac{w^{int}_i}{w_i}$ \\ \hline

Mixing parameter &$\mu_i$ & $\frac{k^{ext}_i}{k_i}$& Weighted mixing parameter &$\mu^{w}_i$ & $\frac{w^{ext}_i}{w_i}$ \\ \hline

 \hline 
    
 \end{tabular}}
\caption{Basic vertex community variables, for unweighted and weighted networks. $A$ and $W$ are the adjacency and the weight matrix, respectively.}
\label{tab:metricsv}
\end{sidewaystable}

\begin{sidewaystable}
\centering
{\renewcommand*{\arraystretch}{2.0}
\begin{tabular}{c|c|c|c||c|c|c|}
\multicolumn{1}{c}{} & 
\multicolumn{3}{c}{{\bf Unweighted networks}} & \multicolumn{3}{c}{{\bf Weighted networks}} \\
\hline
  & Name & Symbol & Definition & Name & Symbol & Definition \\ \hline

 & Internal degree & $k^{int}_C$ & $\sum_{i,j \in C}A_{ij}$ & Internal strength & $w^{int}_C$ & $\sum_{i,j \in C}W_{ij}$ \\ 

 & Average internal degree &$k^{avg\mbox{-}int}_C$ & $\frac{k^{int}_C}{n_C}$& Average internal strength &$w^{avg\mbox{-}int}_C$ & $\frac{w^{int}_C}{n_C}$ \\

\begin{rotate}{90} \hbox{\hspace{0.2em} \large{Internal}} \end{rotate}\hspace*{0.15cm} &  Internal edge density &$\delta^{int}_C$ & $\frac{k^{int}_C}{n_C(n_C-1)}$& Internal weight density &$\delta^{int}_{w,C}$ & $\frac{w^{int}_C}{\bar{w}n_C(n_C-1)}$ \\
 \hline \hline
 &  External degree & $k^{ext}_C$ & $\sum_{i \in C, j \notin C}A_{ij}$ & External strength & $w^{ext}_C$ & $\sum_{i\in C, j\notin C}W_{ij}$ \\ 
&   Average external degree &$k^{avg\mbox{-}ext}_C$ & $\frac{k^{ext}_C}{n_C}$& Average external strength &$w^{avg\mbox{-}ext}_C$ & $\frac{w^{ext}_C}{n_C}$ \\
\begin{rotate}{90} \hbox{\hspace{-0.0em} \large{External}} \end{rotate} \hspace*{0.02cm}  &   External edge density &$\delta^{ext}_C$ & $\frac{k^{ext}_C}{n_C(n-n_C)}$& External weight density &$\delta^{ext}_{w,C}$ & $\frac{w^{ext}_C}{\bar{w}n_C(n-n_C)}$ \\
 \hline \hline    
  &     Total degree & $k_C$ & $\sum_{i \in C, j}A_{ij}$ & Total strength & $w_C$ & $\sum_{i\in C, j}W_{ij}$ \\ 
   &  Average degree &$k^{avg}_C$ & $\frac{k_C}{n_C}$& Average strength &$w^{avg}_C$ & $\frac{w_C}{n_C}$ \\
\begin{rotate}{90} \hbox{\hspace{1.0em} \large{Total}} \end{rotate} \hspace*{0.02cm}  &    Conductance &$C_C$ & $\frac{k^{ext}_C}{k_C}$& Weighted conductance &$C_{w,C}$ & $\frac{w^{ext}_C}{w_C}$ \\
 \hline 
    
 \end{tabular}}
\caption{Basic community variables, for unweighted and weighted networks. $A$ and $W$ are the adjacency and the weight matrix, respectively, 
$n_C$ the number of vertices of the community, $n$ the total number of vertices of the graph, $\bar{w}$ the average weight of the network edges.}
\label{tab:metrics}
\end{sidewaystable}

Finally, we have hybrid measures, combining internal and external connectedness.
Notable examples are:
\begin{itemize}
\item {\it Total degree}, or {\it volume}, $k_{C}$. The sum of the degrees of the vertices of ${C}$. 
Naturally, $k_{C}=k^{int}_{C}+k^{ext}_{C}$. In matrix form, $k_C=\sum_{i \in C, j}A_{ij}$. 
\item {\it Average degree} $k^{avg}_{C}$. Average degree of vertices of ${C}$: 
$k^{avg}_{C}=k_{C}/n_{C}$.
\item {\it Conductance} $C_{C}$. The
ratio between the external degree and the total degree of ${C}$:
\begin{equation} 
C_{C}=\frac{k^{ext}_{C}}{k_{C}}.
\label{eq4}
\end{equation}
\end{itemize}

All definitions we have given hold for the case of undirected and unweighted networks. The extension to weighted graphs is straightforward, as it suffices to
replace the ``number of edges" with the sum of the weights carried by every edge. For instance, the internal degree $k_v^{int}$ of a vertex $v$
becomes the {\it internal strength} $w_v^{int}$, which is the sum of the weights of the edges joining $v$ with the vertices of subgraph $C$.
For the internal and external edge densities of Eqs.~(\ref{eq1}) and (\ref{eq2}) one would have to replace the numerators with their weighted counterparts 
and multiply the denominators by the average edge weight $\bar{w}=\sum_{ij}W_{ij}/2m$, where 
$W_{ij}$ is the element of the {\it weight matrix}, indicating the weight of the edge joining vertices $i$ and $j$ 
($W_{ij}=0$ if $i$ and $j$ are disconnected) and $m$ the total number of graph edges. 
In Tables~\ref{tab:metricsv} and \ref{tab:metrics} we list all variables we have presented along with their extensions to the case of weighted networks. 
In directed networks one would have to distinguish between incoming and outgoing edges. 
Extensions of the metrics are fairly simple to implement, though their usefulness is unclear.

\subsection{Classic view}
\label{sec-defs}

\begin{figure}[h!]
\begin{center}
\includegraphics[width=\columnwidth]{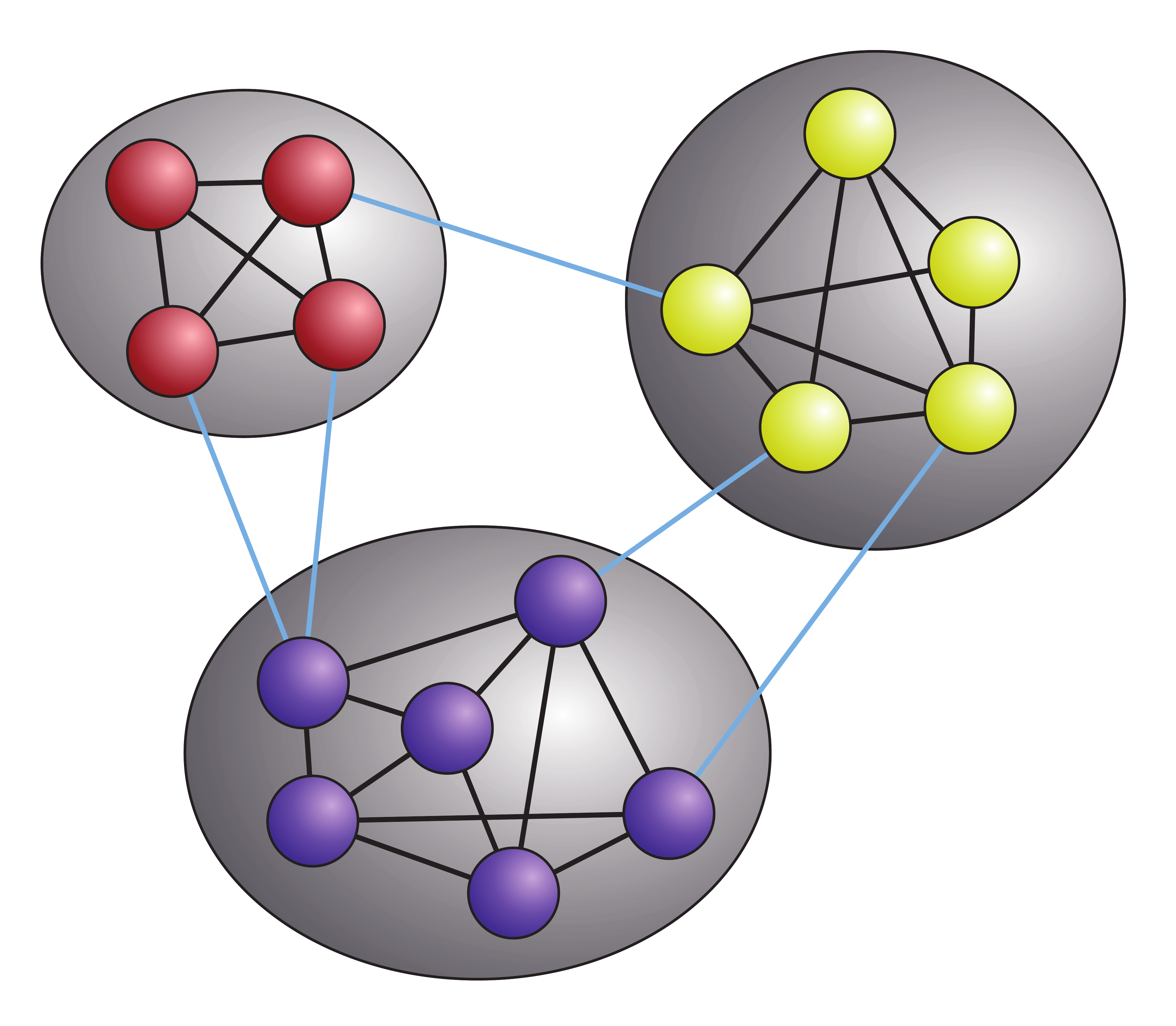}
\end{center}
\caption{Classic view of community structure. Schematic picture of a network with three communities.}
\label{figcom}
\end{figure}

\begin{figure}
\begin{center}
\begin{center}
\includegraphics[width=\columnwidth]{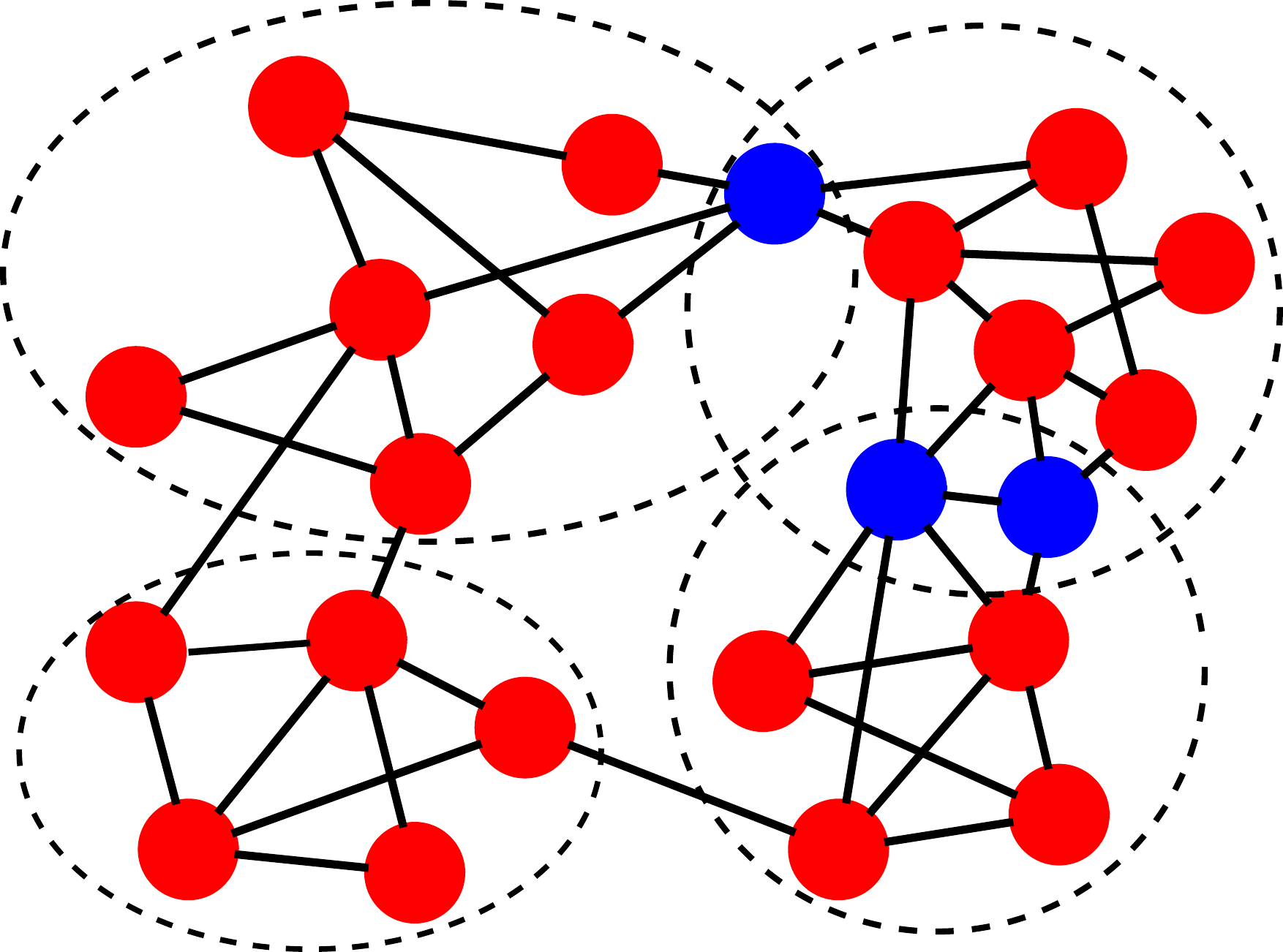}
\end{center}
\caption {Overlapping communities. A network is divided in four communities, enclosed by the dashed contours. 
Three of them share boundary vertices, indicated by the blue dots.}
\label{figoverl}
\end{center}
\end{figure}

Figure~\ref{figcom} shows how scholars usually envision
community structure. The network has three clusters and in each cluster the density of edges is comparatively higher than the density of edges
between the clusters. This can be summarised by saying that communities are dense subgraphs which are well separated from each other.
This view has been challenged, recently~\cite{leskovec09,jeub15}, as we shall see in Section~\ref{ncp}.
Communities may overlap as well, sharing some of the vertices. For instance, in social networks individuals can belong to different circles at the same time, like 
family, friends, work colleagues.
Figure~\ref{figoverl} shows an example of a network with 
overlapping communities. Communities are typically supposed to be overlapping at their boundaries, as in the figure. Recent results
reveal a different picture, though~\cite{yang14} (Section~\ref{ncp}). A subdivision of a network into overlapping communities is called {\it cover} and one speaks of {\it soft clustering}, as opposed to 
{\it hard clustering}, which deals with divisions into non-overlapping groups, called {\it partitions}. The generic term {\it clustering} can be used to indicate both types of subdivisions.
Covers can be {\it crisp}, when shared vertices belong to their communities with equal strength, or {\it fuzzy}, when the strength of their membership can be different
in different clusters\footnote{In the literature the word fuzzy is often used to describe both situations.}.

The oldest definitions of community-like objects were proposed by social network analysts and focused on the 
internal cohesion among vertices of a subgraph~\cite{wasserman94,scott00,moody03}. 
The most popular concept is that of {\it clique}~\cite{luce49}. A clique is a {\it complete graph}, that is, a 
subgraph such that each of its vertices is connected to all the others. It is also a maximal subgraph, meaning that it
is not included in a larger complete subgraph.
In modern network science it is common to call clique any complete graph, not necessarily maximal. 
Triangles are the simplest cliques. Finding cliques is an {\bf NP}-complete problem~\cite{bomze99}; a popular technique is the
Bron--Kerbosch method~\cite{bron73}.

The notion of cliques, albeit useful, cannot be considered a good candidate for a community definition.
While a clique has the largest possible internal edge density, as all internal edges are present, 
communities are not complete graphs, in general.
Moreover, all vertices have identical role in a clique, while in real network communities some vertices are more important than others, due to their
heterogeneous linking patterns.
Therefore, in social network analysis the notion has been relaxed, generating the related concepts of {\it n-cliques}~\cite{luce50,alba73}, 
{\it n-clans} and {\it n-clubs}~\cite{mokken79}. Other definitions are based on the idea that a vertex
must be adjacent to some minimum number of other vertices in the subgraph. 
A {\it $k$-plex} is a maximal subgraph in which each vertex is adjacent to all other vertices of the subgraph except at most $k$ of them~\cite{seidman78}.
Details on the above definitions can be found in specialised books~\cite{wasserman94,scott00}.

For a proper community definition, one should take into account both the internal cohesion of the candidate subgraph and its separation
from the rest of the network. A simple idea that has received a great popularity is that a community is a subgraph such that ``the number of internal edges 
is larger than the number of external edges"\footnote{Here we focus on the case of unweighted graphs, extensions of all definitions to the weighted case are immediate.}. 
This idea has inspired the following definitions.
An {\it $LS$-set}~\cite{luccio69}, or {\it strong community}~\cite{radicchi04}, is a subgraph such that the internal degree of 
each vertex is greater than its external degree. A relaxed condition is
that the internal degree of the subgraph exceeds its external degree 
[{\it weak community}~\cite{radicchi04}]\footnote{The definition of weak community is the natural implementation of the 
na\"{\i}ve expectation that there must be more edges inside than outside. However,
for a subgraph ${C}$ to be a weak community it is not necessary that the number of internal edges 
$m_{C}$ exceeds that of external edges $k^{ext}_{C}$. Since the internal degree $k^{int}_{C}=2m_{C}$
(Section~\ref{sec-var}) the actual condition is $2m_{C}> k^{ext}_{C}$.}.
A strong community is also a weak community, while the converse is not generally true. 

\begin{figure}[h!]
\begin{center}
\includegraphics[width=\columnwidth]{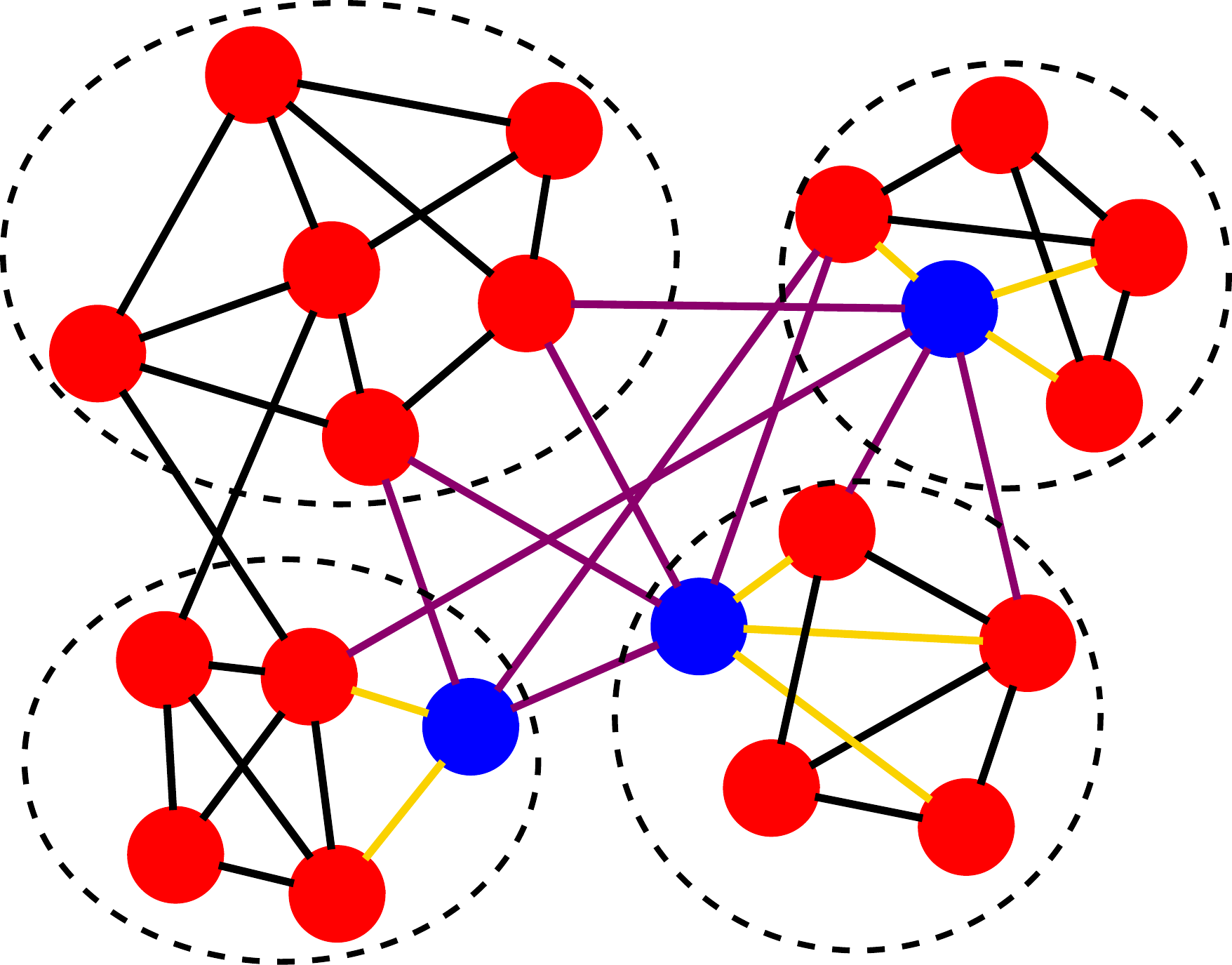}
\caption {Strong and weak communities. The four subgraphs enclosed in the contours are weak communities according to the definitions
of Radicchi et al.~\cite{radicchi04} and Hu et al.~\cite{hu08b}. They are also strong communities according to Hu et al., as the internal degree of each vertex
exceeds the number of edges joining the vertex with the vertices of every other subgraph. However, three of the subgraphs are not strong communities according to Radicchi et al.,
as some vertices (indicated in blue) have external degree larger than their internal degree (the internal and external edges of these vertices are coloured in  
yellow and magenta, respectively).}
\label{figSW}
\end{center}
\end{figure}

A drawback of these definitions is that one separates the subgraph at study from the rest of the network, which is taken as a single object.
But the latter can be in turn divided into communities. If a subgraph ${C}$ is a proper community, it makes sense that each of its vertices 
is more strongly attached to the vertices of ${C}$ than to the vertices of any other subgraph. 
This concept, proposed by Hu et al.~\cite{hu08b}, is more in line (though not entirely) with the modern idea of community that we 
discuss in the following section. It has generated two alternative definitions of strong and weak community. A subgraph ${C}$ is a strong community if
the internal degree of any vertex within ${C}$ exceeds the internal degree of the vertex within any other subgraph, i. e., the number of edges joining
the vertex to those of the subgraph; likewise,
a community is weak if its internal degree exceeds the (total) internal degree of its vertices within every other community.
A strong (weak) community {\`a} la Radicchi et al. is a strong (weak) community also in the sense
of Hu et al.. The opposite is not true, in general (Fig.~\ref{figSW}). In particular, a subgraph can be a strong community in the sense of Hu et al. even though 
all of its vertices have internal degree smaller than their respective external degree.

The above definitions of communities use {\it extensive} variables: their value tends to be the larger, the bigger the community
(e. g., the internal and external degrees).
But there are also variables discounting community size. An example is 
the internal cluster density $\delta_{int}({C})$ of Eq.~(\ref{eq1}). One could assume that a subgraph ${C}$ with $k$ vertices 
is a cluster if $\delta_{int}({C})$ is larger than a threshold $\xi$. Setting the size of 
the subgraph is necessary because otherwise any clique would be among the best possible communities,
including trivial two-cliques (simple edges) or triangles. 

\subsection{Modern view}
\label{sec-MV}

As we have seen in the previous section, traditional definitions of community rely on counting edges (internal, external), in various ways.
But what one should be really focusing on is the {\it probability} that vertices share edges with a subgraph. The existence of communities implies
that vertices interact more strongly with the other members of their community than they do with vertices of the other communities.
Consequently, there is a preferential linking pattern between vertices of the same group. This is the reason why edge densities end up being 
higher within communities than between them. We can formulate that by saying that vertices 
of the same community have a higher probability to form edges with their partners than with the other vertices.

Let us suppose that we estimated the edge probabilities between all pairs of vertices, somehow.
We can define the groups by means of those probabilities. It is a scenario similar to the classic one we have seen in Section~\ref{sec-defs}, where 
we add and compare probabilities, instead of edges. Natural definitions of strong and weak community are:
\begin{itemize}
\item A {\it strong community} is a subgraph each of whose vertices has a higher 
probability to be linked to every vertex of the subgraph than to any other vertex of the graph. 
\item A {\it weak community} is a subgraph 
such that the average edge probability of each vertex with the other members of the group exceeds the average edge probability of the vertex with 
the vertices of any other group\footnote{Since we are comparing average probabilities, which come with a standard error, the definition of weak community 
should not rely on the simple numeric inequality between the averages, but on the statistical significance of their difference. Significance will be discussed in Section~\ref{sec-sign}.}.
\end{itemize}
The difference between the two definitions is that, in the concept of strong community, the inequality between edge probabilities 
holds at the level of every pair of vertices, while 
in the concept of weak community the inequality holds only for averages over groups. Therefore, a strong community is also a weak community, 
but the opposite is not true, in general.

\begin{figure}[h!]
\begin{center}
\includegraphics[width=\columnwidth]{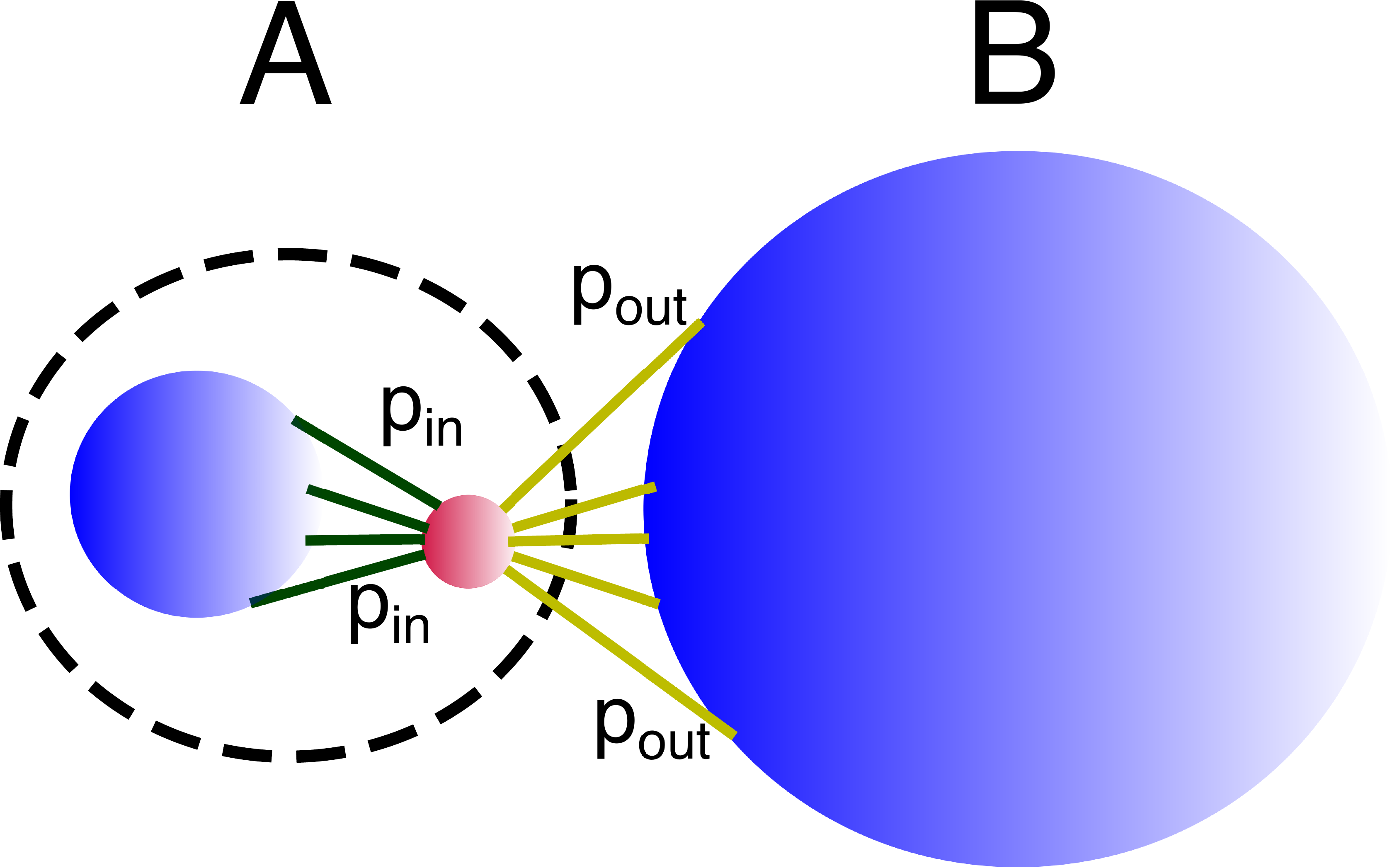}
\caption{Problems of the classic notions of strong and weak communities. A network is generated by the illustrated model, with two subgraphs $A$
and $B$ and edge probabilities $p_{in}$ between vertices of the subgraphs and 
$p_{out} < p_{in}$ between vertices of $A$ and $B$. The red circle is a representative vertex of subgraph $A$, the smaller blue circle represents 
the rest of the vertices of $A$. 
The subgraphs are both strong and weak communities
in the probabilistic sense, but they may be neither strong nor weak communities according to the classic definitions by
Radicchi et al.~\cite{radicchi04} and Hu et al.~\cite{hu08b}, if $B$ is sufficiently larger than $A$.}
\label{figexSW}
\end{center}
\end{figure}

Now we can see why the former definitions of strong and weak community~\cite{radicchi04,hu08b} are not satisfactory. Suppose to have a network with two 
subgraphs $A$ and $B$ of very different sizes, say with $n_A$ and $n_B\gg n_A$ vertices (Fig.~\ref{figexSW}). 
The network is generated by 
a model where the edge probability is $p_{in}$ between vertices of the same group and 
$p_{out}<p_{in}$ for vertices of different groups. The two subgraphs are communities both in the strong and in the weak sense, according to the 
probability-based definitions above. The expected internal degree of a vertex of $A$ is $k_A^{int}=p_{in} n_A$:
since there are $n_A$ possible internal neighbours\footnote{The number of possible community neighbours should actually be $n_A-1$, but 
for simplicity one allows for the formation 
of self-edges, from a vertex to itself. Results obtained with and without self-edges are 
basically undistinguishable, when community sizes are much larger than one.
We shall stick to this setup throughout the paper.}.
Likewise, the expected external degree of a 
vertex of $A$ is $k_A^{ext}=p_{out} n_B$. The expected 
internal and external degrees 
of $A$ are $K_A^{int}=p_{in} n^2_A$ and $K_A^{ext}=p_{out} n_An_B$. For any two values of $p_{in}$ and $p_{out}<p_{in}$ 
one can always choose $n_B$ sufficiently larger than $n_A$ that $k_A^{int} < k_A^{ext}$, which also implies 
that $K_A^{int} < K_A^{ext}$. In this setting the subgraphs are neither strong nor weak communities, according
to the definitions proposed by Radicchi et al. and Hu et al.. 

%
%
How can we compute the edge probabilities between vertices? 
This is still an ill-defined problem, unless one has a model stating how edges are formed.
One can make many hypotheses on the process of edge formation.
\begin{figure*}
\begin{center}
\subfloat[Community structure]{
\label{fig:stylized-hotdog}
\includegraphics[width=0.23\linewidth]{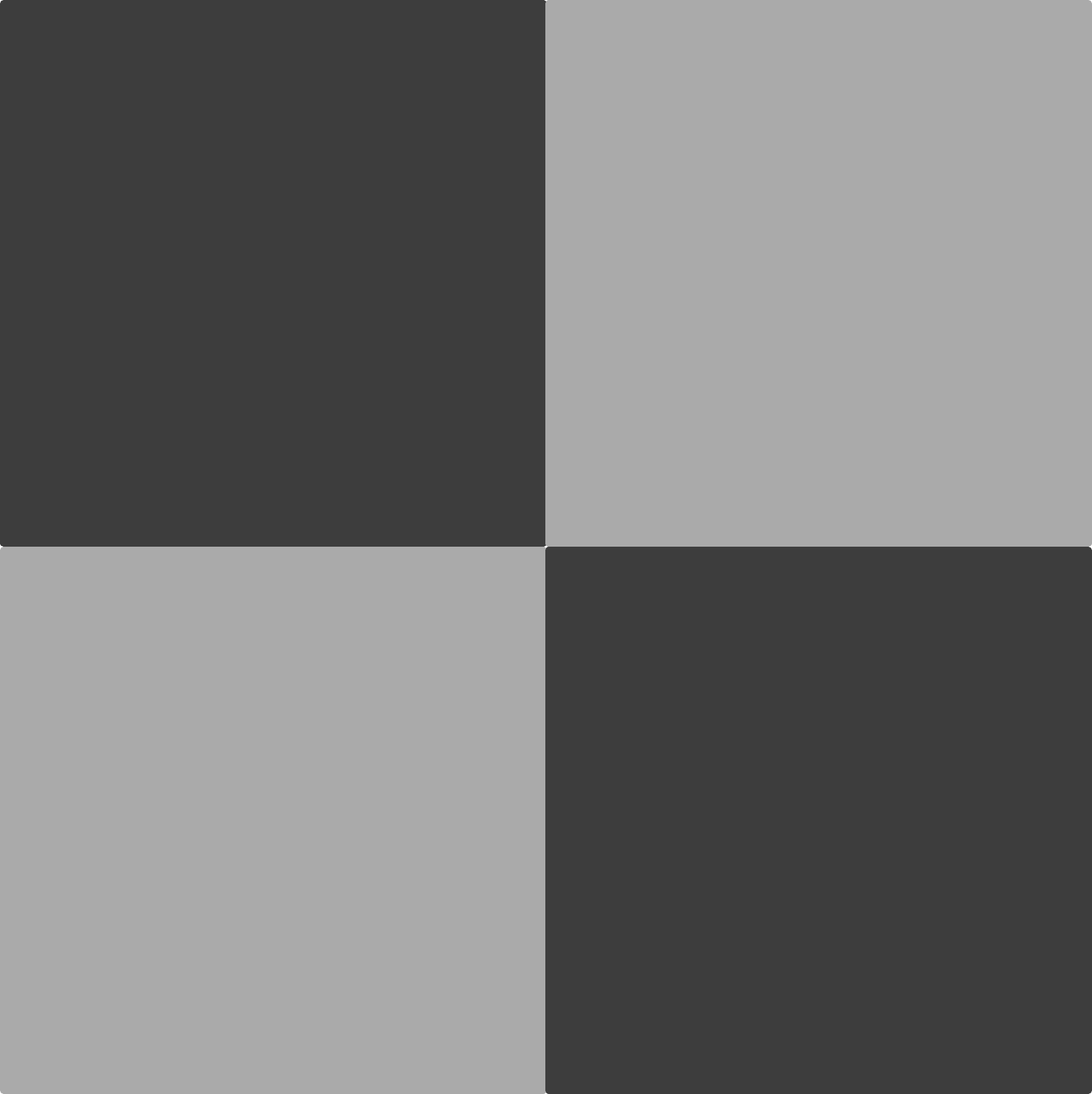}
\includegraphics[width=0.23\linewidth]{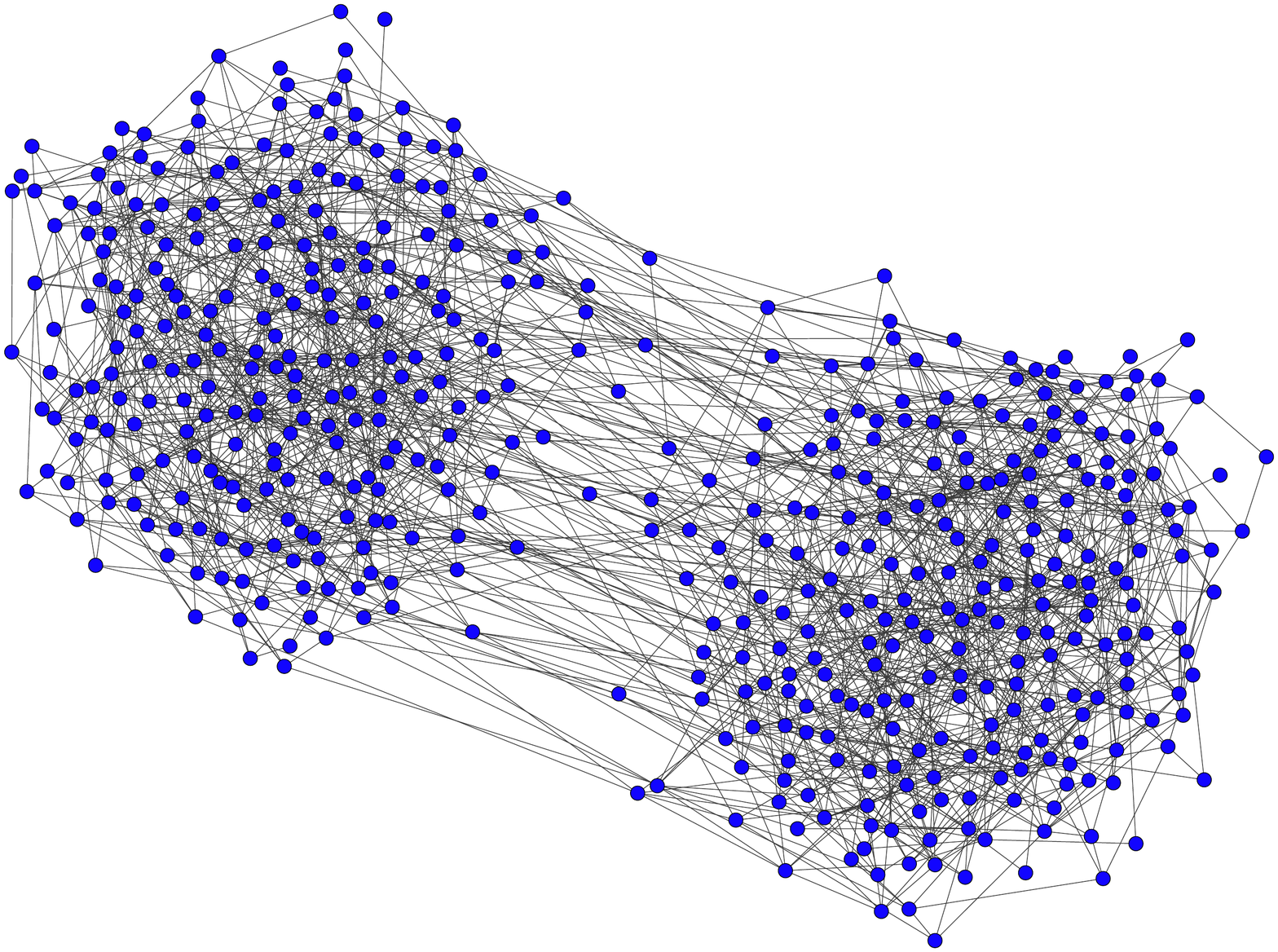}
}
\hfill
\subfloat[Disassortative structure]{
\label{fig:stylized-bipartite}
\includegraphics[width=.23\linewidth]{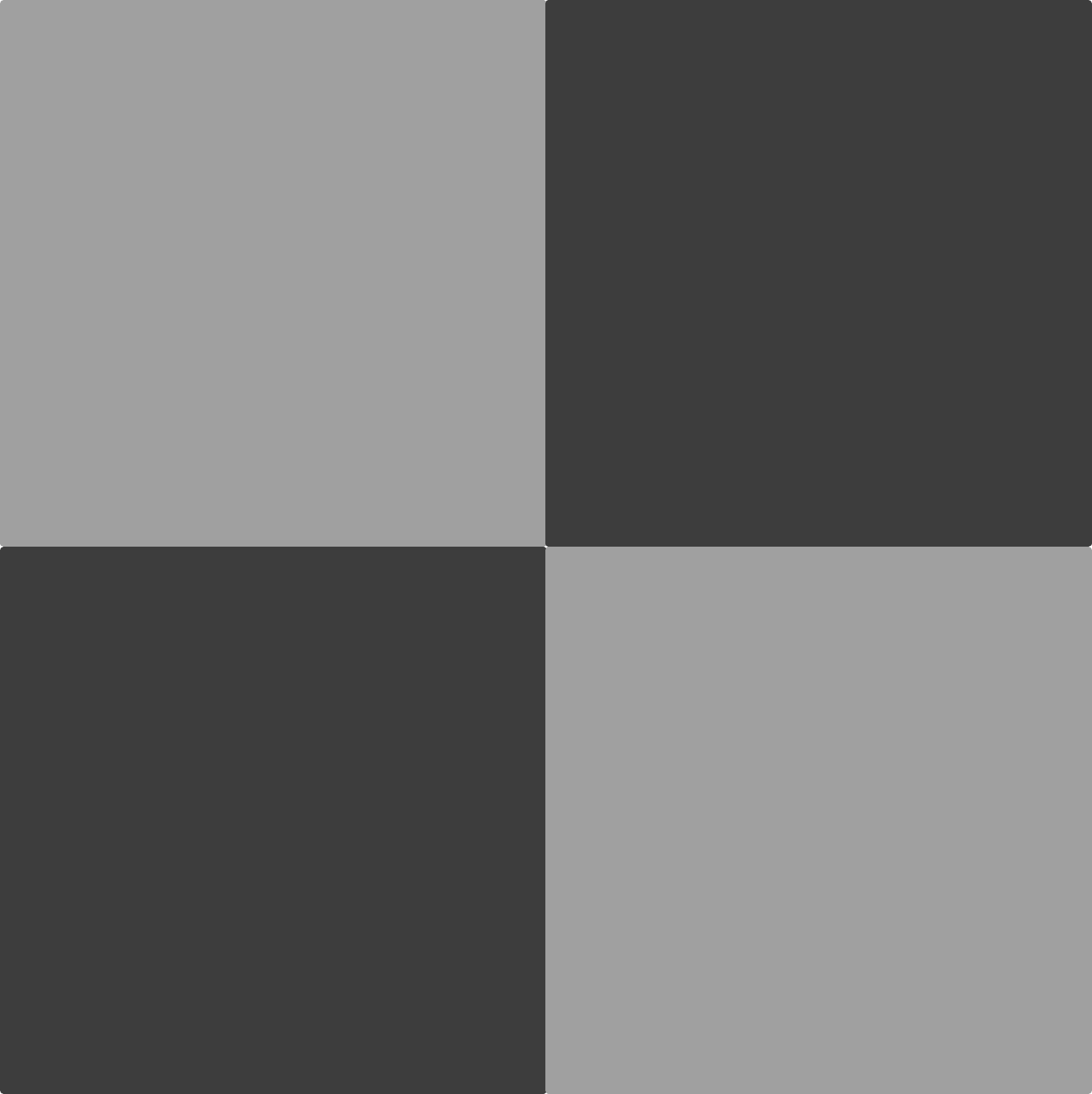}
\includegraphics[width=.23\linewidth]{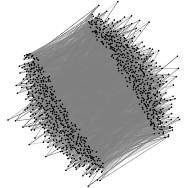}
}
\\
\hfill
\subfloat[Core-periphery structure]{
\label{fig:stylized-coreper}
\hspace*{-0.9cm}
\includegraphics[width=.23\linewidth]{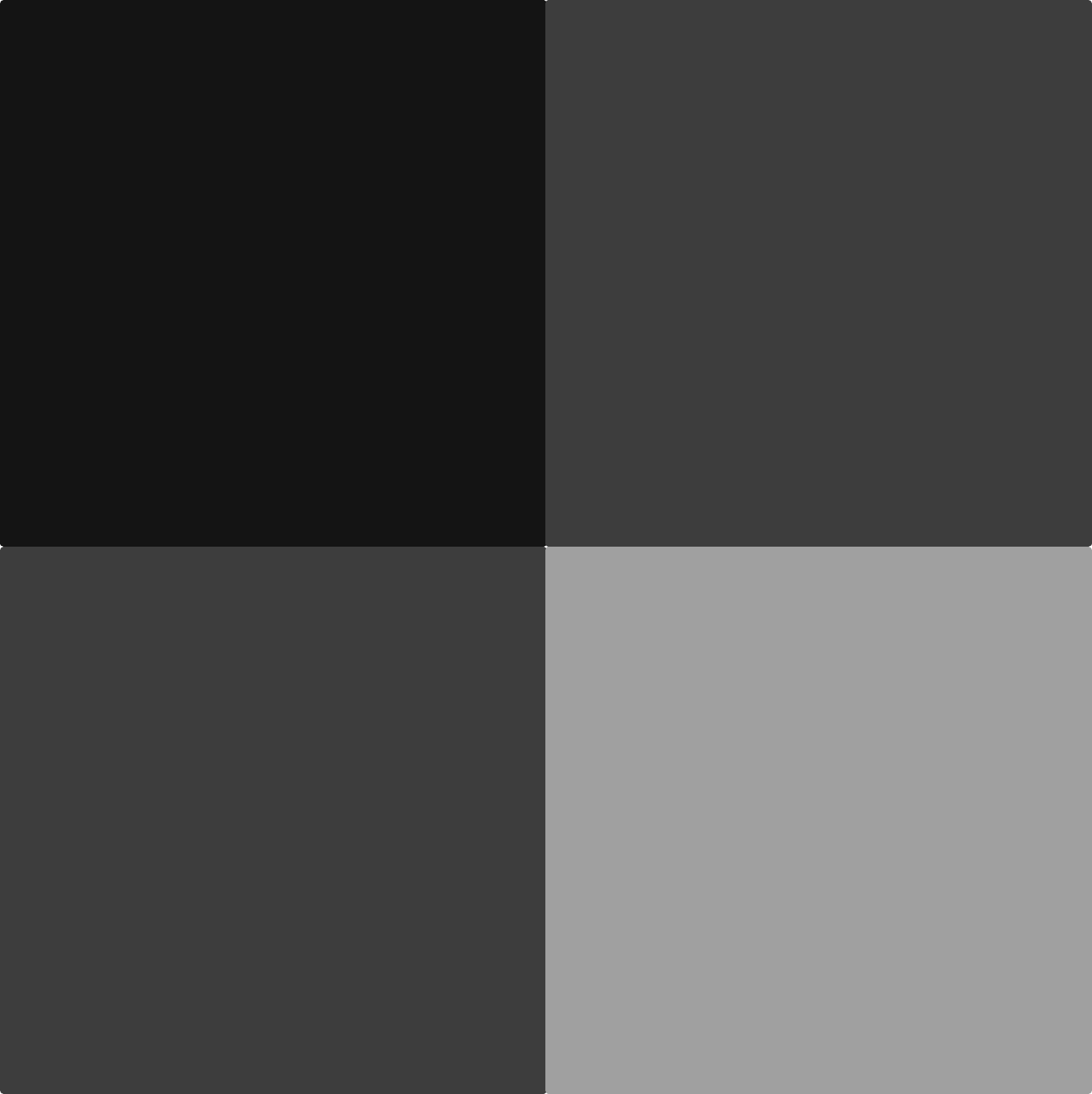}
\includegraphics[width=.23\linewidth]{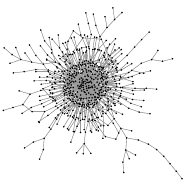}
}
\hfill
\subfloat[Random graph]{
\label{fig:stylized-random}
\includegraphics[width=.23\linewidth]{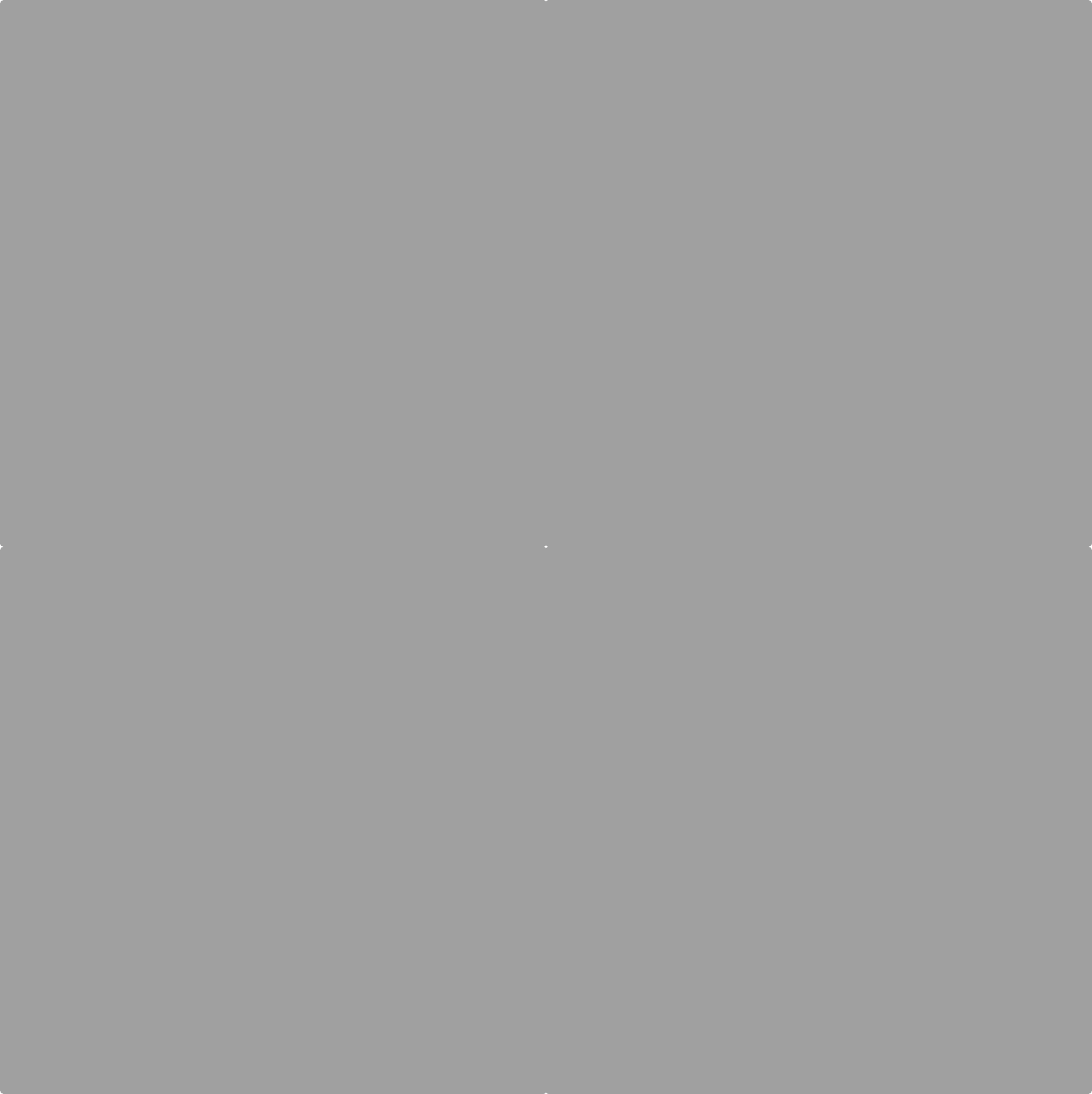}
\includegraphics[width=.23\linewidth]{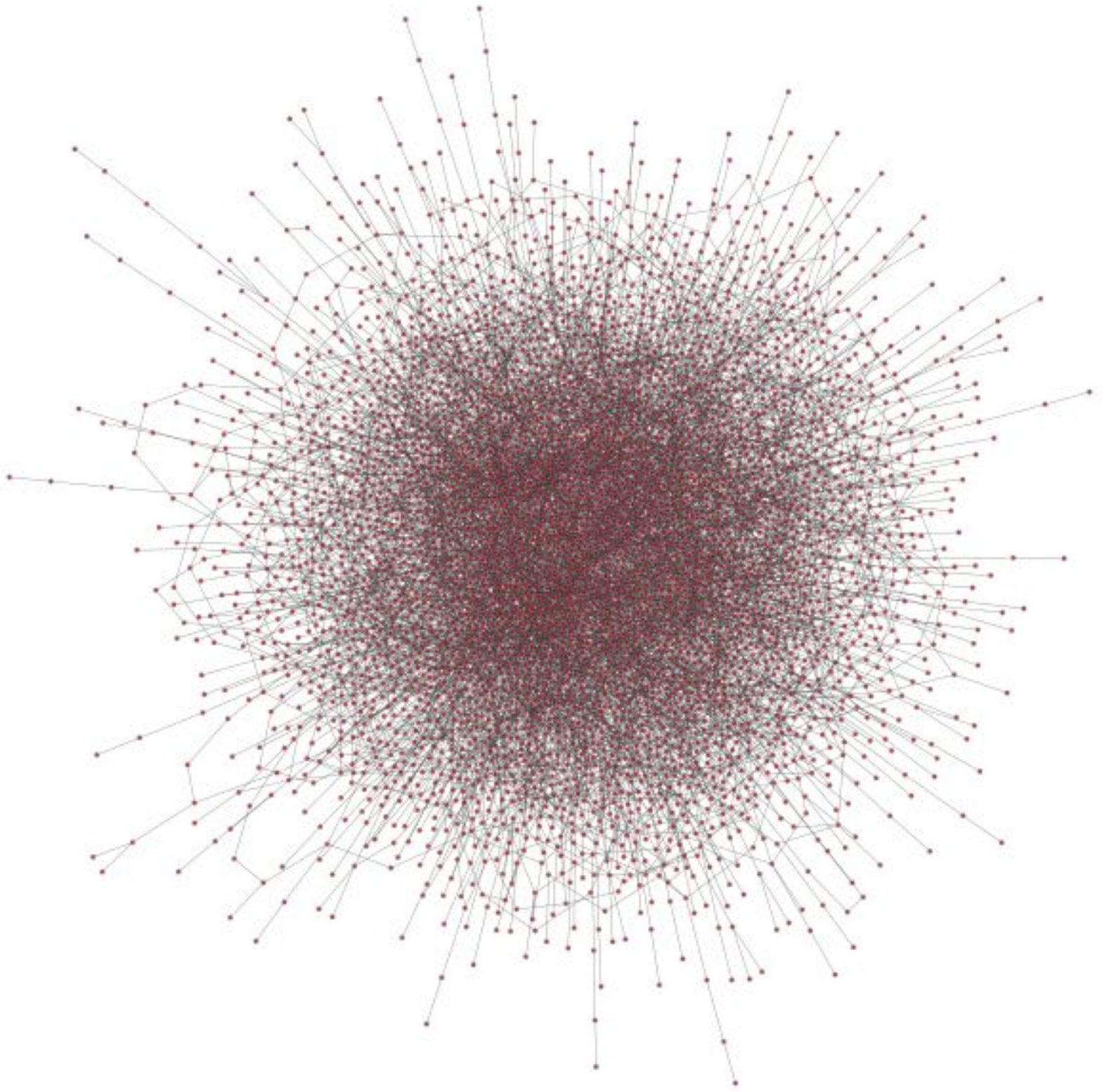}
}
\caption{Stochastic block model. We show the schematic adjacency matrices of network realisations produced by the model 
for special choices of the edge probabilities, along with one representative realisation for each case. For simplicity we show the case of two blocks of equal size.
Darker blocks indicate higher edge probabilities and consequently a larger density of edges inside the block. 
Figure~\ref{fig:stylized-hotdog} illustrates community (or assortative) structure: the probabilities (link densities) are much higher 
inside the diagonal blocks than elsewhere. Figure~\ref{fig:stylized-bipartite} shows the opposite situation (disassortative structure). 
Figure~\ref{fig:stylized-coreper} illustrates a core-periphery structure. Figure~\ref{fig:stylized-random} shows a random graph
{\`a} la Erd\H{o}s and R\'enyi: all edge probabilities are identical, inside
and between the blocks, so there are no actual groups.
Adapted figure with permission from~\cite{jeub15}. \copyright\,2015, by the American Physical Society.}
\label{figSBM}
\end{center}
\end{figure*}
For instance, if we take social networks, we can assume that the probability that two individuals know each other is a decreasing function
of their geographical distance, on average~\cite{liben05}. Each set of assumptions defines a model.
For our purposes, eligible models should take into account the
possible presence of groups of vertices, that behave similarly. 

The most famous model of networks with group structure is the {\it stochastic block model} (SBM)~\cite{fienberg81,holland83,snijders97}.
Suppose we have a network with $n$ vertices, divided in $q$ groups. The group of vertex $i$ is indicated with the 
integer label $g_i=1, 2, \dots, q$.
The idea of the model is very simple:
the probability $P(i\leftrightarrow j)$ that vertices $i$ and $j$ are connected depends exclusively on their group memberships: $P(i\leftrightarrow j)=p_{g_ig_j}$.
Therefore, it is identical for any $i$ and $j$ in the same groups. The probabilities $p_{g_ig_j}$ form a $q\times q$ symmetric matrix\footnote{For directed graphs, 
the matrix is in general asymmetric. The extension of the stochastic block model to directed graphs is straightforward. Here we focus on undirected graphs.}, called the {\it stochastic block matrix}.
The diagonal elements $p_{kk}$ ($k=1, 2, \dots, q$) of the stochastic block matrix are the probabilities that vertices of block $k$ are neighbours, whereas the off-diagonal elements 
give the edge probabilities between different blocks\footnote{In another definition of SBM the number of edges $e_{rs}$ 
between blocks $r$ and $s$ is fixed ($r, s=1, 2, \dots, q$), instead of the edge probabilities. If $e_{rs}\gg 1, \forall \,r,s=1, 2, \dots, q$ the two models are fully equivalent
if the edge probabilities $p_{rs}$ are chosen such that the expected number of edges running between $r$ and $s$ coincides with $e_{rs}$.}.

For $p_{kk}> p_{lm}, \forall k, l, m=1, 2, \dots, q$, with $l\neq m$, we recover community structure, as the probabilities that vertices of the same group are connected
exceed the probabilities that vertices of different groups are joined (Fig.~\ref{fig:stylized-hotdog}). 
It is also called {\it assortative structure}, as it privileges bonds between vertices of the same group. 
The model is very versatile, though, and can generate various types of group structure.
For $p_{kk} < p_{lm}, \forall k, l, m=1, 2, \dots, q$, with $l\neq m$, we have {\it disassortative structure}, 
as edges are more likely between the blocks than inside them (Fig.~\ref{fig:stylized-bipartite}).
In the special case in which $p_{kk}=0, \forall k=1, 2, \dots, q$ we recover {\it multipartite structure}, 
as there are edges only between the blocks.
If $q=2$, $p_{11} \gg p_{12} \gg p_{22}$, we have {\it core-periphery structure}: the vertices of the first block (core) are relatively
well-connected amongst themselves as well as to a peripheral set of vertices that interact very
little amongst themselves (Fig.~\ref{fig:stylized-coreper}).
If all probabilities are equal, $p_{ij}=p$, $\forall i,j$, we recover the classic random graph 
{\`a} la Erd\H{o}s and R\'enyi~\cite{erdos59,erdos60}
(Fig.~\ref{fig:stylized-random}).
Here any two vertices have identical probability of being connected, hence there is no group structure. 
This has become a fundamental axiom in community detection, and has inspired some popular techniques like, e. g., 
modularity optimisation~\cite{newman04b,newman04c} (Section~\ref{sec-modopt}). Random graphs of this type are also useful in the validation
of clustering algorithms (Section~\ref{art-bench}). 


Alternative community definitions are based on the interplay between network topology and dynamics.
Diffusion is the most used dynamics. Random walks are the simplest diffusion processes. 
A simple random walk is a path such that the vertex reached at step $t$ is a random neighbour of the vertex reached at step $t-1$.
A random walker would be spending a long time within communities, due to the 
supposedly low number of routes taking out of them~\cite{rosvall08,delvenne10,rosvall14}. The evolution of random walks does not depend solely on the number or density of edges, in general, 
but also on the structure and distribution of paths formed by consecutive edges, as paths are the routes that walkers can follow. 
This means that random walk dynamics relies on {\it higher-order structures} than simple edges, in general.
Such relationship is even more pronounced when
one considers Markov dynamics of second order or higher, in which the probability of reaching a vertex at step $t+1$ of the walk 
does not depend only on where the walker sits at step $t$, but also on where it was at step $t-1$ and possibly earlier~\cite{rosvall14,persson16}. 
Indeed, one could formulate the network clustering problem by focusing on higher order structures, like {\it motifs} (e. g., triangles)~\cite{arenas08,serrour11,benson16}.
The advantage is that one can preserve more complex features of the network and its communities, which typically get lost when one uses network models solely based on edge probabilities, like 
SBMs\footnote{Since edges are usually placed independently of each other in SBMs, higher order structures like triangles are usually
underrepresented in the model graphs with respect to the actual graph at study.}. The drawback is that calculations become more involved and lengthy.

Is a definition of community really necessary? Actually not, most techniques to detect communities in networks do not require a precise definition of community.
The problem can be attacked from many angles. For instance, one can 
remove the edges separating the clusters from each other, that can be identified via some particular feature~\cite{girvan02,radicchi04}.
But defining clusters beforehand is a useful starting point, that allows one to check the reliability of the final results.

\section{Validation}
\label{sec-valid}

In this section we will discuss the crucial issue of validation of clustering algorithms. Validation usually means 
checking how precisely algorithms can recover the communities in benchmark networks, whose community structure is known.
Benchmarks can be computer-generated, according to some model, or actual networks, whose group structure is supposed to be known
via non-topological features (metadata). The lack of a universal definition of communities makes the search for benchmarks rather arbitrary, in principle.
Nevertheless, the best known artificial benchmarks are based on the modern definition of clusters presented in Section~\ref{sec-MV}.

We shall present some popular artificial benchmarks and show that partition similarity measures have to be handled with care. 
We will see under which conditions communities are detectable by methods,
and expose the interplay between topological information and metadata. We will conclude by
presenting some recent results on signatures of community structure extracted from real networks.

\subsection{Artificial benchmarks}
\label{art-bench}

\begin{figure*}
\begin{center}
\includegraphics[width=\textwidth]{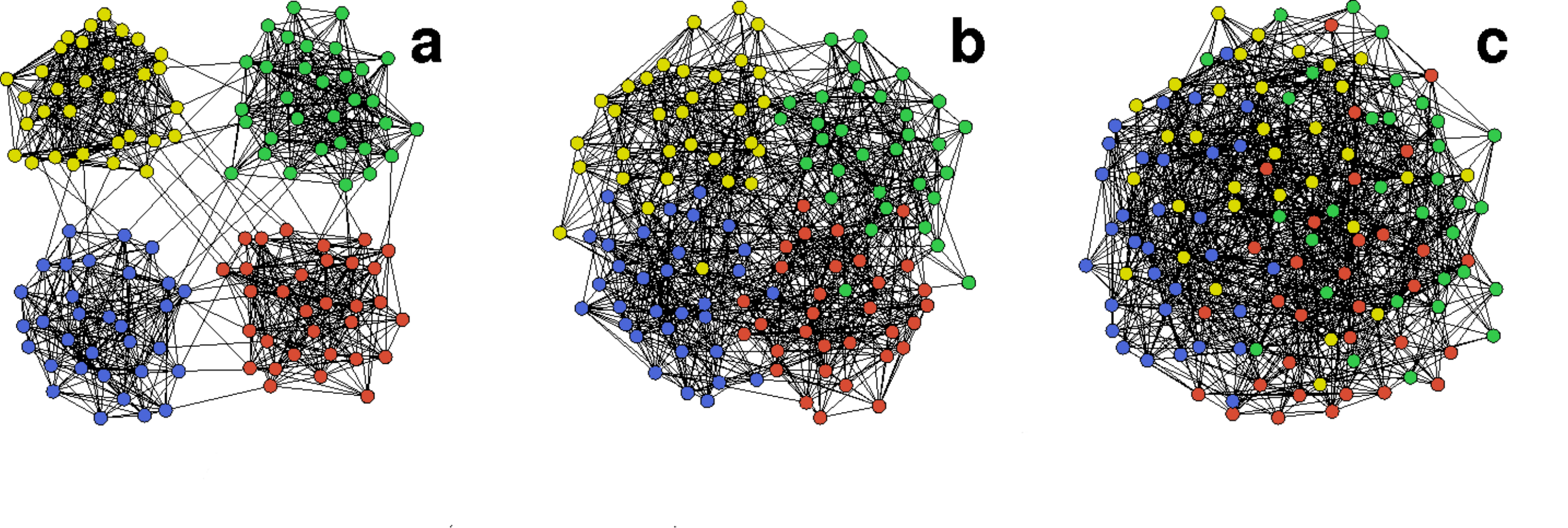}
\caption {Benchmark of Girvan and Newman. The three networks correspond 
to realisations of the model for $\langle k_{out}\rangle=1$ (a), $\langle k_{out}\rangle=5$ (b) and $\langle k_{out}\rangle=8$ (c). In (c) the four groups are hardly 
distinguishable by eye and methods fail to assign many vertices to their groups. 
Reprinted figure with permission from Ref.~\cite{guimera05}. \copyright\,2005, by the
Nature Publishing Group.}
\label{figGN}
\end{center}
\end{figure*}

The principle underneath stochastic block models (Section~\ref{sec-MV}) has inspired many popular benchmark graphs with group structure.
Community structure is recovered in the case in which the probability for two vertices to be joined is larger for vertices of the same group than 
for vertices of different groups (Fig.~\ref{fig:stylized-hotdog}). For simplicity, let us suppose that there are only two values of the edge probability,
$p_{in}$ and $p_{out}< p_{in}$, for edges within and between communities, respectively. Furthermore, we assume that all communities have identical 
size $n_c$, so $qn_c=n$, where $q$ is the number of communities.
In this version, the model coincides with the 
{\it planted l-partition model}, introduced in the context of graph partitioning\footnote{Graph partitioning means
dividing a graph in subgraphs, such to minimise the number of edges joining the subgraphs to each other.
It is related to community detection, as it aims at finding the minimum separation between the parts. However, it usually does not 
consider how cohesive the parts are (number or density of internal edges), except when special measures are used, 
like conductance [Eq.~(\ref{eq4})].}~\cite{bui87,dyer89,condon01}.
The expected internal and external degrees of a vertex are
$\langle k_{in}\rangle=p_{in}n_c$ and
$\langle k_{out}\rangle=p_{out}n_c(q-1)$, respectively, yielding an expected (total) vertex degree
$\langle k\rangle=\langle k_{in}\rangle+\langle k_{out}\rangle=p_{in}n_c+p_{out}n_c(q-1)$.

Girvan and Newman~\cite{girvan02} set $q=4$, $n_c=32$ (for a total number of vertices $n=128$)
and fixed the average total degree $\langle k\rangle$ to $16$. This implies that $p_{in}+3p_{out}=1/2$ and
$p_{in}$ and $p_{out}$ are not independent parameters. The benchmark by Girvan and Newman is still the most popular in the literature
(Fig.~\ref{figGN}).

Performance plots of clustering algorithms typically have, on the horizontal axis, the expected external degree $\langle k_{out}\rangle$.
For low values of $\langle k_{out}\rangle$ communities are well separated\footnote{They are also more cohesive internally, 
since $\langle k_{in}\rangle$ is higher, to keep 
the total degree constant.} and most algorithms do a good job at detecting them. By increasing $\langle k_{out}\rangle$, the performance declines.
Still, one expects to do better than by assigning memberships to the vertices at random, as long as $p_{in} > p_{out}$, which means for 
$\langle k_{out}\rangle< 12$. In Section~\ref{detectab} we will see that the actual threshold is lower, due to random fluctuations.

\begin{figure}
\begin{center}
\includegraphics[width=\columnwidth]{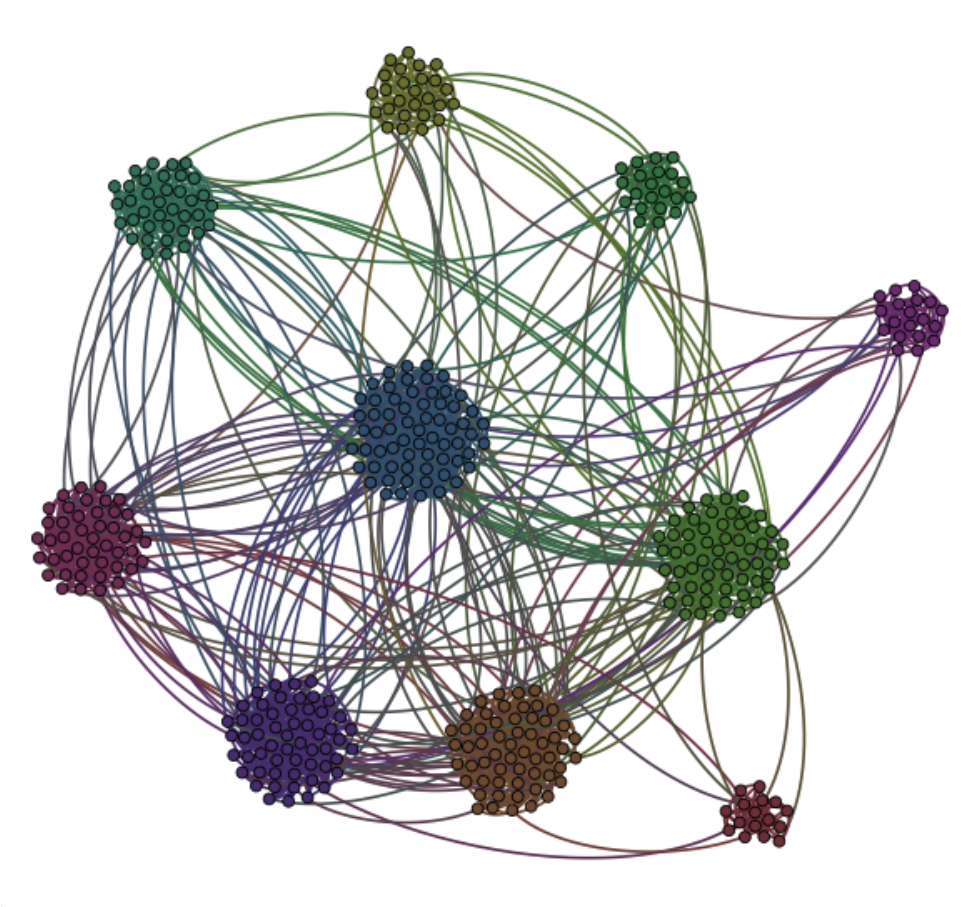}
\caption {LFR benchmark. Vertex degree and community size are power-law distributed, to account for the heterogeneity observed in
real networks with community structure.}
\label{figLFR}
\end{center}
\end{figure}

The benchmark by Girvan and Newman, however, is not a good proxy of real networks with community structure. For one thing,
all vertices have equal degree, whereas the degree distribution of real networks is usually highly heterogeneous~\cite{albert99}.
In addition, most clustering techniques find skewed distributions of community sizes~\cite{palla05,newman04, danon07, clauset04, radicchi04,lancichinetti10b}.
For this reason, Lancichinetti, Fortunato and Radicchi proposed the {\it LFR benchmark}, having power-law distributions
of degree and community size~\cite{lancichinetti08} (Fig.~\ref{figLFR}).

The mixing parameters $\mu_i$ of the vertices (Section~\ref{sec-var}) are set equal to a constant $\mu$, which estimates the 
quality of the partition\footnote{The parameter $\mu$ 
is actually only the average of the mixing parameter over all vertices. In fact, since degrees are integer, it is impossible to tune them such to have exactly
the same value of $\mu$ for each vertex, and keep the constraint on the degree distribution at the same time.}.
LFR benchmark networks are built by joining stubs at random, once 
one has established which stubs are internal and which ones are external with respect to the community of the vertex attached to the stubs. In this respect, it is 
basically a configuration model~\cite{bollobas80,molloy95} with built-in communities. 

Clearly, when $\mu$ is low, clusters are better separated from each other, and easier to detect. When $\mu$ grows, performance starts to decline.
But for which range of $\mu$ can we expect a performance better than random guessing? Let us suppose that 
the group structure is detectable for $\mu\in [0, \mu_c]$. The upper limit 
$\mu_c$ should be such that the network is random for $\mu=\mu_c$.  
The network is random when stubs are combined at random, without distinguishing between internal and external stubs, which yields the 
standard configuration model. There 
the expected number of edges between two vertices with degrees $k_i$ and $k_j$ is $k_ik_j/(2m)$, $m$ being the total number of 
network edges. Let us focus on a generic vertex $i$, belonging to community $C$. We denote with
$K_C$ and $\tilde{K}_C$ the sum of the degrees of the vertices inside and outside $C$, respectively. Clearly $K_C+\tilde{K}_C=2m$.
In a random graph built with the configuration model, vertex $i$ would have an expected internal degree\footnote{The approximation
is justified when the community is large enough that $K_C\gg k_i$.} $k_i^{int\mbox{-}rand} = k_i (K_C-k_i)/(2m)\approx k_i K_C/(2m)$ and 
an expected external degree $k_i^{ext\mbox{-}rand} = k_i \tilde{K}_C/(2m)$. Since, by construction, $k_i^{int}=(1-\mu)k_i$ and $k_i^{ext}=\mu k_i$, 
the community $C$ is not real when $k_i^{ext}=k_i^{ext\mbox{-}rand}$ and $k_i^{int}=k_i^{int\mbox{-}rand}$, which implies $\mu=\mu_C=\tilde{K}_C/(2m)=1-K_C/(2m)$.
We see that $K_C$ depends on the community $C$: the larger the community, the 
lower the threshold is. Therefore, not all clusters are detectable at the same time, in general. For this to happen, $\mu$ must be lower than
the minimum of $\mu_C$ over all communities: $\mu\leq \mu_c=\min_C \mu_C$. If communities are all much smaller than the network as a whole,
$K_C/(2m)\approx 0$ and $\mu_c$ could get very close to the upper limit $1$ of the variable $\mu$.
However, it is possible that the actual threshold is lower than 
$\mu_c$, due to the perturbations of the group structure induced by random fluctuations (Section~\ref{detectab}).
Anyway, in most cases the threshold is going to be quite a bit higher than $1/2$, the value which is mistakenly considered as the threshold by some scholars.

The LFR benchmark turns out to be a special version of the recently introduced 
degree-corrected stochastic block model~\cite{karrer11},
with the degree and the block size distributed according to truncated power laws\footnote{In fact, the correspondence is exact 
for a slightly different parametrisation of the benchmark, introduced in~\cite{peixoto14b}. In this version of the model, 
instead of the mixing parameter $\mu$, which is local, 
a global parameter $c$ is used, estimating how strong the community structure is.}. 

The LFR benchmark has been extended to directed and weighted networks with overlapping communities~\cite{lancichinetti09b}.
The extensions to directed and weighted graphs are rather straightforward. Overlaps are obtained by assigning each vertex to a 
number of clusters and distributing its internal edges equally among them\footnote{A better way to do it would be taking into account the size of the 
communities the vertex is assigned to, and divide the edges proportionally to the (total) degrees of the communities.}. 
Recently, another benchmark with overlapping communities has been introduced by Ball, Karrer and Newman~\cite{ball11}.
It consists of two clusters $A$ and $B$, with overlap $C$. Vertices in the non-overlapping subsets
$A-C$ and $B-C$ set edges only between each other, while 
vertices in $C$ are connected to vertices of both $A$ and $B$. The expected degree of all vertices is set equal to $\langle k \rangle$.
The authors considered various settings, by tuning $\langle k\rangle$, the size of the overlap and the sizes of $A$ and $B$, which may be uneven.
However, the fact that all vertices have equal degree (on average)
makes the model less realistic and flexible than the LFR benchmark.

\begin{figure}
\begin{center}
\includegraphics[width=\columnwidth]{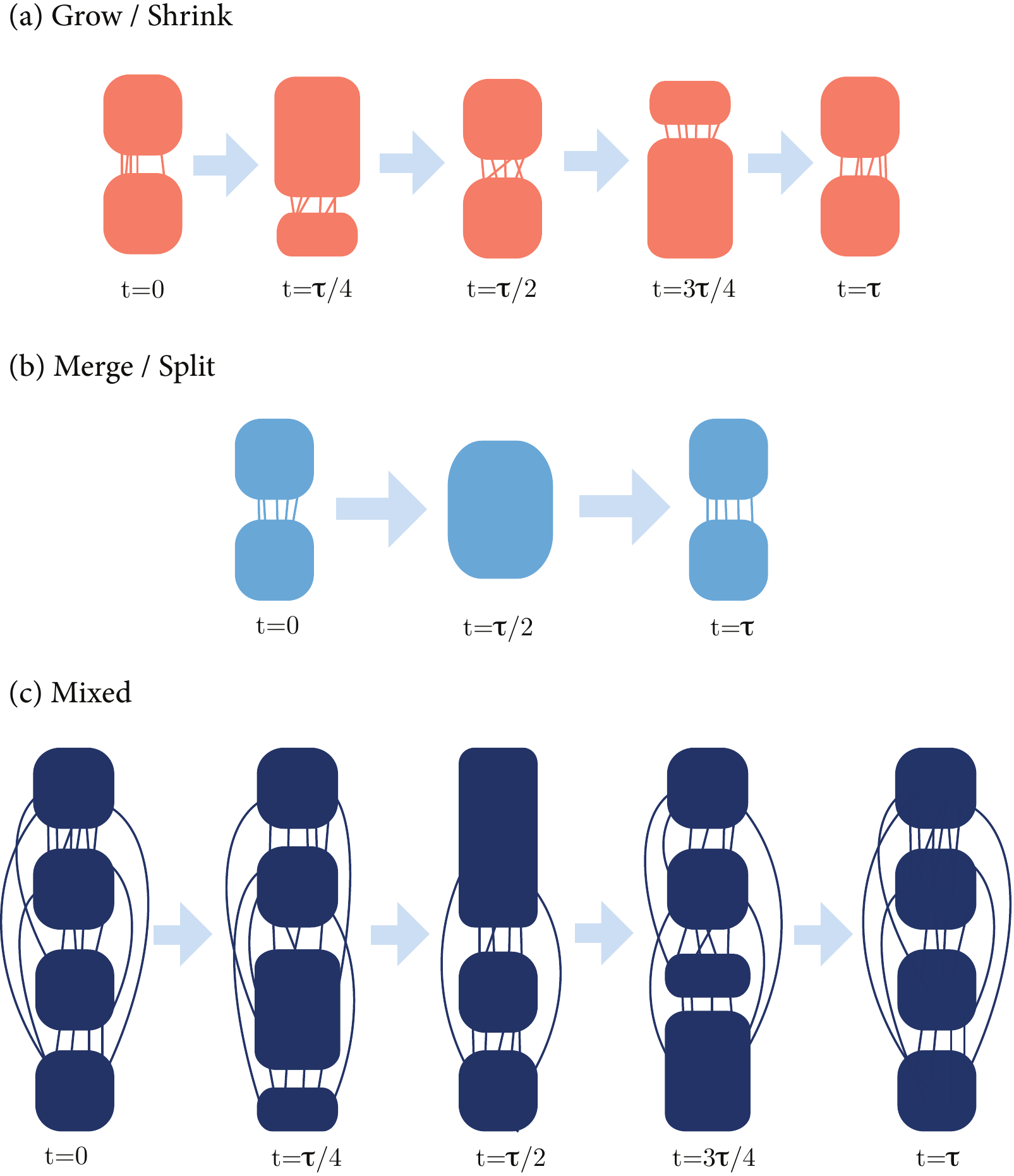}
\caption {Dynamic benchmark. 
(a) Grow-Shrink benchmark. Starting from two
communities of equal size, vertices move from one cluster to the other and back. 
(b) Merge-Split benchmark. It starts with
two communities, edges are added
until there is one community with uniform link density (merge), then
the process is reversed, leading to a fragmentation into two equal-sized clusters. (c) Mixed benchmark. There are four communities: the upper pair undergoes
the grow-shrink dynamics of (a), the lower pair the merge-split dynamics of (b). All processes are periodic with period $\tau$. 
Reprinted figure with permission from~\cite{granell15}. \copyright\,2015, by the American Physical Society.}
\label{figDB}
\end{center}
\end{figure}

Following the increasing availability of evolving
time-stamped network data sets, the analysis and modelling of temporal networks have received a lot of
attention lately~\cite{holme12}. In particular, scholars have started to deal with the problem of detecting evolving 
communities (Section~\ref{sec-dynclus}).
A benchmark designed to model dynamic communities was proposed by Granell et al.~\cite{granell15}. It is based on
the planted $l$-partition model, just like the benchmark of Girvan and Newman, 
where $p_{in}$ and $p_{out}<p_{in}$ are the edge probabilities within communities and between communities, respectively.
Communities may grow and shrink (Fig.~\ref{figDB}a), they may merge with each other or split into smaller clusters (Fig.~\ref{figDB}b),
or do all of the above (Fig.~\ref{figDB}c). The dynamics unfold such that at each time the subgraphs are proper communities in the probabilistic sense
discussed in Section~\ref{sec-MV}. In the merge-split dynamics, clusters actually merge before the inter-community edge probability $p_{out}$ reaches  
the value $p_{in}$ of the intra-community edge probability, due to random fluctuations (Section~\ref{detectab}).

In Section~\ref{sec-MV} we have shown why random graphs cannot have a meaningful group structure\footnote{Here we refer to random graphs where the edge probabilities
do not depend on their membership in groups. Examples are
Erd\H{o}s and R\'enyi random graphs, the configuration model, etc..}. 
That means that they can
be employed as {\it null benchmarks}, to test whether algorithms are capable to recognise the absence of groups.
Many methods find non-trivial communities in such random networks, so they fail the test. 
We strongly encourage doing this type of exam on new algorithms~\cite{lancichinetti09c}.
%
%

\subsection{Partition similarity measures}
\label{sim-me}

The accuracy of clustering techniques depends on their ability to detect the clusters of networks, whose community structure is known.
That means that the partition detected by the method(s) has to match closely the planted partition of the network. How 
can the similarity of partitions be computed? This is an important problem, with no unique solution.
In this section we discuss some issues about partition similarity measures. More information can be found in~\cite{meila07}, \cite{fortunato10} and \cite{traud11}.

Let us consider two partitions ${\cal X}=(X_1, X_2, ... , X_{q_X})$ and ${\cal Y}=(Y_1, Y_2, ... , Y_{q_Y})$ of a network $G$, 
with $q_X$ and $q_Y$ clusters, respectively. Let $n$ be the total number of vertices, 
$n_i^X$ and $n_j^Y$ the number of vertices in clusters $X_i$ and $Y_j$ and 
$n_{ij}$ the number of vertices shared by clusters $X_i$ and $Y_j$: $n_{ij}=|X_i\bigcap Y_j|$. The $q_X\times q_Y$ matrix 
$N_{{\cal XY}}$ whose entries are the overlaps $n_{ij}$ is called {\it confusion matrix}, {\it association matrix} or 
{\it contingency table}.

Most similarity measures can be divided in three categories: measures
based on {\it pair counting}, {\it cluster matching} and {\it information theory}.

Pair counting means computing the number of pairs of vertices which are classified in the same (different) clusters 
in the two partitions. Let $a_{11}$ indicate the number of pairs of vertices which are in the same community in both partitions, $a_{01}$
($a_{10}$) the number of pairs of elements which are in the same community in ${\cal X}$ (${\cal Y}$) and in
different communities in ${\cal Y}$ (${\cal X}$) and $a_{00}$ the number of pairs of vertices that are in different communities in both partitions. 
Several measures can be defined by combining the above numbers in various ways. A famous example is
the {\it Rand index}~\cite{rand71} 
\begin{equation}
R({\cal X},{\cal Y})=\frac{a_{11}+a_{00}}{a_{11}+a_{01}+a_{10}+a_{00}},
\label{eqt02}
\end{equation}
which is the ratio of the number of vertex pairs correctly classified in both partitions (i.~e. either in the same
or in different clusters), by the total number of pairs. 
Another notable option is the {\it Jaccard index}~\cite{ben01b}, 
\begin{equation}
J({\cal X},{\cal Y})=\frac{a_{11}}{a_{11}+a_{01}+a_{10}},
\label{eqt04}
\end{equation}
which is the ratio of the number of vertex pairs classified in the same cluster in both partitions, by the
number of vertex pairs classified in the same cluster in at least one partition. 
The Jaccard index varies over a broader range than the Rand index, 
due to the dominance of $a_{00}$ in $R({\cal X},{\cal Y})$, which typically confines the Rand index to a small interval slightly below $1$.
Both measures lie between $0$ and $1$.

If we denote with ${\cal X}_C$ and ${\cal Y}_C$
the sets of vertex pairs with are members of the same community in partitions ${\cal X}$ and ${\cal Y}$, respectively, the Jaccard index 
is just the ratio between the intersection and the union of ${\cal X}_C$ and ${\cal Y}_C$. Such concept can be used as well to determine the similarity 
between two clusters $A$ and $B$
\begin{equation}
J_{AB}=\frac{|A\bigcap B|}{|A\bigcup B|}.
\label{eqt20}
\end{equation}
The score $J_{AB}$ is also called Jaccard index and is the most general definition of the score, for any two sets $A$ and $B$~\cite{jaccard01}.
Measuring the similarity between communities is very important to determine, given different partitions, which cluster 
of a partition corresponds to which cluster(s) of the other(s). For instance,
the cluster $Y_j$ of ${\cal Y}$ corresponding to cluster $X_i$ of ${\cal Y}$ 
is the one maximising the similarity between $X_i$ and $Y_j$, e. g., $J_{X_iY_j}$. This strategy is also 
used to track down the evolution of communities in temporal networks~\cite{palla07,lancichinetti12}.

\begin{figure*}
\begin{center}
\includegraphics[width=\textwidth]{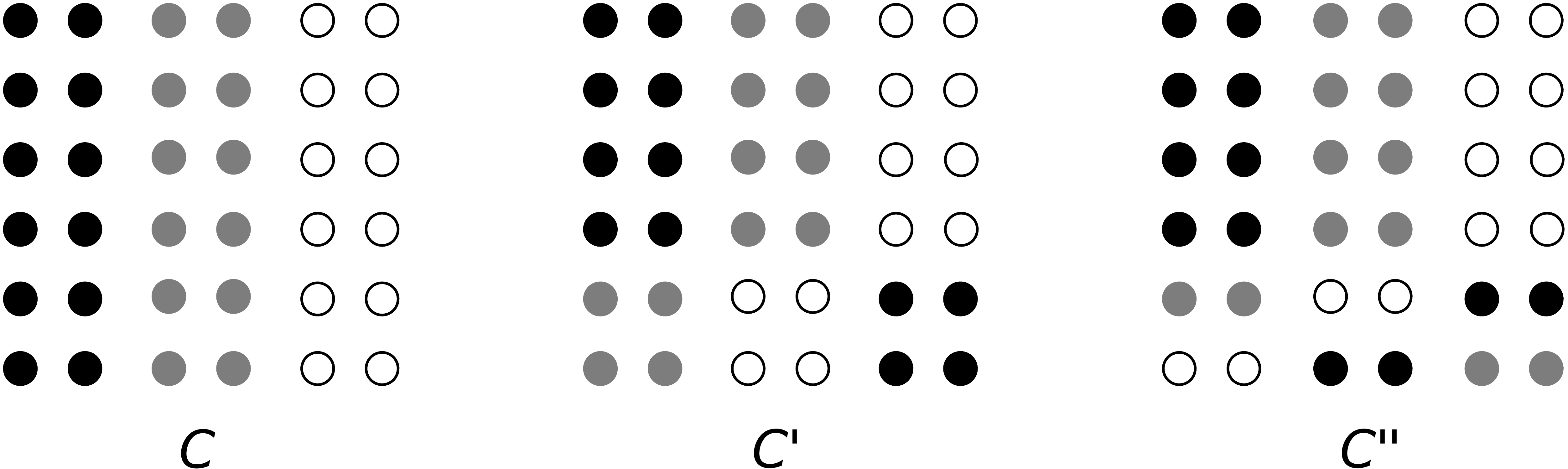}
\caption {Partition similarity measures based on cluster matching. There are three partitions in three clusters: ${\cal C}$, 
${\cal C}^{\prime}$ and ${\cal C}^{\prime\prime}$. The clusters include all elements of columns $1-2$, $3-4$ and $5-6$, which for 
${\cal C}$ are labelled in black, grey and white, respectively.
Partition ${\cal C}^\prime$ is obtained from ${\cal C}$ by reassigning the same fraction of elements from one cluster to the next, 
while ${\cal C}^{\prime\prime}$ is derived from ${\cal C}$ by reassigning the same fraction of elements from each cluster equally between the other clusters.
From cluster matching scores one concludes that ${\cal C}^{\prime}$ and ${\cal C}^{\prime\prime}$ are equally similar to ${\cal C}$, while intuition 
suggests that ${\cal C}^{\prime}$ is closer to ${\cal C}$ than ${\cal C}^{\prime\prime}$. Adapted figure with permission from \cite{meila07}. \copyright\,2007, by Elsevier.}
\label{figsimclus}
\end{center}
\end{figure*}

The Rand and the Jaccard indices, as defined in Eqs.~(\ref{eqt04}) and (\ref{eqt20}), have the disturbing
feature that they do not take values in the entire range $[0, 1]$. For this reason,
adjusted versions of both indices exist, in that a baseline is introduced, yielding the expected values of the score for 
all pairs of partitions $\tilde{\cal X}$ and $\tilde{\cal Y}$ 
obtained by randomly assigning vertices to clusters such that 
$\tilde{\cal X}$ and $\tilde{\cal Y}$ have the same number of clusters and the same size for all clusters of 
${\cal X}$ and ${\cal Y}$, respectively~\cite{hubert85}. The baseline is subtracted from the unadjusted version, 
and the result is divided by the range of this difference, yielding
$1$ for identical partitions and $0$ as expected value for independent partitions. 
But there are problems with these 
definitions as well. The null model used to compute the baseline relies on the assumption that the communities of the independent partitions 
have the same number of vertices as in the partitions whose similarity is to be compared. But such assumption usually does not hold, in practical instances, as
algorithms sometimes need the number of communities as input, but they never impose any constraint on the cluster sizes.
Adjusted indices have also the disadvantage of nonlocality~\cite{meila05}: the similarity between partitions differing only in one
region of the network depends on how the remainder of the network is subdivided.
Moreover, the adjusted scores can take negative values, when the unadjusted similarity lies below the baseline. 

A better option is to use {\it standardised indices}~\cite{brennan74}: for a given score $S_i$ the value of the null model term $\mu_i$ is computed along with its standard deviation 
$\sigma_i$ over many different randomisations of the partitions ${\cal X}$ and ${\cal Y}$. By computing the $z$-score
\begin{equation}
z_i=\frac{S_{i}-\mu_i}{\sigma_i},
\label{eq:zscoreind}
\end{equation}
we can see how non-random the measured similarity score is, and assess its significance. It can be shown that the $z$-scores for the Jaccard, Rand and Adjusted Rand
indices coincide~\cite{traud11}, so the measures are statistically equivalent. Since the actual values $S_i$ of these indices differ for the same pair of partitions, in general, we conclude
that the magnitudes of the scores may give a wrong perception about the effective similarity.

Cluster matching aims at establishing a correspondence between pairs of clusters of different partitions based on the
size of their overlap. A popular measure is the {\it fraction of correctly detected vertices}, introduced by
Girvan and Newman~\cite{girvan02}. A vertex is correctly classified if it is in the same 
cluster as at least half of the other vertices in its cluster in the planted partition.
If the detected partition has clusters given by the merger of two or more groups of the planted partition, all vertices of those clusters
are considered incorrectly classified. The number of correctly classified vertices is then divided by the number $n$ of vertices of the
graph, yielding a number between $0$ and $1$. The recipe to label vertices as 
correctly or incorrectly classified is somewhat arbitrary. The fraction of correctly detected vertices is similar to 
\begin{equation}
H({\cal X},{\cal Y})=\frac{1}{n}\sum_{k^\prime=\mbox{match}(k)}n_{kk^\prime},
\label{eqt05}
\end{equation}
where $k^\prime$ is the index of the best match $Y_{k^\prime}$ of cluster $X_k$~\cite{meila01}.
A common problem of this type of measures is that 
partitions whose clusters have the same overlap would have the same similarity, regardless of what
happens to the parts of the communities which are unmatched. The situation is illustrated schematically in Fig.~\ref{figsimclus}.
Partitions ${\cal C}^\prime$ and ${\cal C}^{\prime\prime}$ are obtained from ${\cal C}$ by reassigning the same fraction of their elements
to the other clusters. Their overlaps with ${\cal C}$ are identical and so are the corresponding similarity scores. However, 
in partition ${\cal C}^{\prime\prime}$ the unmatched parts of the clusters are more scrambled than in ${\cal C}^\prime$, which should be reflected in a
lower similarity score.

Similarity can be also estimated by computing, given a partition, the additional amount of information that one needs to have to 
infer the other partition. If partitions are similar, little information is needed to go from one to the other.
Such extra information can be used as a measure of dissimilarity. To evaluate the Shannon information content~\cite{mackay03} of a partition,
we start from the community assignments $\{x_i\}$ and $\{y_i\}$, where $x_i$ and $y_i$ indicate the cluster labels of 
vertex $i$ in partition ${\cal X}$ and ${\cal Y}$, respectively. The labels $x$ and $y$ are the values of two random
variables $X$ and $Y$, with joint distribution $P(x,y)=P(X=x,Y=y)=n_{xy}/n$, so that $P(x)=P(X=x)=n_x^X/n$ 
and $P(y)=P(Y=y)=n_y^Y/n$. 
The {\it mutual information} $I(X,Y)$ of two random variables is 
$I(X,Y)=H(X)-H(X|Y)$, where $H(X)=-\sum_xP(x)\log P(x)$ is the Shannon entropy of $X$ and $H(X|Y)=-\sum_{x,y}P(x,y)\log P(x|y)$
is the conditional entropy of $X$ given $Y$. 
The mutual information is not ideal as a similarity measure: for a given partition
${\cal X}$, all partitions derived from ${\cal X}$ by splitting (some of) its clusters would all have 
the same mutual information with ${\cal X}$, even though they could be very different from each other. In this case the mutual information
equals the entropy $H(X)$, because the conditional entropy is zero. It is then necessary to introduce an explicit dependence on the other partition, that persists
even in those special cases. This has been achieved by introducing the {\it normalized mutual information} (NMI), obtained by dividing the mutual information
by the arithmetic average\footnote{Strehl and Ghosh introduced an earlier definition of NMI, 
where the mutual information is divided by the geometric average of the entropies~\cite{strehl02}. Alternatively, one could normalise by the larger of the entropies
$H(X)$ and $H(Y)$~\cite{mcdaid11,esquivel12}.} 
of the entropies of ${\cal X}$ and ${\cal Y}$~\cite{fred03}
\begin{equation}
I_{norm}({\cal X}, {\cal Y})=\frac{2I(X,Y)}{H(X)+H(Y)}.
\label{eqt08}
\end{equation}
The NMI equals $1$ if and only if the partitions are identical, whereas it has an expected value of $0$ if they are independent.
Since the first thorough comparative analysis of clustering algorithms~\cite{danon05}, the NMI has been regularly used 
to compute the similarity of partitions in the literature. However, the measure is sensitive to the number of clusters $q_{\cal Y}$ of the detected partition,
and may attain larger values the larger $q_{\cal Y}$, even though more refined partitions are not necessarily closer to the planted one.
This may give wrong perceptions about the relative performance of algorithms~\cite{zhang15}.

A more promising measure, proposed by Meil\u{a}~\cite{meila07} is the {\it variation of information} (VI)
\begin{equation}
V({\cal X}, {\cal Y})=H(X|Y)+H(Y|X).
\label{eqt09}
\end{equation}
The VI defines
a metric in the space of partitions as it has the properties of distance (non-negativity, symmetry and triangle inequality). 
It is a local measure: the VI of
partitions differing only in a small portion of a graph depends on the differences of the clusters in that region, and not 
on how the rest of the graph is subdivided. The maximum value of the VI is $\log n$, 
which implies that the scores of an algorithm on graphs of different sizes cannot be compared with each other, in principle. 
One could divide $V({\cal X}, {\cal Y})$
by $\log n$~\cite{karrer08}, to force the score to be in the range $[0, 1]$, but the actual span of values of the measure
depends on the number of clusters of the partitions. In fact, 
if the maximum number of communities is $q^\star$, with $q^\star \leq \sqrt{n}$,
$V({\cal X}, {\cal Y}) \leq 2\log q^\star$. Consequently, in those cases where it is reasonable to set an upper bound 
on the number of clusters of the partitions, the similarities between 
planted and detected partitions on different graphs become comparable, and it is possible to assess both the 
performance of an algorithm and to compare algorithms across different benchmark graphs. 
We stress, however, that the measure may not be suitable when the partitions to be compared are very dissimilar from each other~\cite{traud11}
and that it shows unintuitive behaviour in particular instances~\cite{delling06b}.

So far we discussed of comparing partitions. What about covers? Extensions of the measures we have presented 
to the case of overlapping communities are usually not straightforward. The {\it Omega index}~\cite{collins88} is an extension of the 
Adjusted Rand index~\cite{hubert85}. Let ${\cal X}$ and ${\cal Y}$ be covers of the same graph to be compared. We denote with
$a_{jj}$ the number of pairs of vertices occurring together in exactly $j$ communities in both covers. 
It is a natural generalisation of the variables $a_{00}$ and $a_{11}$ we have seen above, where $j$ can also be 
larger than $1$ since a pair of vertices can now belong simultaneously to multiple communities. The variable 
\begin{equation}
o({\cal X}, {\cal Y})=\frac{2}{n(n-1)}\sum_{j}a_{jj}
\label{eq_omega}
\end{equation}
is the fraction of pairs of vertices belonging to the same number of communities in both covers (including the case $j=0$, which refers to the pairs 
not being in any community together). The Omega index is defined as
\begin{equation}
\Omega({\cal X}, {\cal Y})=\frac{o({\cal X}, {\cal Y})-o_e({\cal X}, {\cal Y})}{1-o_e({\cal X}, {\cal Y})},
\label{eq_omega1}
\end{equation}
where $o_e({\cal X}, {\cal Y})$ is the expected value of $o({\cal X}, {\cal Y})$ according to the null model discussed earlier, in which vertex labels are
randomly reshuffled such to generate covers with the same number and size of the communities.

The NMI has also been extended to covers by Lancichinetti, Fortunato and Kert\'esz~\cite{lancichinetti09}. The definition is non-trivial:
the community assignments of a cover
are expressed by a vectorial random variable, as each vertex may belong to multiple clusters at the same time. 
The measure overestimates the similarity of two covers, in special situations, where intuition suggests much lower values.
The problem can be solved by using an alternative normalisation, as shown in~\cite{mcdaid11}.
Unfortunately neither the definition by Lancichinetti, Fortunato and Kert\'esz nor the one by McDaid, Greene and Hurley
are proper extensions
of the NMI, as they do not coincide with the classic definition of Eq.~(\ref{eqt08}) when partitions in non-overlapping clusters are compared.
However, the differences are typically small, and one can rely on them in practice. Esquivel and Rosvall have proposed an actual extension~\cite{esquivel12}. Following the comparative 
analysis performed in~\cite{lancichinetti09c}, the NMI by Lancichinetti, Fortunato and Kert\'esz has been regularly used in the literature, also in the case of 
regular partitions, without overlapping communities\footnote{Recently the NMI has been extended to handle the comparison of hierarchical partitions as well~\cite{perotti15}.}.

If covers are fuzzy (Section~\ref{sec-defs}), the similarity measures above cannot be used, as they do not take into account the degree of membership
of vertices in the communities they belong to. A suitable option is the {\it Fuzzy Rand index}~\cite{hullermeier09}, which is an extension of the Adjusted Rand index.
Both the Fuzzy Rand index and the Omega index coincide with the Adjusted Rand index when communities do not overlap.

For temporal networks, a na\"{\i}ve approach would be comparing partitions (or covers) corresponding 
to configurations of the system in the same time window, and to see how this score varies across different time windows.
However, this does not tell if the clusters are evolving in the same way, as there would be no connection between 
clusterings at different times. A sensible approach is comparing sequences of clusterings, by building a confusion matrix that 
takes into account multiple snapshots. This strategy allows one to define dynamic versions of various indices, like the NMI and the VI~\cite{granell15}.

In conclusion, while there is no clear-cut criterion to establish which similarity measure is best, we recommend to use measures 
based on information theory. In particular, the VI seems to have more potential than others, for the reasons we explained, modulo the caveats in Refs.~\cite{delling06b,traud11}.
There are currently no extensions of the VI to handle the comparison of covers, but it would not be difficult to engineer one, 
e. g., by following a similar procedure as in~\cite{lancichinetti09,mcdaid11}, though this might cost the sacrifice of some of its nice features.

\begin{figure}[h!]
\includegraphics[width=0.49\columnwidth]{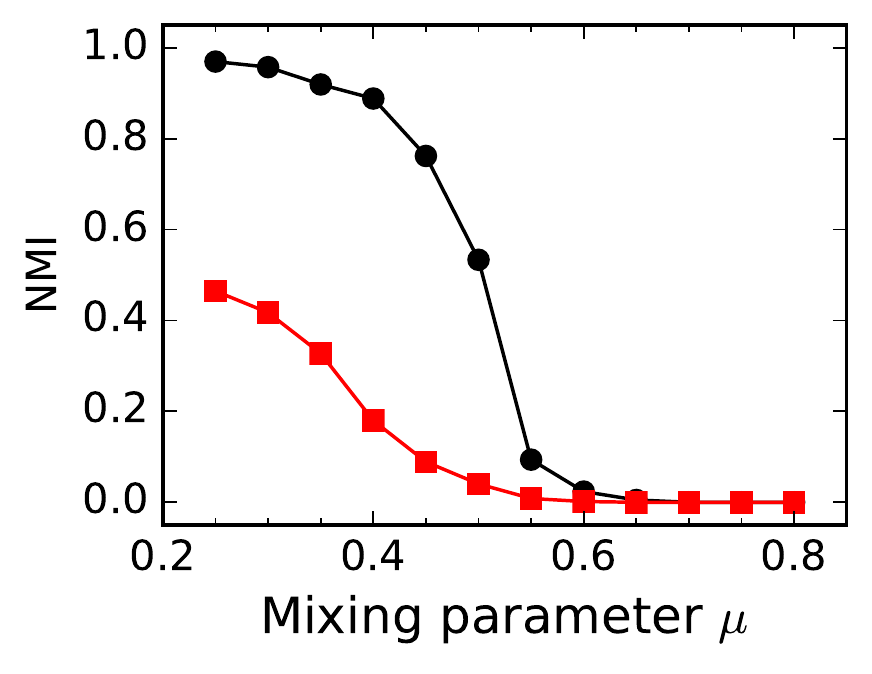}
\includegraphics[width=0.49\columnwidth]{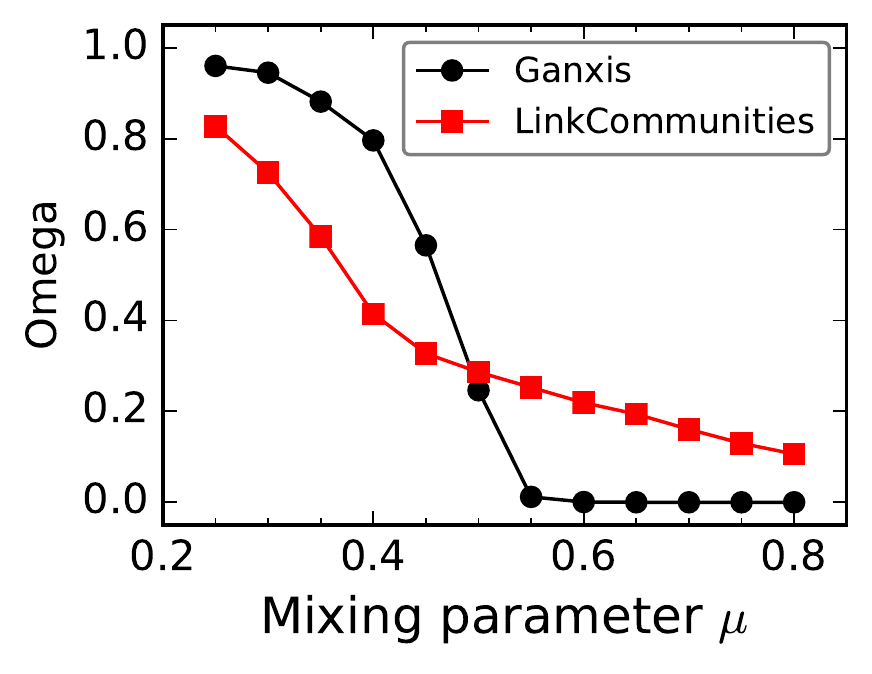}
\caption{Importance of choice of partition similarity measures. The plots show the comparison between the planted partition of LFR benchmark graphs and the
ones found by two algorithms: {\it Ganxis} and {\it LinkCommunities}. In the left diagram similarity was computed with the NMI, in the right one with the Omega index. 
The performances of the algorithms appear much closer 
when the Omega index is used. The LFR benchmark graphs used in the analysis have $1\,000$ vertices, average degree $15$, maximum degree $50$, 
exponents $2$ and $1$ for the degree
and community size distributions and range $[10, 50]$ for the community size.}
\label{figcompsim}
\end{figure}

One should keep in mind that the choice of one similarity index or another is a sensitive one, and warped conclusions may be drawn
when different measures are adopted. In Fig.~\ref{figcompsim} we show the accuracy of two algorithms on the LFR benchmark
(Section~\ref{art-bench}): {\it Ganxis}, a method based on label propagation~\cite{xie12} and {\it LinkCommunities}, a method
based on grouping edges instead of vertices~\cite{ahn10} (Section~\ref{sec-LC}). The accuracy is estimated with the NMI by Lancichinetti, Fortunato and Kert\'esz~\cite{lancichinetti09}
(left diagram) and with the Omega index [Eq.~(\ref{eq_omega1})] (right diagram). From the left plot one would think that Ganxis clearly outperforms 
LinkCommunities, whereas from the right plot Ganxis still prevails for $\mu$ until about $0.5$ (though the curves are closer to each other 
than in the NMI plot) 
and LinkCommunities is better for larger values of $\mu$.

\subsection{Detectability}
\label{detectab}


In validation procedures one assumes that, if the network has clusters, there must be a way to identify them. Therefore, if we do not manage, we have to blame the 
specific clustering method(s) adopted.
But are we certain that clusters are always detectable?

Most networks of interest are {\it sparse}, i. e., their average degree is much smaller than the 
number of vertices. This means that the number of edges of the graph is much smaller than the number of possible edges $n(n-1)/2$.
A more precise way to formulate this is by saying that a graph is sparse when, in the limit of infinite size, the average degree of the graph remains finite.
A number of analytical calculations can be carried out by using network sparsity. Many algorithms for community detection only work on sparse graphs.

On the other hand, sparsity can also give troubles. Due to the very low density of edges, small amounts of noise could perturb considerably the structure of the system.
For instance, random fluctuations in sparse graphs could trick algorithms into finding groups that do not really exist (Section~\ref{sec-sign}).
Likewise, they could make actual groups undetectable. Let us consider the simplest version of the assortative stochastic block 
model, which matches the planted partition model (Section~\ref{art-bench}). There are $q$ communities of the same size $n/q$, 
and only two values for the edge probability: $p_{in}$ for pairs of vertices in the same group and
$p_{out}$ for pairs of vertices in different groups. Since the graphs are sparse, $p_{in}$ and $p_{out}$ vanish in the limit of infinite graph size.
So we shall use the expected internal and external degrees $\langle k_{in}\rangle=np_{in}/q$ and $\langle k_{out}\rangle=np_{out}(q-1)/q$, 
which stay constant in that limit.
By construction, the groups are communities so long as $p_{in}>p_{out}$ or, equivalently, for $\langle k_{in}\rangle>\langle k_{out}\rangle/(q-1)$.
But that does not mean that they are always detectable.

In principle, dealing with the issue of detectability involves examining all conceivable
clustering techniques, which is clearly impossible. Fortunately, it is not necessary, because 
we know what model has generated the communities of the graphs we are considering. 
The most effective technique to infer the groups is then fitting the stochastic block model on the data ({\it a posteriori block modelling}).  
This can be done via the maximum likelihood method~\cite{gelman14}. 
In recent work~\cite{decelle11}, Decelle et al. have shown that, in the limit of infinite graph size,  
the partition obtained this way is correlated with the planted partition whenever
\begin{equation}
\langle k_{in}\rangle-\frac{\langle k_{out}\rangle}{q-1} > \sqrt{\langle k_{in}\rangle+\langle k_{out}\rangle},
\label{eqdetect}
\end{equation}
which implies
\begin{equation}
\langle k_{in}\rangle > \frac{\langle k_{out}\rangle}{q-1}+\frac{1}{2}\left(1+\sqrt{1+\frac{4q\langle k_{out}\rangle}{q-1}}\right).
\label{eqdetect1}
\end{equation}
So, given a value of $\langle k_{out}\rangle$, when $\langle k_{in}\rangle$ is in the range 
$\left[\frac{\langle k_{out}\rangle}{q-1},  \frac{\langle k_{out}\rangle}{q-1}+\frac{1}{2}\left(1+\sqrt{1+\frac{4q\langle k_{out}\rangle}{q-1}}\right)\right]$
the probability $p_c$ of classifying a vertex correctly is not larger than the probability $1/q$ of 
assigning the vertex to a randomly chosen group, although 
the groups are communities, according to the model. We stress that this result only holds when the graphs are sparse: if 
$p_{in}$ and $p_{out}$ remain non-zero in the large-n limit (dense graph), the classic detectability threshold $p_{in} > p_{out}$ 
is correct.

A fortiori, no clustering technique can detect the clusters better than random assignment 
when the inference of the model parameters fails to do so. 
If communities are searched via the spectral optimisation of Newman-Girvan's modularity~\cite{newman06},
one obtains the same threshold of Eq.~(\ref{eqdetect1})~\cite{nadakuditi12}, provided the network is not too sparse.

\begin{figure}
\begin{center}
\includegraphics[width=\columnwidth]{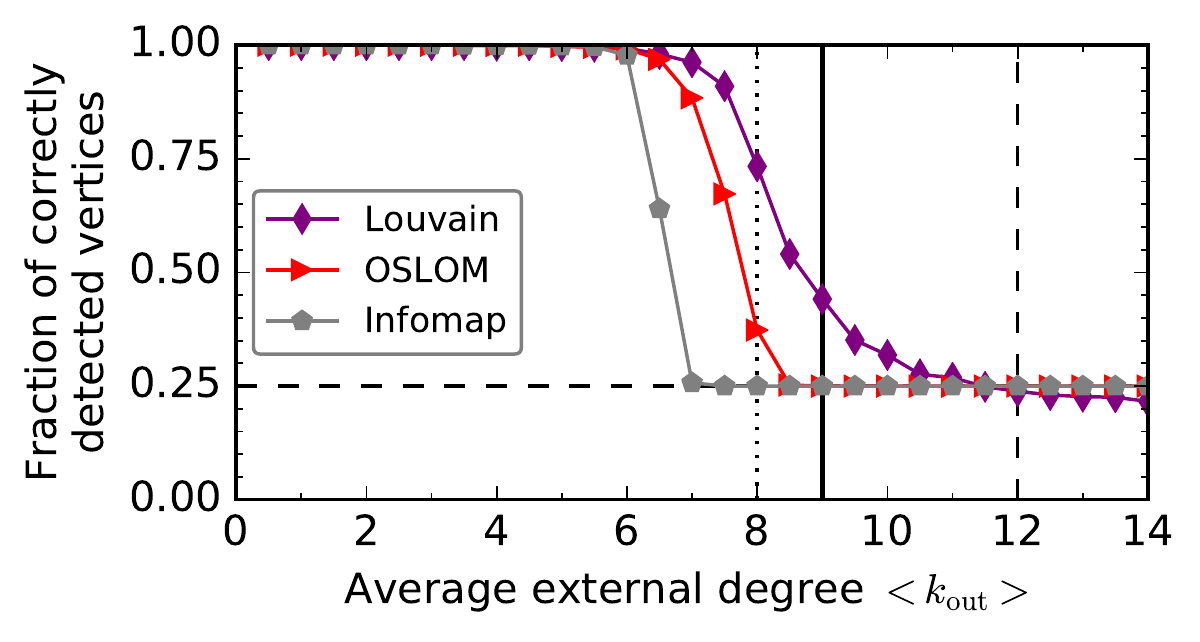}
\caption {Detectability of communities. Performances of three popular algorithms on the benchmark by Girvan and Newman. The dotted vertical line at $\langle k_{out}\rangle=8$ 
indicates the threshold corresponding to the concept of strong community {\'a} la Radicchi et al., the dashed line at $\langle k_{out}\rangle=12$ the threshold according to the  
probability-based definition of strong community we have given in Section~\ref{sec-MV}. The baseline of random assignment is $1/4$ (horizontal dashed line). All algorithms do not do better than random assignment already before $\langle k_{out}\rangle=12$. The theoretical detectability limit is at $\langle k_{out}\rangle=9$, in the limit of groups of infinite size.}
\label{figdet}
\end{center}
\end{figure}

For the benchmark of Girvan and Newman (Section~\ref{art-bench})~\cite{girvan02} it has long been unclear
where the actual detectability limit sits. Girvan-Newman benchmark graphs are not infinite, their size being set to $128$, so there is no proper detectability transition, but rather a smooth 
crossover from a regime in which clusters are frequently detectable to a regime where they are frequently undetectable. For this reason there cannot be a sharp threshold separating the two regimes.
Still it is useful to have an idea of where the pattern changes. In the following we shall still use the term threshold to refer to the crossover point.
In the beginning, scholars thought that clusters are detectable as long as they satisfy the definition of strong community by Radicchi et al.~\cite{radicchi04} (Section~\ref{sec-defs}), i. e., 
as long as the expected internal degree exceeds the expected external degree, yielding a threshold $\langle k_{in} \rangle = \langle k_{out}\rangle$~\cite{girvan02}. Since
the expected total degree of a vertex is set to $16$, communities are detectable as long as $\langle k_{out}\rangle < k_d^{strong}=8$.
It soon became obvious that the actual threshold should be the one of the ``modern" definition of community we have presented in Section~\ref{sec-MV}, according to which the condition\footnote{In the setting of the Girvan-Newman benchmark, where edge probabilities are identical for all vertices, the strong and weak definitions 
we presented in Section~\ref{sec-MV} coincide.} is $p_{in} > p_{out}$, that is $\langle k_{out}\rangle< k_d^{standard}=12$. However, numerical calculations reveal that algorithms tend to fail
long before that limit. From Eq.~(\ref{eqdetect1}) we see that for the case of four infinite clusters and total expected degree $\langle k_{tot} \rangle=16$, 
the theoretical detectability limit is $k_d^{theor}=9$. 
In Fig.~\ref{figdet} we see the performance on the benchmark of three well-known algorithms:
Louvain~\cite{blondel08}, a greedy optimisation technique of Newman--Girvan modularity~\cite{newman04b} (Section~\ref{sec-modopt}); Infomap, which is 
based on random walk dynamics~\cite{rosvall08} (Section~\ref{sec-dynmet}); OSLOM, that searches for clusters via a local optimisation 
of a significance score~\cite{lancichinetti11}. The accuracy is estimated via the fraction of correctly detected vertices
(Section~\ref{sim-me}). The three thresholds $k_d^{strong}$, $k_d^{standard}$ and $k_d^{theor}$ are represented by vertical lines.
The performance of all methods becomes comparable with random assignment well before $k_d^{standard}$. The theoretical 
limit $k_d^{theor}$ appears to be compatible with the performance curves.

Graph sparsity is a necessary condition for clusters to become undetectable, but it is not sufficient. 
The symmetry of the model we have considered plays a major role too.
Clusters have equal size and vertices have equal degree. This helps to ``confuse" algorithms. If communities have unequal sizes
and the degree of vertices are correlated with the size of their communities, so that vertices have larger degree, the bigger their clusters, 
community detection becomes easier, as the degrees can be used as proxy for group membership. In this case, 
the non-trivial detectability limit disappears when there are four clusters or fewer, while it persists up to a given extent of group size
inequality when there are more than four clusters~\cite{zhang16}. Other types of block structure, like core-periphery, do not suffer from detectability issues~\cite{zhang15b}.

LFR benchmark graphs are more complex models than the one studied in~\cite{zhang16} and it is not clear 
whether there is a non-trivial detectability limit, though it is unlikely, due to the big heterogeneity in the distribution of 
vertex degree and community size.

\subsection{Structure versus metadata}
\label{topmet}

Another standard way to test clustering techniques is using real networks with known community structure. 
Knowledge about the memberships of the vertices typically comes from {\it metadata}, i. e., non-structural information.
If vertices are annotated communities 
are assumed to be groups of vertices with identical tags. Examples are user groups in social networks like 
LiveJournal and product categories for co-purchasing networks of products of 
online retailers such as Amazon. 

\begin{figure}[h!]
\begin{center}
\includegraphics[width=\columnwidth]{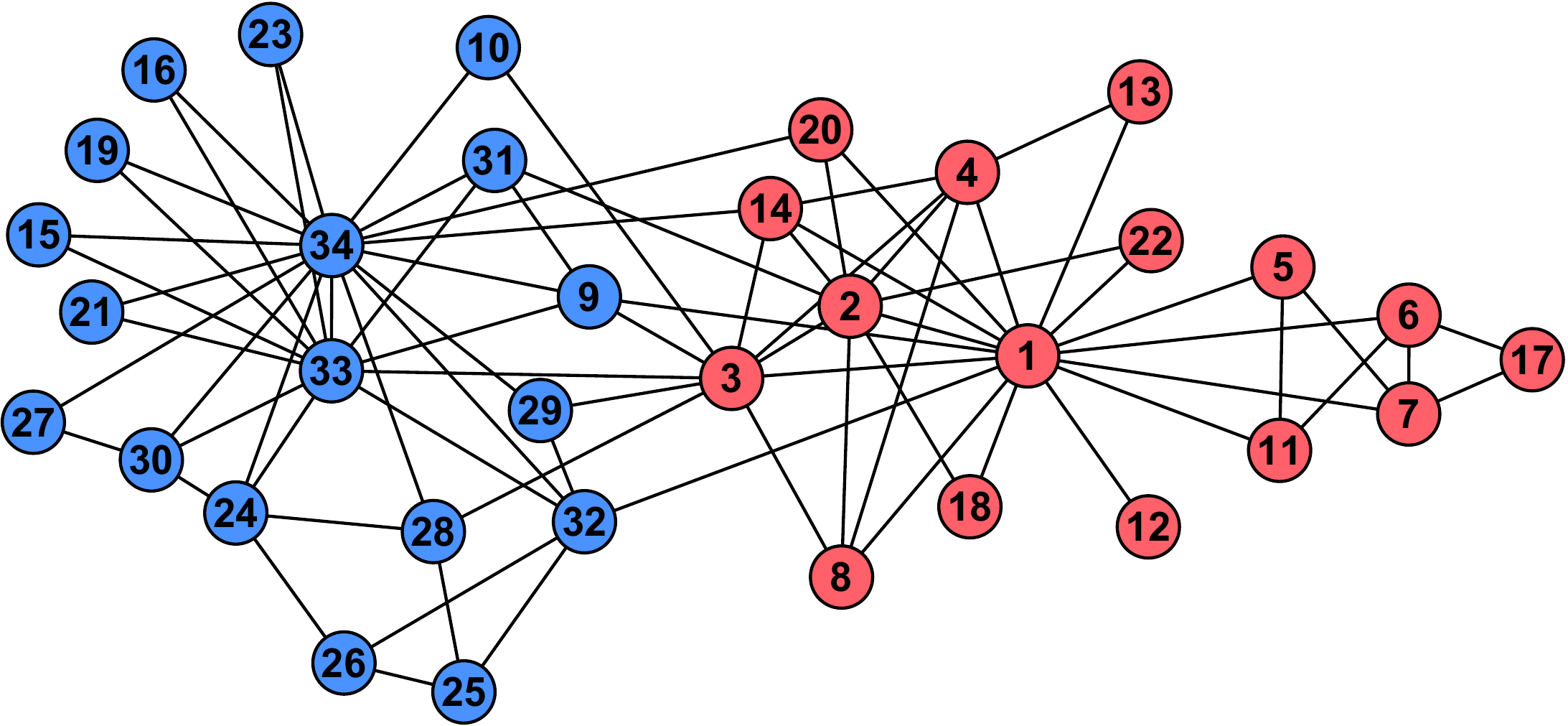}
\caption{Zachary karate club network. Symbols of different colours indicate the two groups generated by the fission of the network, following the disagreement
between the club's instructor (vertex 1) and its president (34).}
\label{zach}
\end{center}
\end{figure}

In Fig.~\ref{zach} we show the most popular of such benchmark graphs, Zachary karate club network~\cite{zachary77}. 
It consists of $34$ vertices, the
members of a karate club in the United States, who were
observed over a period of three years. Edges connect
individuals interacting outside the activities
of the club. Eventually a conflict between
the club president (vertex $34$) and the instructor (vertex $1$) led to the fission of
the club in two separate groups, whose members supported the instructor
and the president, respectively (indicated by the colours). 
Indeed, the groups make sense topologically: vertices $1$ and $34$ are hubs, and most members are directly 
connected to either of them.

Most algorithms of community detection have been tested on this network, as well as others, e. g., the American college football network~\cite{girvan02,evans10} or
Lusseau's network of bottlenose dolphins~\cite{lusseau03}.
The idea is that the method doing the best
job at recovering groups with identical annotations would also be the most reliable in applications.

Such idea, however, is based on a questionable principle, i. e., that the groups corresponding to the metadata are also communities in the 
topological sense we have discussed in Section~\ref{sec-comm}. Communities exist because their vertices are supposed to be similar to each other, in some way.
The similarity among the vertices is then revealed topologically through the higher edge probability among pairs of members of the same group
than between pairs of members of different groups, whose similarity is lower. Hence, when one is provided with annotations or other sources of information that allows 
to classify vertices based on their similarity, one expects that such similarity-based classes are also the best communities that structure-based algorithms may detect.
\begin{figure*}
\begin{center}
\includegraphics[width=\textwidth]{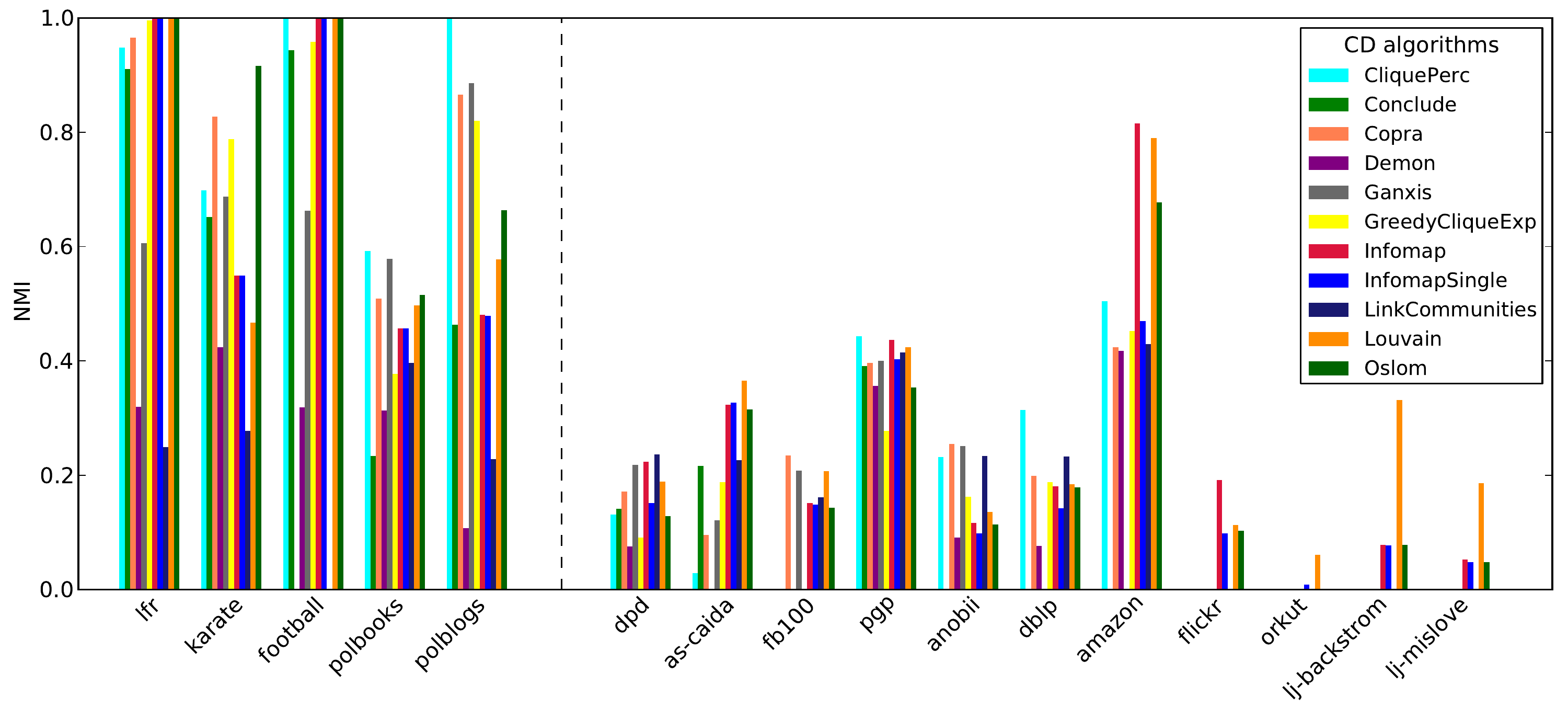}
\caption{Mismatch between structural communities and metadata groups. The vertical axis reports the similarity between communities detected 
via several clustering methods and groups of vertices with identical attributes for several networks, estimated via the NMI.
Scores are grouped by data sets on the horizontal axis. The vertical dashed line separates small classic data sets from large ones, recently compiled.
The low scores obtained for the latter indicate that the correspondence between structural and annotated communities is poor.
Reprinted figure with permission from~\cite{hric14}. \copyright\,2014, by the American Physical Society.}
\label{tmhric}
\end{center}
\end{figure*}

Indeed, for some small networks like Zachary's karate club this seems to be the case. But for quite some time scholars could not test this 
hypothesis, due to the limited number of suitable data sets. Over the past few years this has finally become possible, due to the
availability of several large network data sets with annotated vertices~\cite{yang12c,yang13,yang14,hric14}.
It turns out that the alignment between the communities found by standard clustering algorithms and the annotated groups is not good, in general.
In Fig.~\ref{tmhric} we show the similarity between the topological partitions found by different methods and the annotated partitions, for several 
social, information and technological networks~\cite{hric14}. The heights of the vertical bars are 
the values of the normalised mutual information (NMI)~\cite{lancichinetti09}. Groups of contiguous bars represent the scores for a given data set.
To the left of the vertical dashed line we see the results for classic benchmarks, like LFR graphs (Section~\ref{art-bench})~\cite{lancichinetti08}, 
Zachary karate club, etc., and the scores are generally good. But for the large data sets to the right of the line the scores are rather low, signalling a significant
mismatch between topological and annotated communities. 
For Amazon co-purchasing network~\cite{yang12}, in which vertices are products and edges are set between products 
often purchased by the same customer(s), the similarity is quite a bit higher than for the other networks. This is because the classification of Amazon products 
is hierarchical (e. g., \texttt{Books/Fiction/Fantasy}), so there are different levels of annotated communities, and the reported scores refer to the one which is most similar 
to the structural ones detected by the algorithms, while the other levels would give lower similarity scores.
Low similarity at the partition level does not rule out that
some communities of the structural partition significantly overlap with their annotated counterparts, 
but precision and recall scores show that this is not the case.
Results depend more on the network than on the specific method adopted, none of
which appears to be particularly good on any (large) data set. 

So the hypothesis that structural and annotated clusters are aligned is not warranted, in general. There can be multiple reasons for that. 
The attributes could be too general or too specific, yielding communities which are too large or too small to be interesting.
Moreover, while the best partition of the network delivered by an algorithm
can be poorly correlated with the metadata, there may be
alternative topological divisions that also belong to a set of valid solutions, according to the algorithm\footnote{For instance, for clustering methods based on
optimisation, there are many partitions corresponding to values of the quality function very close to the searched optimum~\cite{good10}. The algorithm will return one (or some) of them, but the others are comparably good solutions.}, but happen to be better correlated
with the annotations~\cite{newman16}.

The fact that structural and annotated communities may not be strongly correlated has important consequences.
Scholars have been regularly testing their algorithms on small annotated graphs, like Zachary's karate club, by tuning
parameters such to obtain the best possible performance on them. This is not justified, in general, as it makes sense only when there is
a strong correspondence, which is a priori unknown. Also, forcing an alignment with 
annotations on one data set does not guarantee that there is going to be a good alignment with the annotations of a different network.
Besides, one of the reasons why people use clustering algorithms is to provide 
an improved classification of the vertices, by using structure. If one obtained the same thing, why bother? 

The right thing to do is using structure {\it along with} the annotations, instead of insisting on matching them. This way the information coming from 
structure and metadata can be combined and we can obtain more accurate partitions, if there is any correspondence between them.
Recent approaches explicitly assume that the metadata
(or a portion thereof) are either exactly or approximately correlated
with the best topological partition~\cite{moore11,leng13,peel15,yang13b,bothorel15}.
A better approach is not assuming a priori that the metadata correlate with the structural
communities. The goal is quantifying the relationship between metadata and community
and use it to improve the results. If there is no correlation, the metadata would be ignored, leaving us
with the partition derived from structure alone. 

\begin{figure}
    \begin{minipage}{\columnwidth}\centering
      \begin{tikzpicture}[inner sep=0, minimum size=0.1]
        \begin{scope}[yshift=0,every node/.append style={yslant=0.5,xslant=-1},yslant=0.5,xslant=-1]
          \node[draw, rectangle, dashed, thin, fill={rgb:black,0;white,5}, fill opacity=.2,text opacity=1,label={[label distance=-12.5em]35:{{Data, $\bm{A}$}}}] at (0,0) {\includegraphics[width=.45\columnwidth, trim=0cm 0cm 0cm 0cm,clip=True]{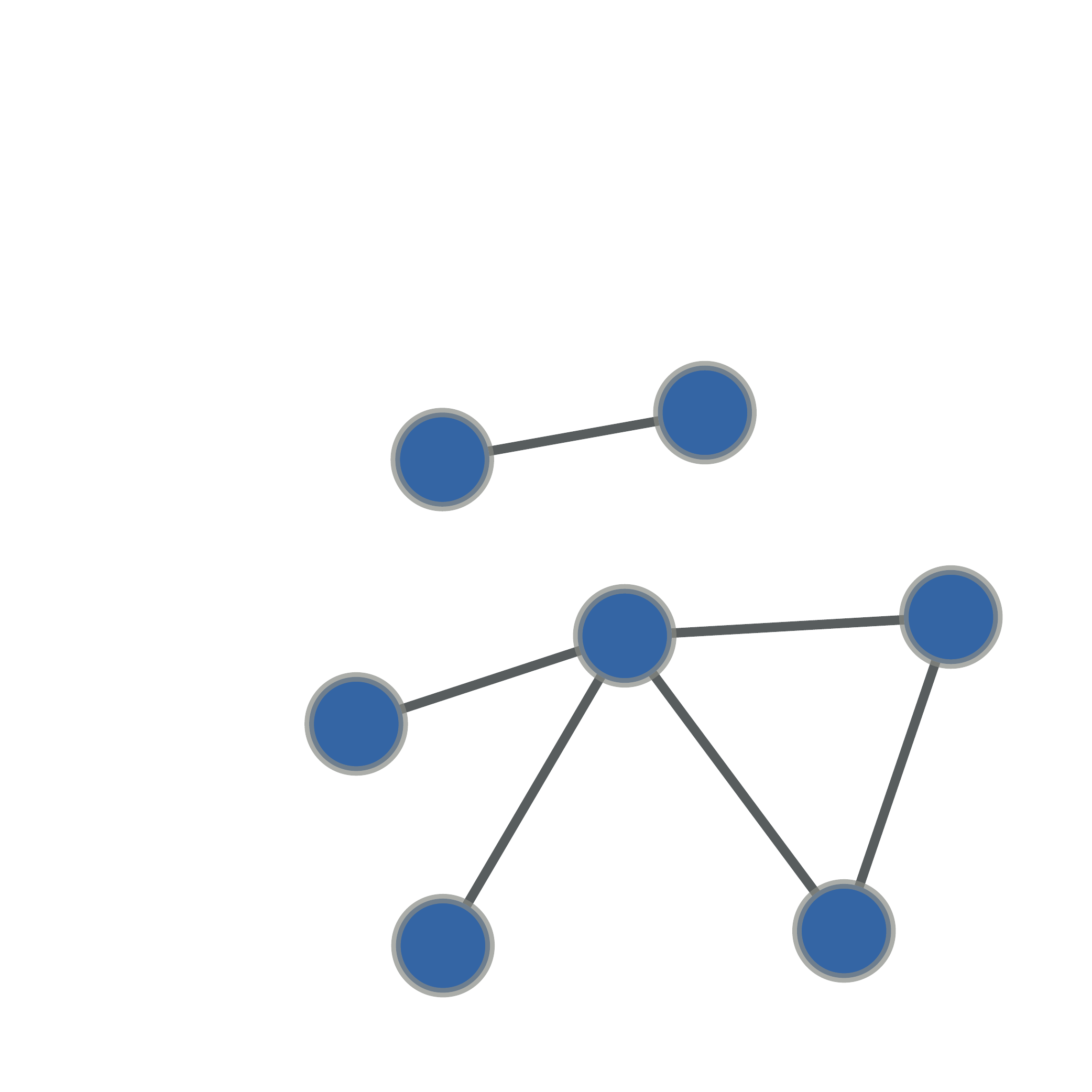}};
        \end{scope}
        \begin{scope}[yshift=85,every node/.append style={yslant=0.5,xslant=-1},yslant=0.5,xslant=-1]
          \node[draw, rectangle, dashed, thin, fill={rgb:black,0;white,5}, fill opacity=.2,text opacity=1,label={[label distance=-12.4em]30:{{Metadata, $\bm{T}$}}}] at (0,0) {\includegraphics[width=.45\columnwidth, trim=0cm 0cm 0cm 0cm,clip=True]{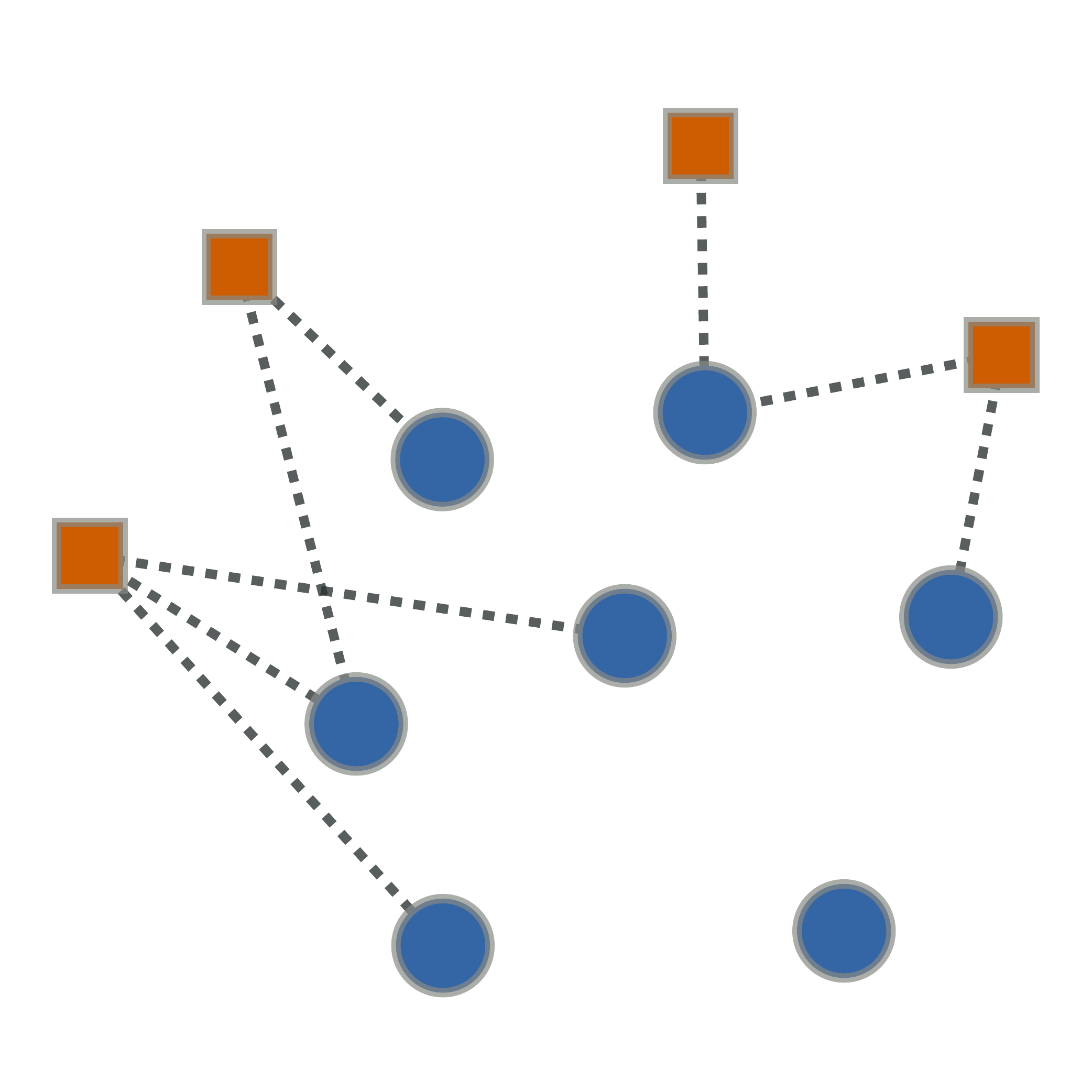}};
        \end{scope}
      \end{tikzpicture}
    \end{minipage}
    \caption{Data-metadata stochastic block model by Hric et al.. One layer consists of the network itself, with its vertices and edges.
    The other layer is composed of the graph vertices and vertices representing the annotations, the edges indicating which vertices are associated to which tag.
    The presence of network vertices on both layers induces a coupling
    between them. Reprinted figure with permission from~\cite{hric16}. \copyright\,2016, by the American Physical Society.}
    \label{dameta}
\end{figure}

Methods along these lines have been developed, using stochastic block models.
Newman and Clauset~\cite{newman16} have proposed a model in which vertices are initially assigned to clusters based on metadata, 
and then edges are placed between vertices according to the degree-corrected stochastic block model~\cite{karrer11}.
Hric et al. have designed a similar model~\cite{hric16}, in which the interplay between structure and metadata is represented by a multilayer network (Fig.~\ref{dameta}).
The generative model is an extension of the hierarchical stochastic block
model (SBM)~\cite{peixoto14b} with degree-correction for the case with edge layers~\cite{peixoto15b}.
%
Here the metadata is not supposed to correspond simply to a
partition of the vertices. The majority of data sets contain
rich metadata, with vertices being annotated multiple
times, and often few vertices possess the exact same annotations and can be thus associated to the same group.
In addition, while the number of
communities is required as input by the method of Newman and Clauset,
here it is inferred from the data. Finally, it is also possible
to assess the metadata in its power to predict the network
structure, not only their correlation with latent
partitions. This way it is possible to predict {\it missing vertices} of the network, i. e., to infer 
the connections of a vertex from its annotations only. We stress that neither method requires that all vertices
are annotated.

Applications of the method by Hric et al.~\cite{hric16} reveal that 
in many data sets there are statistically significant correlations between
the annotations and the network structure, while in some cases the metadata 
seems to be largely uncorrelated with structural clusters.
We conclude that network metadata should not be used indiscriminately 
as ground truth for community detection methods. 
Even when the metadata is strongly predictive of
the network structure, the agreement between the annotations
and the network division tends to be complex, and
very different from the one-to-one mapping that is more
commonly assumed. Moreover, data sets usually
contain considerable noise in their annotations, and some
metadata tags are essentially random, with no relationship to structure.

\subsection{Community structure in real networks}
\label{ncp}

Artificial benchmark graphs are certainly very useful to assess the performance of clustering algorithms. 
However, one could always question whether the model of community structure they propose is reliable.
How can we assess this? In order to characterise ``real" communities we have to find them first. But that can only be done 
via some algorithm, and different algorithms search for different types of objects, in general.  
Still, one may hope that general properties of communities can be consistently uncovered across different methods and data sets, while 
other features are more closely tied to the specific method(s) used to detect the clusters and (or) the specific data set at study (or classes thereof).

\begin{figure}[h!]
\begin{center}
\includegraphics[width=\columnwidth]{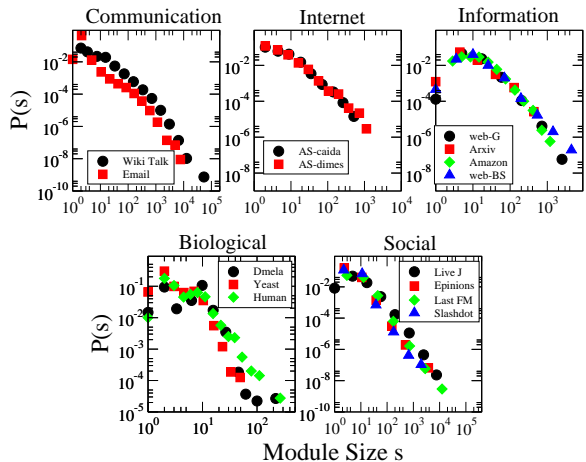}
\caption {Distribution of community sizes in real networks. The clusters were detected with Infomap~\cite{rosvall08}, but 
other methods yield qualitatively similar results. Various classes of networks are considered. All distributions are broad, 
spanning several orders of magnitude. 
Reprinted figure with permission from~\cite{lancichinetti10b}.}
\label{figcomsize}
\end{center}
\end{figure}

A seemingly robust feature of communities in real networks is the heterogeneity of their size distribution.
Most clustering techniques find skewed distributions of cluster sizes in many networks.
So, there appears to be no characteristic
size for a community: small communities usually coexist with large ones. This feature is rather independent of the type of network (Fig.~\ref{figcomsize}).
It may signal a hierarchy among communities, with small clusters included in large ones. Methods unable to distinguish between hierarchical levels might 
find ``blended" partitions, consisting of communities of different levels and hence of very different sizes. The LFR benchmark 
was the first graph model to take explicitly into account the heterogeneity of community sizes (Section~\ref{art-bench}).

\begin{figure*}
\begin{center}
\includegraphics[width=0.46\textwidth]{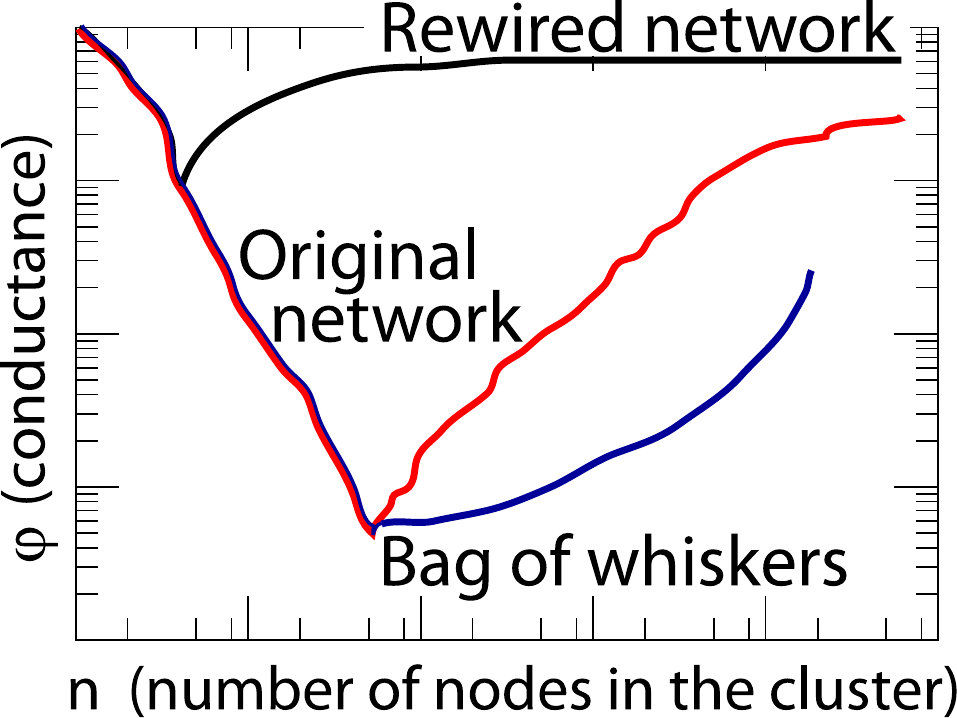}
\includegraphics[width=0.46\textwidth]{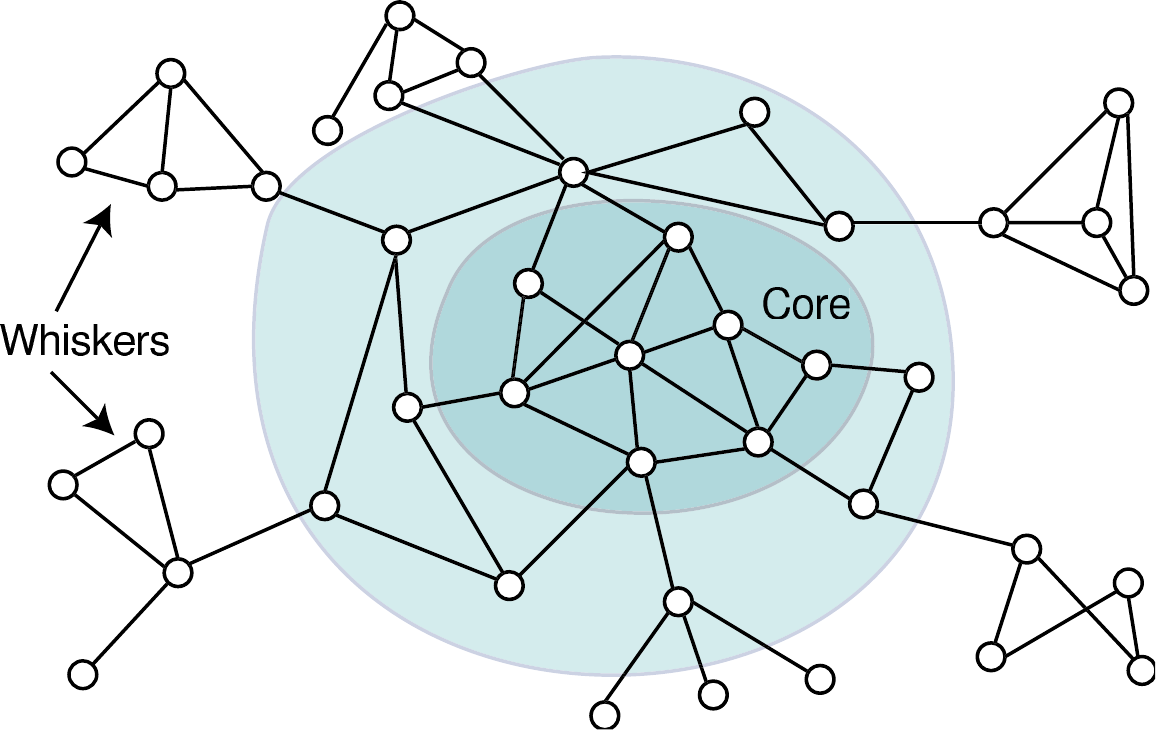}
\caption {Core-periphery structure of real networks.
(Left) Schematic shape of the network community profile (NCP), showing how the minimum conductance of subgraphs of size $k$ 
varies with $k$. This pattern is very common in large social and information networks.
The ``best'' communities in terms of conductance have a size of about $100$ vertices (minimum of the curve), whereas communities of larger
sizes have lower quality. The curve labeled {\it Rewired network}
is the NCP of a randomised version of 
the network, where edges are randomly rewired by keeping the degree distribution; the one labeled {\it Bag of whiskers} 
gives the minimum conductance scores of communities composed of disconnected pieces.
(Right) Scheme of the core-periphery 
structure in large social and information networks associated to the NCP above. 
Most vertices are in a central core, where large communities are blended together, whereas the
best communities, which are rather small, are weakly connected to the core. 
Reprinted figure with permission from Ref.~\cite{leskovec09}. \copyright\,2009, by Taylor and Francis.}
\label{figlesk}
\end{center}
\end{figure*}

Another interesting question is how the quality of 
communities depends on their size. Leskovec et al.~\cite{leskovec09}
carried out a systematic analysis of clusters in many large networks, including traditional and on-line
social networks, technological, information networks and web graphs. 
Instead of considering partitions, they focused on individual communities, which are derived by optimising conductance (Section~\ref{sec-var})
around seed vertices. 
We remind that the conductance of a cluster is the ratio between the number of external edges and the total degree
of the cluster. Minimising conductance effectively combines the two main community demands, i. e., good separation 
from the rest of the graph (low numerator) and large number of internal edges (high denominator). The measure is also relatively insensitive to the size of the clusters, as 
both the numerator and the denominator are typically proportional to the number of vertices of the community\footnote{This is 
exactly true when the ratio between the external and the total degree 
(mixing parameter) is the same for all community vertices.}. 
Therefore one could use it to compare the quality of 
clusters of different sizes. For any given size $k$ Leskovec et al. identified the subgraph with $k$ vertices with the lowest conductance.
This way, for each network one can draw the {\it network community profile} (NCP), showing the minimum conductance 
as a function of community size. The NCPs of all networks studied by Leskovec et al.
have a characteristic shape: they go downwards till $k\sim 100$ vertices, and then they rise monotonically for larger subgraphs
[Fig.~\ref{figlesk} (left)]. 
Alternative shapes have been recently found for other data sets~\cite{jeub15}.

For networks characterised by NCPs like the one in Fig.~\ref{figlesk} (left) the most pronounced communities are fairly small in size.
Such small clusters are weakly connected to the rest of the network,
often by a single edge (in this case they are called {\it whiskers}), and form the {\it periphery} of the graph. 
Large clusters have comparatively lower quality and melt into a big {\it core}. Large communities 
can often be split in parts with lower conductance, so they can be considered conglomerates of 
smaller communities. A schematic picture of the resulting network structure is shown in Fig.~\ref{figlesk} (right). The shape of the NCP is fairly independent of the specific technique adopted to 
identify the subgraphs with minimum conductance. The different shapes of the NCPs encountered in data suggest that
core-periphery is not the only model of group structure of real networks~\cite{jeub15}.

The NCP is a signature that can be used to select
generative mechanisms of community structure. Indeed, many standard models typically yield NCP sloping steadily downwards, at odds with
the ones encountered in many social and information networks. Stochastic block models (Section~\ref{sec-MV}) are sufficiently versatile that they can
reproduce the NCP shape of Fig.~\ref{figlesk} (left), by suitably tuning the parameters. 
In the standard LFR benchmark (Section~\ref{art-bench}) the mixing parameters are tightly concentrated  
about a value $\mu$ by construction, hence all clusters have 
approximately conductance $\mu$, yielding a roughly flat NCP\footnote{If all vertices of a subgraph have mixing parameter equal to $\mu$, 
it can be easily shown that the conductance of the subgraph is exactly equal to $\mu$.}~\cite{jeub15}. However, the model can be easily modified 
by making $\mu$ community-dependent and a large variety of NCPs are attainable, including the one of Fig.~\ref{figlesk} (left).

The main problem of working with NCPs is that they are based on extreme statistics, as one systematically reports the minimum conductance for a given cluster size.
How representative is this extremal subgraph of the population of subgraphs with the same size? There may be just a few clusters of a given size with low conductance.
It may happen that many subgraphs 
have conductance near the minimum corresponding to their size(s), which would then be representative. Alternatively most subgraphs might have
much larger conductance
than the minimum but low enough that they can be still considered
communities. In this case one should conclude that communities of that size are not of very high quality, on average.
The above scenarios might 
lead to different conclusions about the actual community structure of the system. 
In general, even if one could produce a version of the NCP where the trend refers to 
representative samples of communities of equal size (whatever that means), the actual values of the conductance are as important as the shape of the curve.
If conductance is sufficiently low for all cluster sizes, it means that there are good communities of any size.
The fact that small clusters could be of higher quality does not undermine the role of large clusters. The observation that large clusters 
consist of smaller clusters of higher quality may just be evidence of hierarchical structure in the network, which is 
a trademark of many complex systems~\cite{simon62}. In that case high levels of the hierarchy are not less important than low ones, a priori.
In fact, the actual relative importance of communities should not only come from
the sheer value of specific metrics, like conductance, but also from their statistical 
significance (Section~\ref{sec-sign}).

That notwithstanding, we strongly encourage analyses like the one by Leskovec et al., as they provide a statistical characterisation of community structure, in a way 
that is only weakly algorithm-dependent. One has to define operationally what a cluster is, but in a simple intuitive way that allows us to draw
conclusions about the structure of the graph. In principle one could do the same by analysing the clusters delivered by any algorithm,
but there would be two important drawbacks. First, the clusters may not be easy to interpret, as most clustering algorithms usually do not 
require a clear-cut definition of community. Second, one would have to handle a partition of the network in communities, instead of probing locally the group structure of the 
network. Therefore, for a given vertex one would have only one cluster (or a handful, if communities overlap), while a local exploration allows to 
analyse a whole population of candidate subgraphs, which gives more information. The local subgraphs recovered this way do not need to 
be strongly matching the clusters delivered by any algorithm, but they provide useful signatures that allow to restrict the set of possible model explanations 
for the network's group structure. Such investigation can be replicated on any model graph to check whether the results match (e. g., whether the NCPs coincide).

\begin{figure}
\begin{center}
{
\label{fig:noover}
\includegraphics[width=0.45\columnwidth]{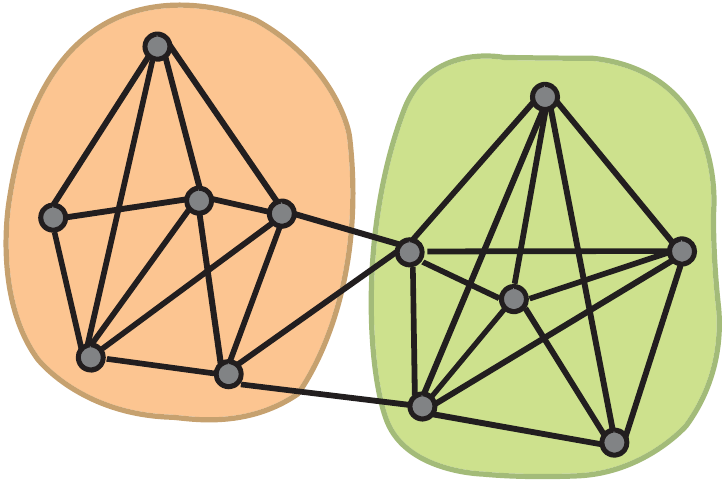}
}
\hfill
{
\label{fig:over}
\includegraphics[width=0.45\columnwidth]{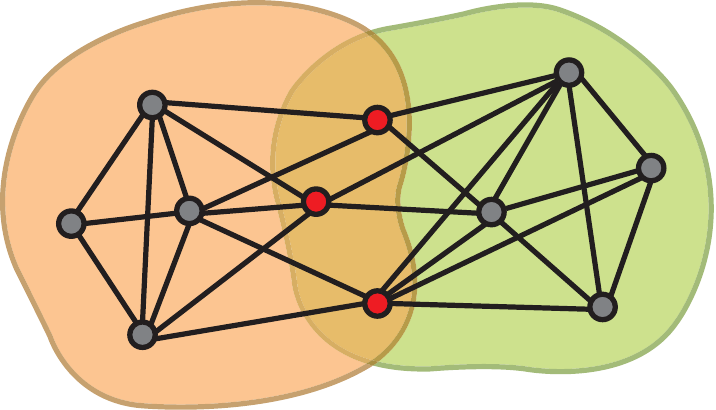}
}
\\
\hfill
\subfloat[No overlaps]{
\label{fig:adnoover}
\includegraphics[width=0.45\columnwidth]{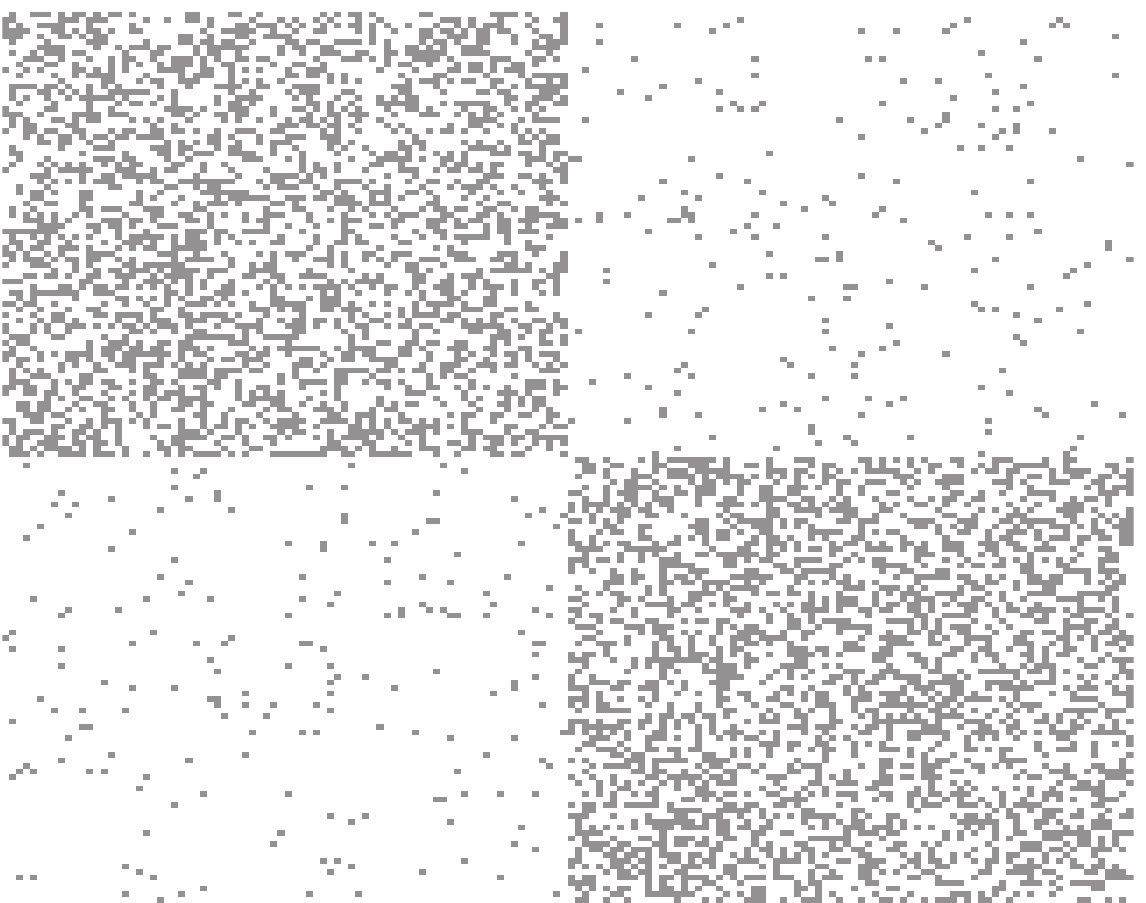}
}
\hfill
\subfloat[Sparse overlaps]{
\label{fig:adover}
\includegraphics[width=0.45\columnwidth]{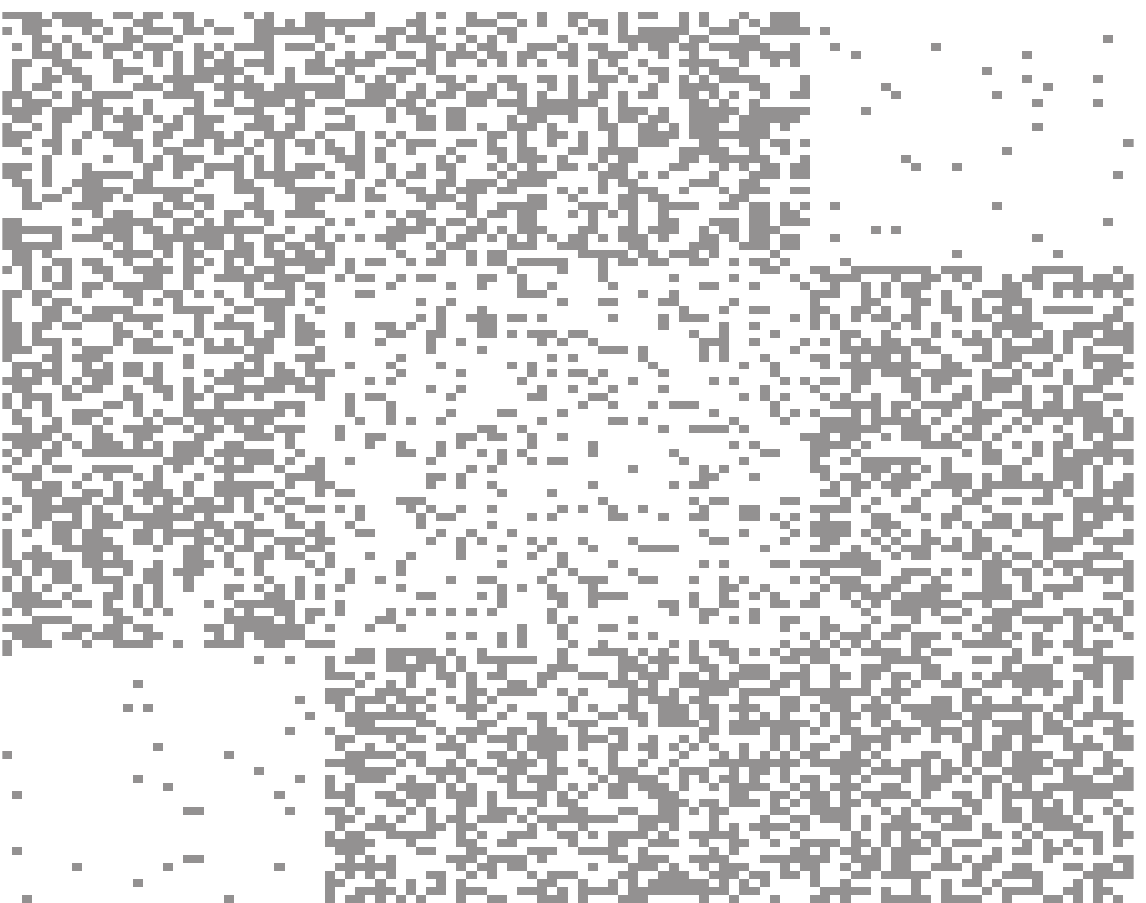}
}
\\
\hfill
\subfloat[Dense overlaps]{
\label{fig:adnoover1}
\includegraphics[width=0.45\columnwidth]{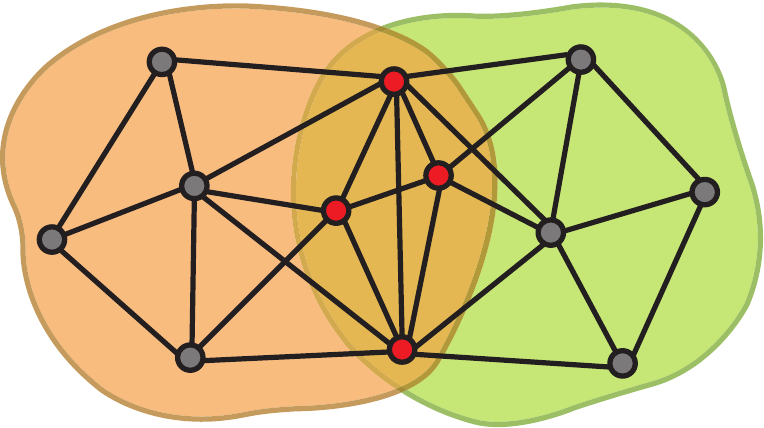}
}
\hfill
\subfloat[Adjacency matrix of (c)]{
\label{fig:adover1}
\includegraphics[width=0.45\columnwidth]{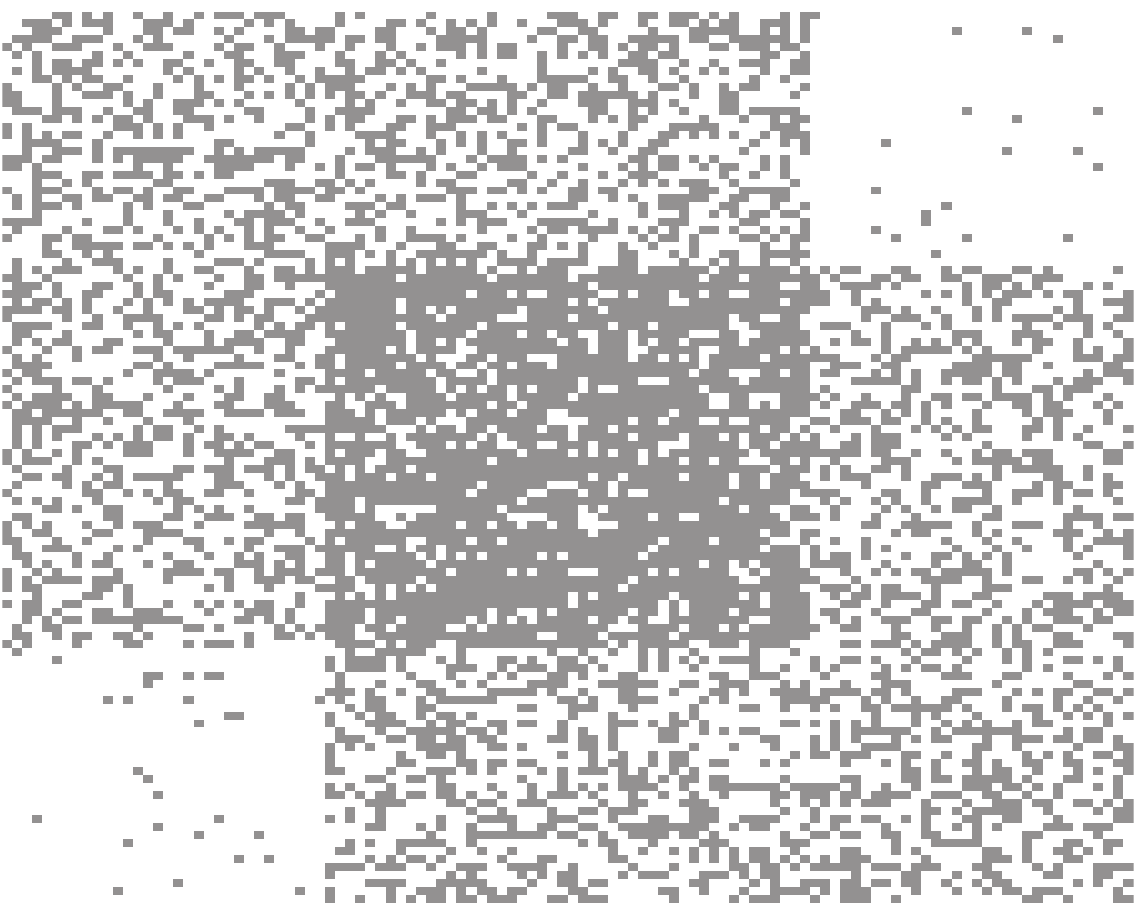}
}
\caption{Stylised views of community structure. In (a) and (b) we show the conventional 
pictures of non-overlapping and overlapping clusters, respectively. Under the network diagrams   
we see the corresponding adjacency matrices. The overlaps have a lower edge density than the rest of the communities.
The analysis by Yang and Leskovec suggests that a more realistic model could be the one 
shown in (c, d), where the overlaps are denser than the non-overlapping parts.
Reprinted figure with permission from~\cite{yang14}. \copyright\,2014, by the Association for Computing Machinery, Inc.}
\label{fig:AGM}
\end{center}
\end{figure}

Another approach to infer properties of clusters of real networks is using annotations. While we have shown that
annotated clusters do not necessarily coincide with structural ones (Section~\ref{topmet}), 
general features can be still derived, provided they are consistently found across different data sets and annotations.
A recent analysis by Yang and Leskovec has questioned the common picture of networks with overlapping communities~\cite{yang14}. 
Scholars usually assume that clusters overlap at their boundaries, hence edge density should be larger in the non-overlapping parts (Fig.~\ref{figoverl}).
Instead, by analysing the overlaps of annotated clusters in large social and information networks, Yang and Leskovec found that the probability that two vertices are connected 
is larger in the overlaps, and grows with the number of communities sharing that pair of vertices. 
In addition, {\it connector vertices}, i. e., vertices with the largest number of neighbours within a community, are more likely to be found in the overlaps.
These findings suggest that the overlaps may play an important role in the community structure of networks. In Fig.~\ref{fig:AGM} we compare 
the conventional view with the one resulting from the analysis. The {\it Community-Affiliation Graph Model} (AGM)~\cite{yang14} and the {\it Cluster Affiliation Model for Big Networks}
(BIGCLAM)~\cite{yang13} are clustering techniques based on
generative models of networks featuring communities with dense overlaps. The models are based on the principle that vertices are more likely to be neighbours 
the more the communities sharing them, in line with the empirical finding of~\cite{yang14}. 

Actual overlapping communities exist in many contexts. However, it is unclear whether soft clustering is statistically founded. 
A recent analysis
aiming at identifying suitable stochastic block models to describe real network data indicate that in many cases hard partitions ought to be preferred, as they give 
simpler descriptions of the group structure of the data than soft partitions~\cite{peixoto15}. This could be due to the fact that the underlying models 
are based on placing edges independently of each other, neglecting
higher order structures between vertices, like motifs.
By adopting approaches that take into account higher-order structures
things may change and community overlaps might become a statistically robust feature.  
The pervasive overlaps found by Yang and Leskovec in annotated data can be found if higher order effects are considered~\cite{rosvall14,persson16}, without ad hoc hypotheses.


%

\section{Methods}
\label{sec-meth}

There are many algorithms to detect communities in graphs. They can be grouped in categories,
based on different criteria, like the actual operational method~\cite{fortunato10}, or the underlying concept of community~\cite{coscia11}.
In most applications, however, just a few popular algorithms are employed. In this section we present a critical analysis of these methods.
We show the advantages of knowing the number of clusters before-hand 
and how it is possible to derive robust solutions from partitions delivered by stochastic clustering techniques.
We discuss approaches to the problem of detecting communities in evolving networks and how to assess the significance of the detected clustering.
We conclude by suggesting the methods that currently appear to be most promising.

\subsection{How many clusters?}
\label{sec-tools}

In general, the only preliminary information available
to any algorithm is the structure of the network, i. e., which
pairs of vertices are connected to each other and which are
not (possibly including weights). Any insight about community structure
is supposed to be given as output of
the procedure. Naturally, it would be valuable to have some information
on the unknown division of the network beforehand, as
one could reduce considerably the huge space of possible
solutions, and increase the chance of successfully
identifying the communities. 

\begin{figure}[h!]
\begin{center}
\includegraphics[width=\columnwidth]{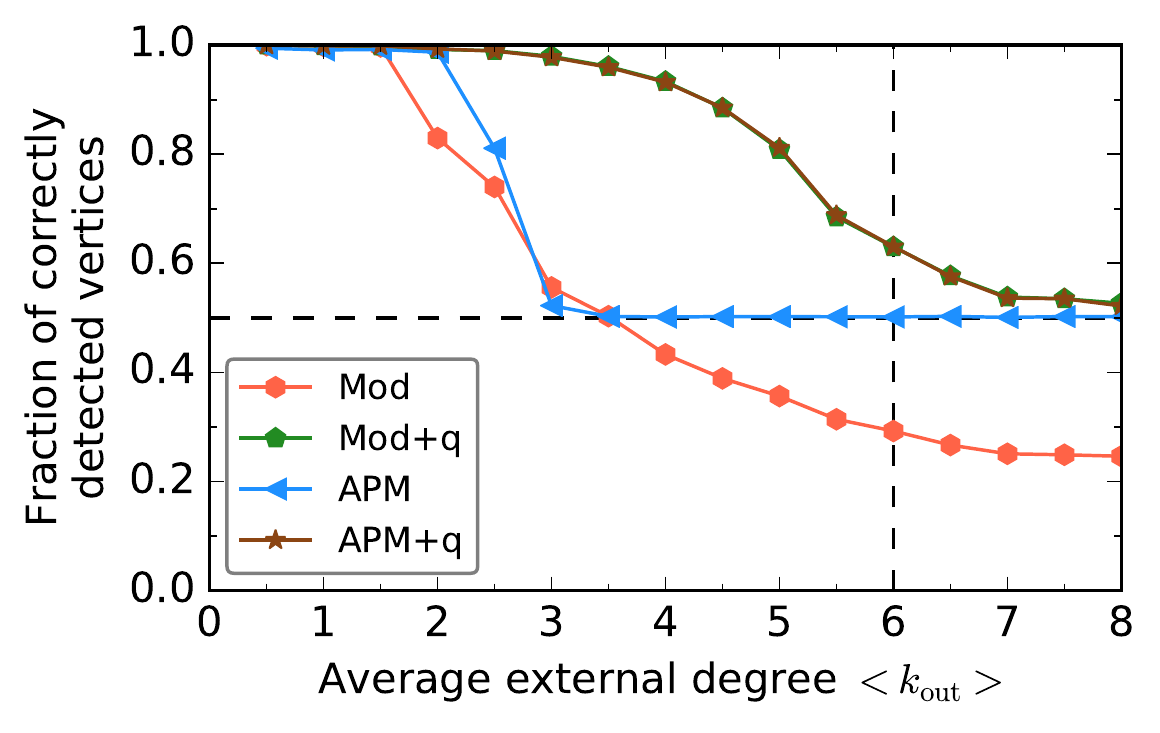}
\caption {Knowing the number of clusters beforehand improves community detection. 
The diagram shows the performance on the planted partition model of two methods: modularity optimisation 
via simulated annealing~\cite{guimera05} and the Absolute Potts Model (APM)~\cite{ronhovde10}. 
Networks have $400$ vertices, which are grouped in two equal-sized communities. 
The accuracy is measured via the fraction of correctly detected vertices (Section~\ref{sim-me}). 
The horizontal line indicates the accuracy of random guessing, the dashed vertical line the theoretical detectability limit (Section~\ref{detectab}).
For each algorithm we show two curves, referring to the results of the method
in the absence of any information on the number of clusters, and when such information is fed
into the model as initial input. In both cases, knowing the
number of clusters beforehand leads to a much better performance.}
\label{fignumclus}
\end{center}
\end{figure}

Among all the possible pre-detection inputs, the
number $q$ of clusters plays a prominent role. Many popular classes 
of algorithms require the specification of $q$ before they run, like methods 
imported from data clustering or parametric statistical inference approaches
(Section~\ref{sec-inference}). Other methods are capable to 
infer $q$ as they can choose among partitions into different numbers of communities.
But even such methods could benefit from a preliminary knowledge of $q$~\cite{darst14}. In Fig.~\ref{fignumclus}
we report standard accuracy plots of two algorithms on the planted partition model (Section~\ref{art-bench}) with two clusters of equal size.
The algorithms are modularity optimisation 
via simulated annealing~\cite{guimera05} and the Absolute Potts Model (APM)~\cite{ronhovde10} (Section~\ref{sec-dynmet}). 
There are two performance curves for each method: one comes from the standard application of the method, without constraints; the other
is obtained by forcing the method to explore only the subset of partitions with the correct number of clusters $q=2$.

We see that the accuracy improves considerably when $q$ is known. This is particularly striking in the case of modularity optimisation, which 
is known to have a limited resolution, preventing the method from identifying
the correct scale of the communities, even when the
latter are very pronounced~\cite{fortunato07,good10} (Section~\ref{sec-modopt}). Knowing $q$ and constraining the optimisation of the
measure to partitions with fixed $q$, the problem can be alleviated~\cite{nadakuditi12}.

But how do we know how many clusters there are? Here we briefly discuss some heuristic techniques, for statistically principled methods we defer the reader to
Section~\ref{sec-inference}.
It has been recently shown that 
in the planted partition model $q$ can be
correctly inferred all the way up to the detectability limit from the spectra of two matrices: the
{\it non-backtracking matrix} $\bf B$~\cite{krzakala13} and the {\it flow matrix}
$\bf F$ ~\cite{newman13c}.
They are $2m\times 2m$ matrices, where $m$ is the number of edges of
the graph. Each edge is considered in both directions, yielding $2m$
directed edges and indicated with the notation $i\rightarrow j$,
meaning that the edge goes from vertex $i$ to vertex $j$. 
Their elements read
\begin{equation}
B_{i\rightarrow j, r\rightarrow s}=\delta_{is}(1-\delta_{jr})
\label{eqNBTM}
\end{equation}
and
\begin{equation}
F_{i\rightarrow j, r\rightarrow s}=\frac{\delta_{is}(1-\delta_{jr})}{k_i-1}.
\label{eqFM}
\end{equation}
In Eq.~(\ref{eqFM}) $k_i$ is the degree of vertex $i$. So the
elements of $\bf F$ are basically the elements of $\bf B$, normalised
by vertex degrees. This is done to account for the heterogeneous degree
distributions observed in most real networks. Both matrices have non-zero elements only for each
pair of edges forming a directed path from the first vertex of
one edge to the second of the other edge. To do that, edges have to
be incident at one vertex. As a matter of fact, the non-backtracking
matrix $\bf B$ is just the adjacency matrix of the (directed) edges of
the graph.

The name of the matrix ${\bf B}$ is due to a
connection with the properties of non-backtracking
walks. A {\it non-backtracking walk}~\cite{angel15} is a
path across the edges of a graph that is allowed to return to
a vertex visited previously only after at least
two other vertices have been visited; immediate returns
like $1 \rightarrow 2 \rightarrow 1$ are forbidden. The elements of the $k$-th power of ${\bf B}$
yield the number of non-backtracking walks of length $k$ from a (directed) edge of the graph to another and the trace of the power matrix
the number of closed non-backtracking walks of length $k$ starting from any given (directed) edge.

A remarkable property of both matrices is that on networks with homogeneous groups (i. e., of similar size and internal edge density) most eigenvalues, which
are generally complex, are enclosed by a circle centred at the
origin, and that the number of
eigenvalues lying outside of the circle is a good proxy of the number
of communities of the network~\cite{krzakala13,newman13c}. For $\bf B$ the circle's radius is
given by the square root $\sqrt{c}$ of the leading eigenvalue $c$, which may diverge for networks with heterogeneous degree distributions
(e. g., power laws); for
$\bf F$ it equals $\sqrt{\langle k/(k-1)\rangle/\langle k\rangle}$,
which is never greater than $1$. 

Unfortunately, computing the eigenvalues of the
non-backtracking or the flow matrix is lengthy. Both are 
$2m\times 2m$ matrices. The adjacency
matrix $\bf A$ has $n\times n$ elements, so $\bf B$ and $\bf F$ are
larger by a factor of
$\langle k\rangle^2$, where $\langle k\rangle$ is the average
degree of the network. An approximate but reliable computation of the spectra requires a time which
scales superlinearly (approximately quadratic) with the network size $n$. So the problem is intractable for
graphs with number of edges of the order of millions or higher. Also, if communities have diverse sizes and edge densities, as it happens in most networks encountered in
applications, the bulk of eigenvalues 
may not have a circular shape, and it may become problematic to identify eigenvalues falling outside of the bulk.

Besides, non-backtracking walks must contain cycles, hence trees\footnote{We remind that trees are connected acyclic graphs.} dangling off the graph do not 
affect the spectrum of ${\bf B}$, which remains unchanged if all dangling trees are removed.
This is a disturbing feature, as tree-like regions of the graph may play a role in the network's community structure, and most methods would find
different partitions if trees are kept or removed\footnote{Singh and Humphries showed that the problem can be solved via {\it reluctant backtracking walks}, in which the walker has a small but non-zero probability of returning to the vertex immediately~\cite{singh15}.}. The spectrum of the flow matrix, instead, changes when dangling trees are kept or removed~\cite{newman13c}.
In the limiting case in which the network itself is a tree, all eigenvalues of ${\bf B}$ and ${\bf F}$ are zero
and even if there were a community structure one gets no relevant information.

The number of clusters can also be deduced by studying how the eigenvectors of graph matrices rotate 
when the adjacency matrix of the graph is subjected to random perturbations~\cite{sarkar16}. On stochastic block models this approach
infers the correct value of $q$ up to a threshold preceding the detectability limit. The method is also 
computationally expensive. 

In general, if one can identify a set (range) of promising $q$-values, from preliminary information or via calculations like the ones described above or in Section~\ref{sec-inference}, it 
is better to run constrained versions of clustering methods, searching for solutions only among partitions with those numbers of communities, than letting the methods 
discover $q$ by themselves, which may lead to solutions of lower quality.

\subsection{Consensus clustering}
\label{sec-consensus}

Many clustering techniques are stochastic in character and do not deliver a
unique answer. A common scenario is when the desired solution
corresponds to extrema of a cost function, that can only be found
via approximation techniques, with
results depending on random seeds and on the choice of initial conditions. 
Techniques not based on optimisation sometimes have the same feature, when tie-break rules are adopted 
in order to choose among multiple equivalent options encountered along the calculation.

What to do with all these partitions? Sometimes there are objective criteria
to sort out a specific partition and discard all others. For instance, in algorithms based on optimisation,
one could pick the solution yielding the largest (smallest) value of the function to optimise.
For other techniques there is no clear-cut criterion. 

\begin{figure}[h!]
\begin{center}
\includegraphics[width=\columnwidth]{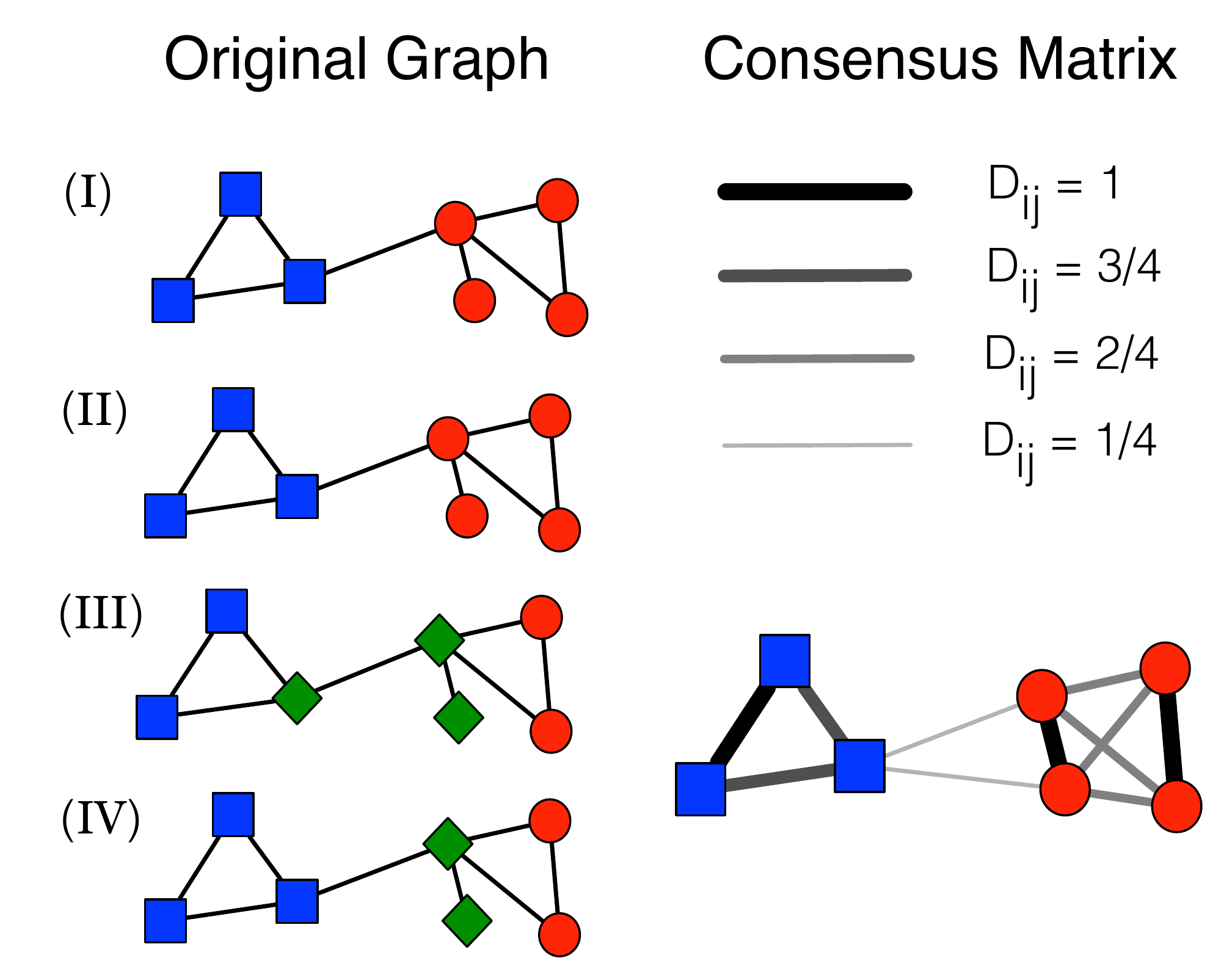}
\caption {Consensus clustering in networks. A simple graph has a natural partition in two communities
[squares and circles on diagrams (I) and (II)].
The combination of (I), (II), (III) and (IV) yields the
weighted consensus matrix illustrated on the right. The thickness of
each edge is proportional to its weight. 
In the consensus matrix the community structure of the original network
is more visible: the two communities have become cliques, with
``heavy'' edges, whereas inter-community edges are rather weak. 
This improvement is obtained despite the presence of two inaccurate
partitions in three clusters (III and IV). Reprinted figure with permission from~\cite{lancichinetti12}.
\copyright\,2012, by the Nature Publishing Group.}
\label{figcons}
\end{center}
\end{figure}

A promising approach is combining the information of the different outputs into a new
partition. Consensus clustering~\cite{strehl02,topchy05,goder08} is based on this idea. The goal is
searching for a {\it consensus} {\it partition}, 
that is better fitting than the input partitions.  
Consensus clustering is a difficult combinatorial
optimisation problem. An alternative greedy strategy~\cite{strehl02} relies on the {\it consensus
  matrix}, which is a matrix based on
the co-occurrence of vertices in communities of the input
partitions (Fig.~\ref{figcons}). The consensus matrix is used as an input for the
graph clustering technique adopted, leading to a new set of
partitions, which produce a new consensus matrix, etc., until 
a unique partition is finally reached, which is not changed by
further iterations. 
The steps of the procedure are listed below. 
The starting point is a network $G$ with $n$ vertices and a clustering
algorithm ${\bf A}$.
\begin{enumerate}
\item{Apply ${\bf A}$ on $G$ $n_P$ times, yielding $n_P$ partitions.}
\item{Compute the consensus matrix ${\bf D}$: $D_{ij}$ is the
number of partitions in which vertices $i$ and $j$ of $G$ are
assigned to the same community, divided by $n_P$.}
\item{All entries of ${\bf D}$ below a chosen threshold $\tau$ are set to zero\footnote{Without thresholding the consensus matrix would quickly turn into a dense matrix, 
rendering the application of clustering algorithms computationally expensive. However, the method does not strictly require thresholding, and if the network is not too large one can 
skip step $3$ and use the full matrix all along~\cite{bruno15}.}.}
\item{Apply ${\bf A}$ on ${\bf D}$ $n_P$ times, yielding $n_P$ partitions.}
\item{If the partitions are all equal, stop\footnote{The consensus matrix
    is block-diagonal in this case.}. Otherwise go back to 2.}
\end{enumerate}
Since the consensus matrix is in general weighted, the algorithm 
${\bf A}$ must be able to handle weighted networks, even if the graph at study is binary. Fortunately many popular algorithms
have natural extensions to the weighted case.

The integration of consensus
clustering with popular existing techniques leads
to more accurate partitions than those delivered by
the methods alone on LFR benchmark graphs~\cite{lancichinetti12}. 
Interestingly, this holds even for methods whose
direct application gives poor results on the same graphs, like modularity optimisation (Section~\ref{sec-modopt}).
The variability of the partitions, rather
than being a problem, becomes a factor of performance
enhancement. 
The outcome of the procedure depends on the choice of the threshold parameter $\tau$ and the number of 
input partitions $n_P$, which can be selected by testing the performance on benchmark networks~\cite{lancichinetti12}.
Consensus clustering is also a promising technique to detect communities in evolving networks (Section~\ref{sec-dynclus}).

\subsection{Spectral methods}
\label{sec-spectral}

Spectral graph clustering is an approach to detect clusters 
using spectral properties of the graph~\cite{luxburg06,fortunato10}. 
The eigenvalue spectrum of several graph matrices (e. g., the adjacency matrix, the Laplacian, etc.) typically consists of a dense bulk of closely
spaced eigenvalues, plus some outlying eigenvalues
separated from the bulk by a significant gap. The
eigenvectors corresponding to these outliers contain information
about the large-scale structure of the network, like community structure\footnote{Typically each such eigenvector is {\it localised}, in that 
its entries are markedly different from zero in correspondence of the vertices of a community, while the other entries are close to zero.}. 
Spectral clustering consists in
generating a projection of the graph vertices in a metric space, by using the entries 
of those eigenvectors as coordinates. 
The $i$-th entries of 
the eigenvectors are the coordinates of vertex $i$ in a $k$-dimensional Euclidean space, where $k$ is the number of eigenvectors used.
The resulting points can be grouped in clusters by using standard partitional clustering techniques like $k$-means clustering~\cite{macqueen67}.

Spectral clustering is not always reliable, however. When the network is very sparse (Section~\ref{sec-MV})
the separation between the eigenvalues of the community-related eigenvectors and the bulk is not sharp.
Eigenvectors corresponding to eigenvalues outside of the bulk may be correlated to high-degree vertices (hubs), instead of 
group structure. Likewise, community-related eigenvectors can be associated to eigenvalues ending up inside the bulk.
In these situations, selecting eigenvectors based on whether their associated eigenvalues are inside or outside the bulk 
yields a heterogeneous set, containing information both on communities and on other features (e. g., hubs).
Using those eigenvectors for the spectral clustering procedure renders community detection more difficult, sometimes impossible.
Unfortunately, many of the networks encountered
in practical studies are very sparse and can lead to this type of problems.

Indeed on sparse networks constructed with the planted partition model spectral methods relying on standard matrices [adjacency matrix, Laplacian, modularity matrix~\cite{newman06}, etc.]
fail before the theoretical detectability limit (Section~\ref{detectab})~\cite{krzakala13}.
The non-backtracking matrix ${\bf B}$ of Eq.~(\ref{eqNBTM}) was introduced to address this problem~\cite{krzakala13}.
On the planted partition model the associated eigenvalues of the community-related eigenvectors of ${\bf B}$ are 
separated from the bulk until the theoretical detectability limit, so spectral methods using the top eigenvectors of ${\bf B}$ are capable to find
communities as long as they are detectable, modulo the caveats we expressed in Section~\ref{sec-tools}.

\subsection{Overlapping communities: Vertex or Edge clustering?}
\label{sec-LC}

Soft clustering, where communities may overlap, is an even harder problem than hard clustering, where there is no
community overlap. The possibility of having multiple memberships for the vertices introduces additional degrees 
of freedom in the problem, causing a huge expansion of 
the space of possible solutions. It has been pointed out that overlapping communities, especially in social networks, 
reflect different types of associations between people~\cite{evans09,ahn10}. Two actors could be co-workers, friends, relatives, sport mates, etc..
Actor $A$ could be a work colleague of $B$ and a friend of $C$, so she would sit in the overlap between the community of colleagues of $B$ and the 
community of friends of $C$. For this reason, it has been suggested that an effective way to recover overlapping clusters is to group edges, rather than vertices.
In the example above, the edges connecting $A$ with $B$ and $A$ with $C$ would be placed in different groups, and since they both have $A$ as endpoint,
the latter turns out to be an overlapping vertex.

Moreover, edge clustering is claimed to have the additional advantage of reconciling soft clustering with hierarchical community structure~\cite{ahn10}. 
If there is hierarchy, communities are nested within each other as many times as there are hierarchical levels. 
Hierarchical structure is often represented via {\it dendrograms}\footnote{A dendrogram, or {\it hierarchical tree}, is a tree diagram used to represent 
hierarchical partitions. At the bottom there are as many nodes as there are vertices in the graph,
representing the singleton clusters. At the top there is one node (root), standing for 
the partition grouping all vertices in a single cluster. 
Horizontal lines indicate 
mergers of a pair of clusters or, equivalently, splits of one cluster. 
Each vertex of the tree identifies one cluster, whose elements can be read by following all bifurcations starting from the vertex 
all the way down to the leaves of the tree.}, with the network being divided in clusters, which are in turn divided in clusters, and so on until one ends up with 
singleton clusters, consisting of one vertex each. But this can be done only if communities do not share vertices.
Overlapping vertices should be assigned to multiple clusters of lower hierarchical levels, yielding multiple copies of them in the dendrogram.
Instead, one could build a dendrogram
displaying edge communities, where each edge is assigned to a single cluster, but clusters can still overlap because edges in different clusters
may share one endpoint~\cite{ahn10}.

Some remarks are in order. First, there may still be overlapping communities
even if there were a single type of association between the vertices.
For instance, if we keep only the friendship relationships within a given population of actors, there are
many social circles and there could be active actors with multiple ties within two circles, or more. Second, in the traditional picture of networks with community structure
(Fig.~\ref{figcom}), the edges connecting two different groups may be assigned to one of the communities they join or they could be put together in a separate group. Either way, they
would signal an overlap between the communities, which is artificial. This happens even in the extreme case of a single edge connecting vertices $A$ and $B$ of two groups, as that edge will have to be 
assigned to a group, which inevitably forces $A$ and $B$ into a common cluster. Third, if we rely on the picture emerging from the analysis by Yang and Leskovec~\cite{yang14} (Section~\ref{ncp})
overlaps between clusters could be much denser than we expect, hence not only vertices but also edges may be shared among different groups, and edge dendrograms would have 
the same problem as classic vertex dendrograms\footnote{In this case, if we used edge clustering, each edge would be placed in one cluster only. 
However, when one turns edge communities into vertex communities, multiple relationships can be still recovered~\cite{ahn10}.}.
Fourth, the computational complexity of the calculation can rise substantially, as in networks of interest 
there are typically many more edges than vertices.
Finally, there is nothing revealing that there is a conceptual or algorithmic advantage in grouping edges versus vertices,
other than works showing that a specific edge clustering technique outperforms some vertex clustering techniques on a specific set of networks.

To shed some light on the situation, we performed the following test. We took some network data sets with annotated vertices, 
giving an indication about what the communities of those networks could be\footnote{We have seen in Section~\ref{topmet} that metadata are not necessarily
correlated with topological clusters. We used data sets for which there is some correlation.}.  For each network $G$ we derived the corresponding {\it line graph} $L(G)$, 
which is the graph whose vertices are the edges 
of $G$, while edges are set between pairs of vertices of $L(G)$ whose corresponding edges in $G$ are adjacent at one of their endpoints.
Vertex communities of $L(G)$ are then edge communities of the original network $G$. The question is whether by working on $L(G)$ the detection improves or not.
We searched for overlapping communities with
OSLOM~\cite{lancichinetti11}.
We applied OSLOM on the original graphs and on their line graphs. The covers found on the line graphs were turned into covers of the vertices of $G$, by replacing each 
vertex of $L(G)$ with the pair of vertices of the corresponding edge of $G$.

\begin{figure}[h!]
\begin{center}
\includegraphics[width=\columnwidth]{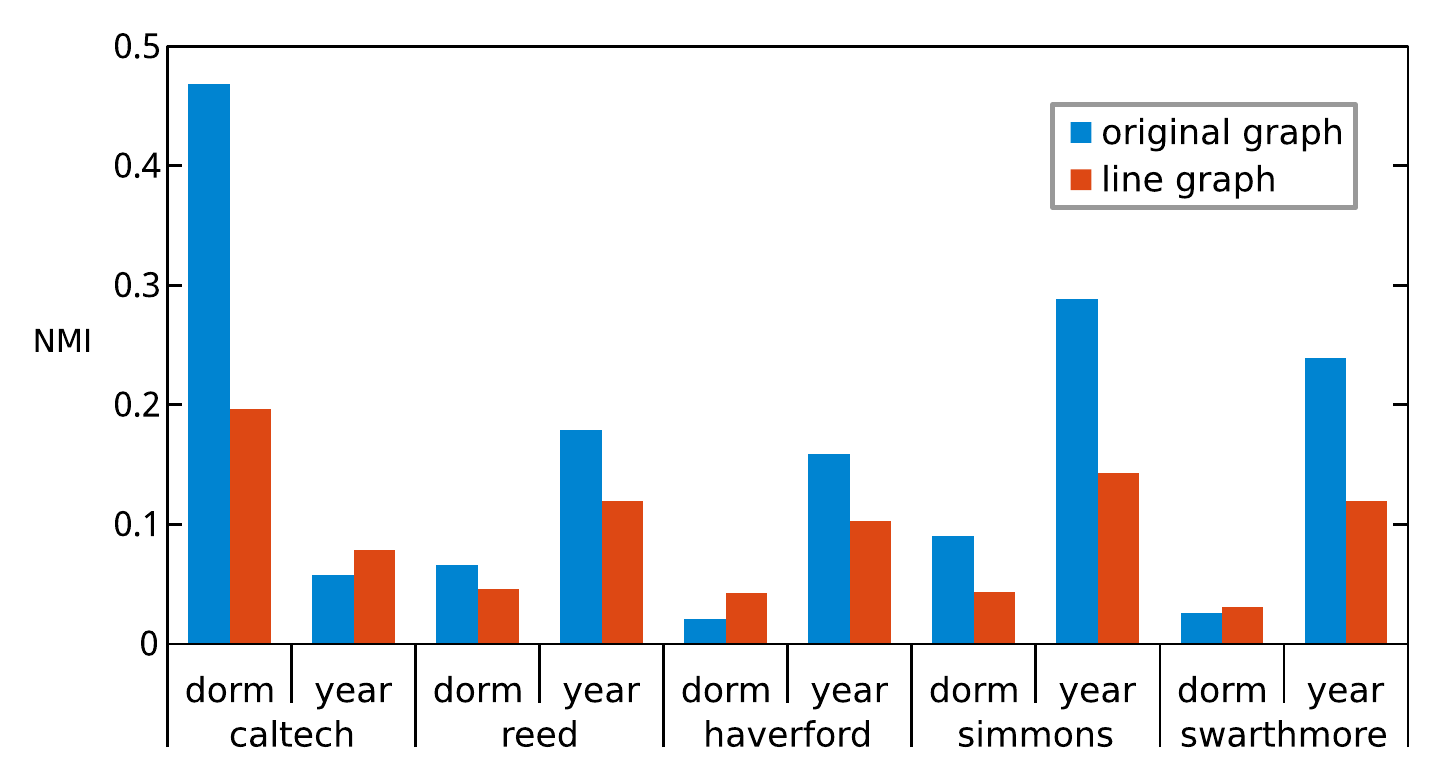}
\caption{Comparison between edge communities and vertex communities. The diagram shows the similarity between the covers of annotated vertices of five social networks and
the topological covers found by OSLOM on them (left bar of each pair) and on their line graphs (right bar). 
The networks represent FaceBook friendships between students of US universities~\cite{traud12}. There are several annotations for each
student, we selected those which are more closely related to topological groups: year of study and dormitory. 
The similarity was computed by using the normalised mutual information (NMI), in the version for covers proposed by Lancichinetti, Fortunato and Kert\'esz~\cite{lancichinetti09}.
Vertex communities detected in the original graphs are overall better 
correlated with the annotated clusters than edge communities.}
\label{figLG}
\end{center}
\end{figure}

The results can be seen in Fig.~\ref{figLG}, showing how similar the covers found on the original networks and on the line graphs are with respect to the covers of annotated vertices.
Neither approach is very accurate, as expected (see Section~\ref{topmet} and Fig.~\ref{tmhric}), 
but vertex communities show a greater association to the annotated clusters than edge communities, except in a few instances where the similarity is very low. 
Analyses carried out on LFR benchmark graphs (not shown) lead to the same conclusion. 
We stress that traditional line graphs have the problem that edges adjacent to a hub vertex in the original graph
turn into vertices who are all connected to each other, forming giant cliques, which might dominate the structure of the line graph, misleading clustering techniques. 
The procedure can be refined by introducing weights for the edges of the line graphs, that can be 
computed in various ways, e. g., based on the similarity of the neighbourhoods of adjacent edges in the original network~\cite{ahn10}. 

Still we believe that our tests provide some
evidence that edge clustering is no better than vertex clustering, in general. 
The superiority 
of algorithms based on either approach should be assessed a posteriori, case by case, and the answer may depend 
on the specific data sets under investigation.

\subsection{Methods based on statistical inference}
\label{sec-inference}

Statistical inference provides a powerful set of tools to tackle the problem of community detection. The standard approach is to 
fit a generative network model on the data~\cite{hastings06,newman07,guimera09,karrer11,ball11,peixoto14b}. 
The stochastic block model (SBM) is by far the most used generative model of graphs with communities 
(see Section~\ref{sec-MV} and references therein). We have seen that it can describe other types of group structure,
like disassortative and core-periphery structure (Fig.~\ref{figSBM}). The unnormalised maximum log-likelihood that a given partition $g$ in $q$ groups 
of the network $G$ is reproduced by the standard SBM reads~\cite{karrer11}
\begin{equation} 
{\cal L}_S(G|g)=\sum_{r,s=1}^q e_{rs}\log \left(\frac {e_{rs}}{n_rn_s}\right),
\label{eqLLSBM}
\end{equation}
where $e_{rs}$ is the number of edges running from group $r$ to group $s$, $n_r$ ($n_s$) the number of vertices in $r$ ($s$) and the sum runs over
all pairs of groups (including when $r=s$). This version of the model, however, does not account for the degree heterogeneity of most real networks, 
so it does a poor job at describing the group structure of many of them. Therefore, Karrer and Newman proposed the {\it degree-corrected stochastic block model} (DCSBM)~\cite{karrer11}, 
in which the degrees of the vertices are kept constant, on average, via the introduction of additional suitable parameters\footnote{The authors were inspired by modularity maximisation, 
which gives far better results when the null model consists of rewiring edges by preserving the degree sequence of the network (on average), than 
by preserving only the total number of edges.}. 
The unnormalised maximum log-likelihood for the DCSBM is
\begin{equation} 
{\cal L}_{DC}(G|g)=\sum_{r,s=1}^q e_{rs}\log \left(\frac {e_{rs}}{e_re_s}\right),
\label{eqLLDC}
\end{equation}
where $e_r$ ($e_s$) is the sum of the degrees of the vertices in $r$ ($s$).

The most important drawback of this type of approach is the need to specify the number $q$ of groups beforehand, which is usually unknown
for real networks.
This is because a straight maximisation of the likelihoods of Eqs.~(\ref{eqLLSBM}) and (\ref{eqLLDC}) over the whole set of possible solutions
yields the trivial partition in which each vertex is a cluster ({\textit{overfitting}}). 
In Section~\ref{sec-tools} we have seen ways to extract $q$ from spectral properties of the graph. But it would be better to have 
statistically principled methods, to be consistent with the approach used to perform the inference. 

A possibility is {\it model selection}, for instance
by choosing the model that best compresses the data~\cite{rissanen78,grunwald05}. The extent of the compression can be estimated via 
the total amount of information necessary to
describe the data, which includes not only the fitted model, but also the information necessary
to describe the model itself, which is a growing function of the number of blocks $q$~\cite{peixoto13}. This quantity, that we indicate with $\Sigma$, is called the
{\it description length}. Minimising the description length naturally avoids overfitting. Partitions with large $q$ are associated to 
``heavy" models in terms of their information content, and do not represent the best compression. On the other hand, partitions with low $q$
have high information content, even if the model itself is not loaded with parameters. Hence the minimum description length 
corresponds to a non-trivial number of groups and it makes sense to minimise $\Sigma$ to infer the block structure of the graph.

It turns out that this approach has a limited resolution on the standard SBM: the maximum number of blocks that can be 
resolved scales as $\sqrt{n}$ for a fixed average degree $\langle k\rangle$, where $n$ 
is the number of vertices of the network. This means that the minimum size of detectable blocks
scales as $\sqrt{n}$, just as it happens for modularity maximisation (Section~\ref{sec-modopt}).
A more refined method of model selection, 
consisting in a nested hierarchy of stochastic block models,
where an upper level of the hierarchy serves as prior
information to a lower level, brings the resolution limit down to $\log n$,
enabling the detection of much
smaller blocks~\cite{peixoto14b}.

Other techniques to extract the number of groups 
have been proposed~\cite{handcock07,daudin08,latouche12,come15,newman16c}.

\subsection{Methods based on optimisation}
\label{sec-modopt}

Optimisation techniques have received the greatest attention in the literature. The goal is finding 
an extremum, usually the maximum, of a function indicating the quality of a clustering, over the space of all possible clusterings.
Quality functions can express the goodness of a partition or of single clusters. 

The most popular quality function is the {\it modularity} by Newman and Girvan~\cite{newman04b}.
It estimates the quality of a partition of the network in communities.
The general expression of modularity is 
\begin{equation}
Q=\frac{1}{2m}\sum_{ij}\left(A_{ij}-P_{ij}\right)\delta(C_i,C_j),
\label{eq:mod0}
\end{equation}
where $m$ is the number of edges of the network, the sum runs over all pairs of vertices $i$ and $j$, $A_{ij}$ is the element of the adjacency matrix, 
$P_{ij}$ is the {\it null model term} and
in the Kronecker delta at the end $C_i$ and $C_j$ indicate the communities of $i$ and $j$. The term $P_{ij}$ indicates the average adjacency matrix of an ensemble of networks,
derived by randomising the original graph, such to preserve some of its features. Therefore, modularity measures how different the original graph is from 
such randomisations. The concept was inspired by the idea that by randomising the network structure communities are destroyed, so the comparison between
the actual structure and its randomisation reveals how non-random the group structure is. A standard choice is $P_{ij}=k_ik_j/2m$, 
$k_i$ and $k_j$ being the degrees of $i$ and $j$, and 
corresponds to the expected number of edges 
joining vertices $i$ and $j$ if the edges of the network were rewired such to preserve the degree of all vertices, on average. This yields the classic form of modularity
\begin{equation}
Q=\frac{1}{2m}\sum_{ij}\left(A_{ij}-\frac{k_ik_j}{2m}\right)\delta(C_i,C_j).
\label{eq:mod}
\end{equation}
Other choices of the null model term allow us to incorporate specific features of network structure, like bipartiteness~\cite{barber07}, correlations~\cite{macmahon15}, signed edges~\cite{traag09}, space
embeddedness~\cite{expert11}, etc.. The extension of Eq.~(\ref{eq:mod}) and of its variants to the case of weighted networks is straightforward~\cite{newman04d}. For simplicity we focus on 
unweighted graphs here, but the issues we discuss are general. 

\begin{figure}[h!]
\begin{center}
\includegraphics[width=0.47\columnwidth]{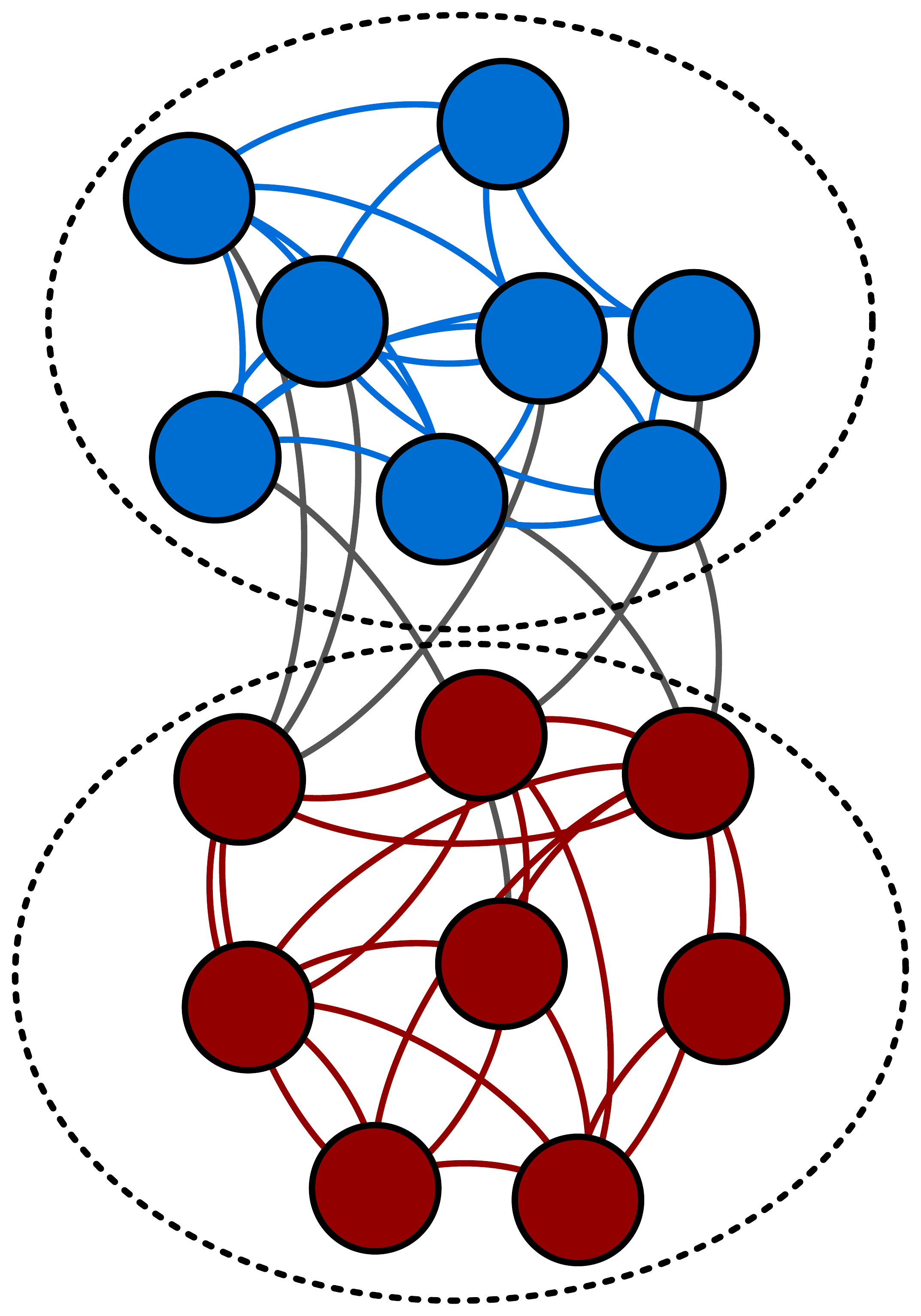}
\includegraphics[width=0.47\columnwidth]{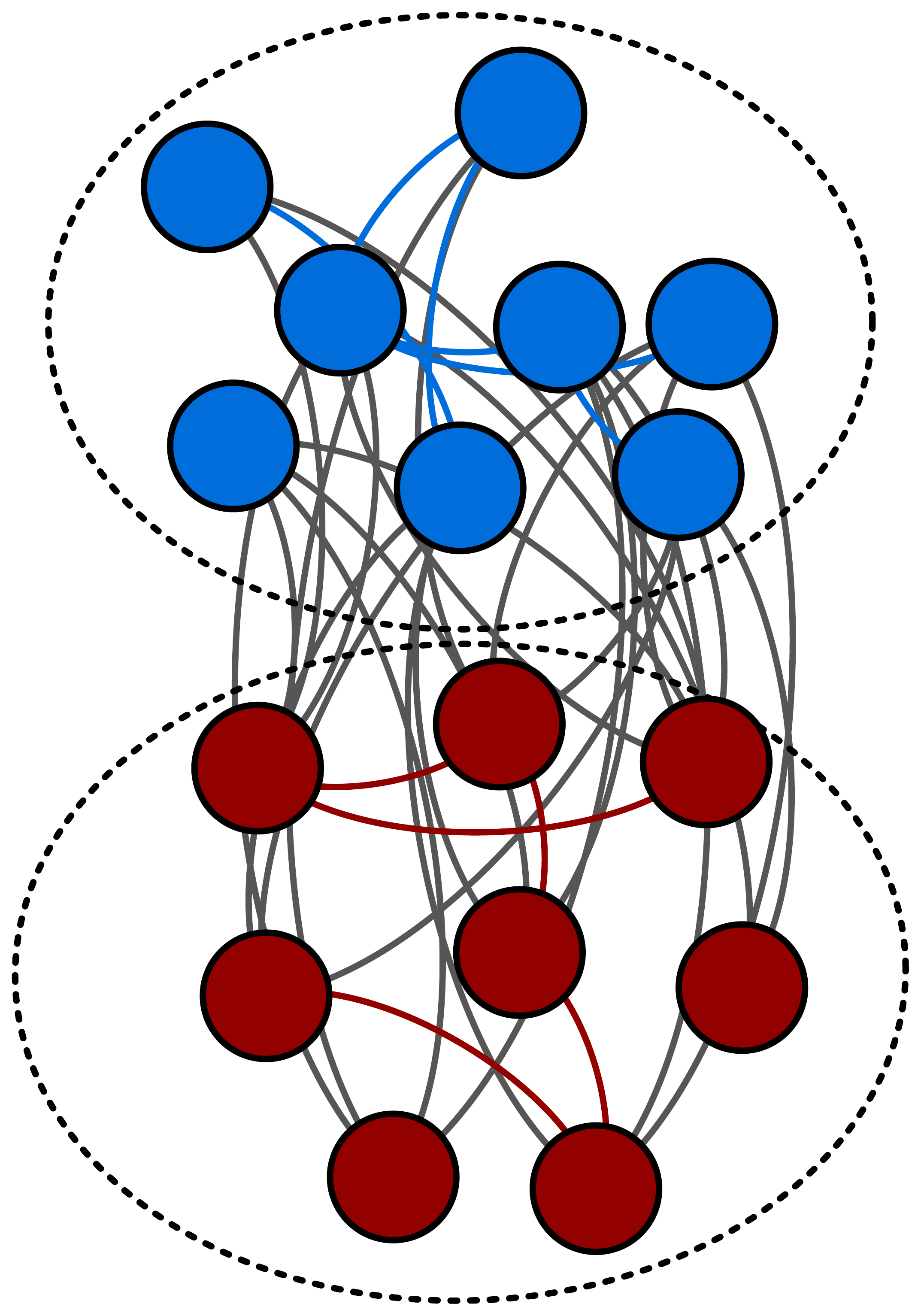}
\caption {Modularity by Newman and Girvan. The network on the left has a visible community structure, with two clusters, whose vertices are
highlighted in blue and red, respectively. Modularity measures how different the clusters of the partition are from the corresponding 
clusters of the ensemble of random graphs obtained by randomly joining the vertices, such to preserve their degrees, on average. The picture on the right shows
the result of one such randomisation. The internal edges are coloured in blue and red. They are just a handful compared to the number of edges
joining the same groups of vertices in the original network (blue and red lines in the left picture), while there are now many more edges running between
the subgraphs (black lines): the randomisation has destroyed the community structure of the 
graph, as expected. The value of modularity for the bipartition on the left is expected to be large.}
\label{mod-ill}
\end{center}
\end{figure}

Because of the delta, the only contributions to the sum come from vertex pairs belonging to the same
cluster, so we can group these contributions together and rewrite the sum over the vertex pairs 
as a sum over the clusters\footnote{A partition quality function that can be formulated as a sum over the clusters is called {\it additive}.}
\begin{equation}
Q=\sum_{C}\Big[\frac{l_C}{m}-\left(\frac{k_C}{2m}\right)^2\Big].
\label{eq:mod1}
\end{equation}
Here $l_C$ the total number of edges joining vertices of community $C$ and 
$k_C$ the sum of the degrees of the vertices of $C$ (Section~\ref{sec-var}). 
The first term of each summand in Eq.~(\ref{eq:mod1}) is the 
fraction of edges of the graph falling within community $C$, whereas the second term is the expected fraction of 
edges that would fall inside $C$ if the graph were taken from the ensemble of
random graphs preserving the degree of each vertex of the original network, on average. 
The difference in the summand would then indicate how ``non-random" subgraph $C$ is. The larger the difference
the more confident we can be that the placement of edges within $C$ is not random (Fig.~\ref{mod-ill}). Large values of $Q$ are then supposed to 
indicate partitions with high quality.

Modularity maximisation is NP-hard~\cite{brandes08b}. Therefore one can realistically hope to find only 
decent approximations of the modularity maximum and a wide variety of approaches has been proposed. Due to its simplicity, the prestige of its
inventors and early results on the benchmark of Girvan and Newman (Section~\ref{art-bench}) and on small real benchmark networks, like Zachary karate club network
(Fig.~\ref{zach}), modularity has become the best known and most studied object in network clustering. In fact, soon after its introduction, it seemed 
to represent the essence of the problem, and the key to its solution.

However, it became quickly clear that the measure is not as good as it looks. 
For one thing, there are high-modularity partitions even in random graphs without groups~\cite{guimera04}. This seems counterintuitive,
given that modularity has been designed to capture the difference between random and non-random structure. 
Modularity is a sort of distance between the actual network and {\it an average} over random networks, ignoring altogether the distribution
of the relevant community variables, like the fractions of edges within the clusters, over all realisations generated by the configuration model. 
If the distribution is not strongly peaked, the values of the community variables measured on the original graph may be found in 
a large number of randomised networks, even though the averages look far away from them. In other words, we should pay more attention to the {\it significance} 
of the maximum modularity value $Q_{max}$, 
than to the value itself. How can we estimate the significance of $Q_{max}$?
A natural way is maximising $Q$ over all partitions of every randomised graph. 
One then computes the average $\langle Q_{rand}\rangle$ and the standard deviation $\sigma_Q^{rand}$ of the resulting values.
The statistical significance of $Q_{max}$ is indicated 
by the distance of $Q_{max}$ from the null model average $\langle Q_{rand}\rangle$
in units of the standard deviation $\sigma_Q^{rand}$, i. e., by the $z$-score
\begin{equation}
z=\frac{Q_{max}-\langle Q_{rand}\rangle}{\sigma_Q^{rand}}.
\label{eq:zscore}
\end{equation}
If $z\gg 1$, $Q_{max}$ indicates strong community structure. This approach has problems, though.
The main drawback is that the distribution of $Q_{rand}$ over the ensemble of null model random graphs, though peaked, 
is not Gaussian. Therefore, one cannot attribute to the values of the $z$-score the significance  
corresponding to a Gaussian distribution, and one ought to compute the statistical significance for the correct distribution. Also, the $z$-score depends on 
the network size, so the same values may indicate different levels of significance for networks differing considerably in size. 

\begin{figure}
\begin{center}
\includegraphics[width=\columnwidth]{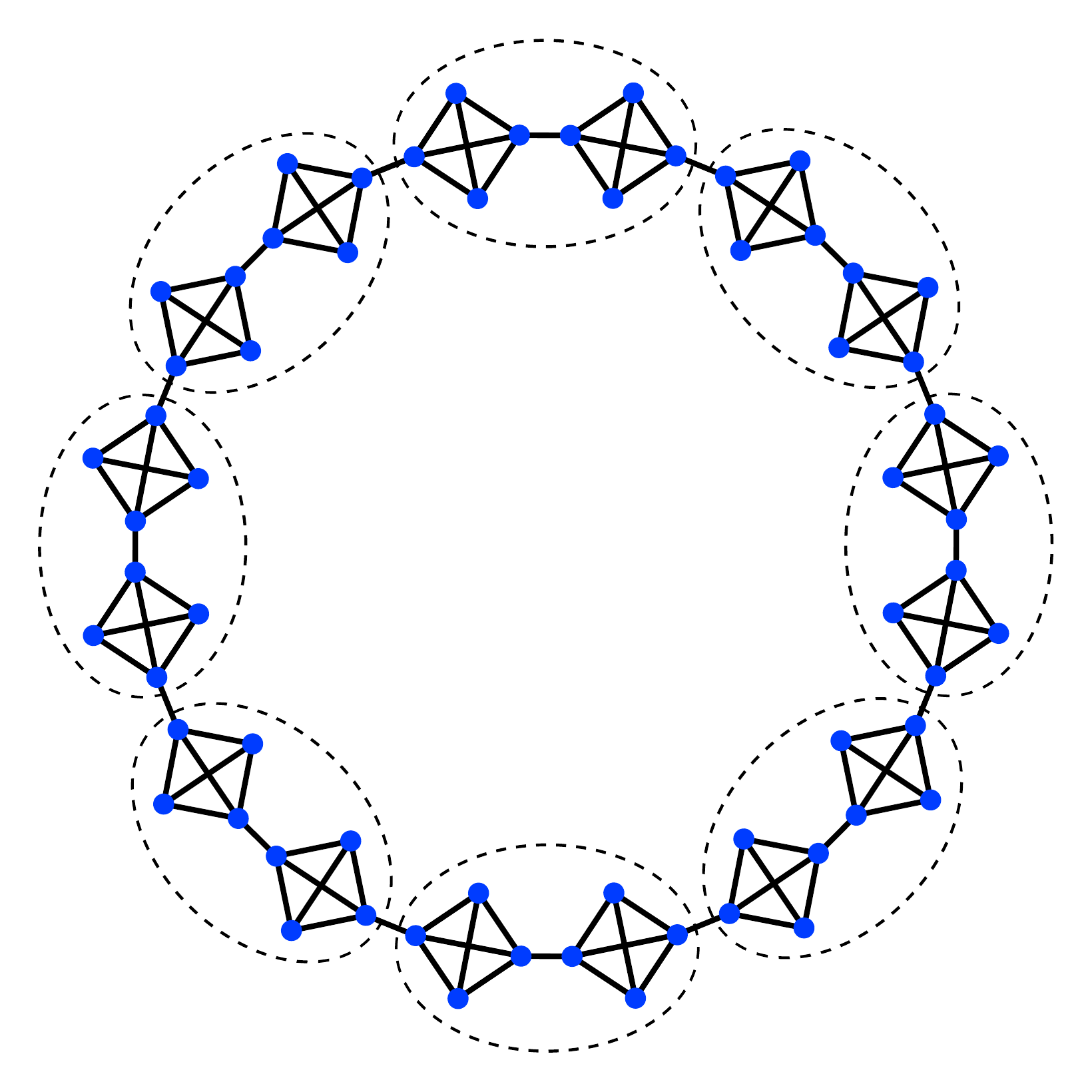}
\caption {Resolution limit of modularity optimisation. The network in the figure is made of cliques of four vertices, arranged such to form a ring-like structure, with 
each clique joined to two other cliques by a single edge. The na\"{\i}ve expectation is that modularity would reach its maximum for the partition whose communities are the cliques, 
which appears to be the natural partition of the network.
However, it turns out that there are partitions with higher modularity, whose clusters are combinations of cliques, like the partition indicated by the dashed contours.}
\label{reslim}
\end{center}
\end{figure}

Next, it is not true that the modularity maximum always corresponds to the most pronounced community structure of a network.
In Fig.~\ref{reslim} we show the well-known example of the ring of cliques~\cite{fortunato07}. The network consists of $16$ cliques with four vertices each. Every clique has two neighbouring 
cliques, connected to it via a single edge. Intuition suggests that the graph has a natural community structure, with $16$ communities, each corresponding to one clique. Indeed, 
the $Q$-value of this partition is $Q_1=89/112\approx 0.79464...$, pretty close to $1$, which is the upper bound of modularity. However, there are partitions with larger values, like the partition in $8$ clusters 
indicated by the dashed contours, whose modularity is $Q_2=90/112\approx 0.80357> Q_1$.

This is due to the fact that $Q$ has a preferential scale for the communities, deriving from the underlying null model and revealed by
its explicit dependence on the number of edges $m$ of the network [Eq.~(\ref{eq:mod1})]. According to the configuration model, 
the expected number $l_{AB}$ of edges running between two subgraphs $A$ and $B$ with total degree $k_A$ and $k_B$, respectively, is approximately
$k_Ak_B/2m$. Consequently, if $k_A$ and $k_B$ are of the order of $\sqrt{m}$ or smaller, $l_{AB}$ could become smaller than $1$. This means that
in many randomisations of the original graph $G$, subgraphs $A$ and $B$ are disconnected and even a single edge joining them in $G$ 
signals a non-random association. In these cases, modularity is larger when $A$ and $B$ are put together than when they are 
treated as distinct communities, as in the example of Fig.~\ref{reslim}.
The modularity scale depends only on the number of edges $m$, and it may have nothing to do with the size of the actual communities of the network.
The resolution limit questions the usefulness of modularity in practical applications~\cite{fortunato07}.

Many attempts have been made to mitigate the consequences of this disturbing feature. One approach consists in introducing a resolution parameter $\gamma$ into
modularity's formula~\cite{reichardt06,arenas08b}.
By tuning $\gamma$ it is possible to arbitrarily vary the resolution scale of the method, going from very large to very small communities.
We shall discuss such multi-resolution approaches in Section~\ref{sec-dynmet}.
Here we emphasize that 
multi-resolution versions of modularity do not provide a reliable solution to the problem. This is because modularity maximisation has an additional bias: large subgraphs are usually split in smaller pieces~\cite{lancichinetti11b}. This problem has the same source as the resolution limit, namely the choice of the null model. Since modularity has a preferential scale for the communities, when a subgraph is too large it is convenient to break it down, to increase the modularity of the partition. So, when there is no characteristic scale
for the communities, like when there is a broad cluster size distribution, large communities are likely to be broken, and small communities are likely to be merged.
Since multi-resolution versions of modularity can only shift the resolution scale of the measure back and forth, they are unable to correct 
both effects at the same time\footnote{More promising results can be obtained with hierarchical multi-level methods, in which multi-resolution modularity is applied 
iteratively on every cluster with independent resolution parameters, so that a coexistence of very diverse scales is permitted~\cite{granell12}. Such approaches, however, deviate
from the original idea of modularity maximisation, which is based on a global null model valid for the network as a whole.}~\cite{lancichinetti11b}. 
In addition, tuning the resolution parameter in the search for 
good partitions is usually computationally very demanding, as in many cases the optimisation procedure has to be repeated over and over for all $\gamma$-values one 
desires to investigate.

\begin{figure}[h!]
\begin{center}
\includegraphics[width=\columnwidth]{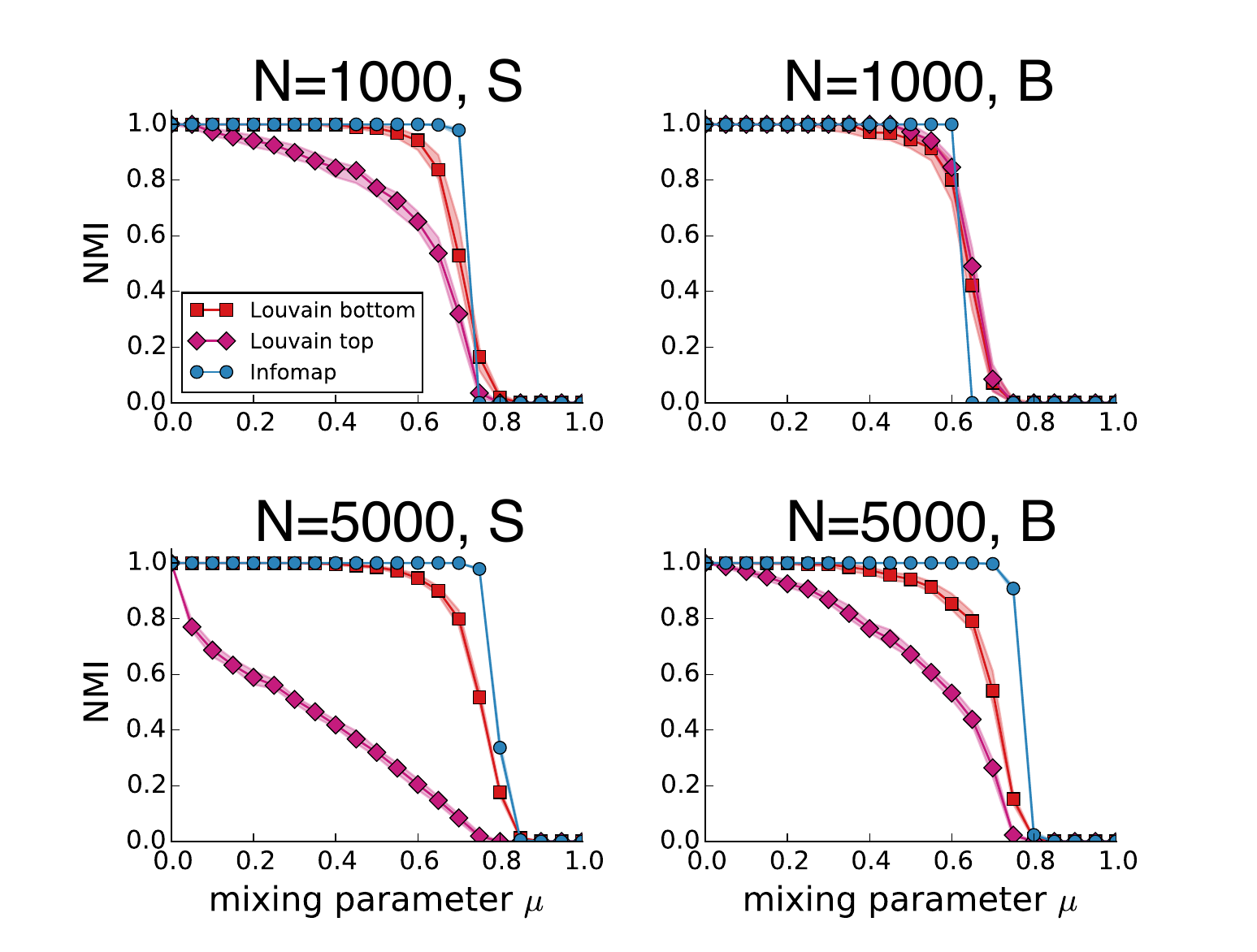}
\caption {Performance of the Louvain method. The panels indicate the accuracy of the algorithm to detect the planted partition of the LFR benchmark as a function of the mixing parameter, for different
choices of the network size ($1000$ and $5000$ vertices) and of the range of community sizes (label $S$ indicates that communities have between $10$ and $50$ vertices, label
$B$ that they have between $20$ and $100$ vertices). Accuracy is calculated via the normalised mutual information (NMI), in the version by Lancichinetti, Fortunato and Kert\'esz~\cite{lancichinetti09}.
Results are heavily depending on the hierarchical level one chooses at the end of the procedure. When one picks the top level (diamonds), which is the one 
with largest modularity, the accuracy is poor, as expected, especially when communities are smaller. 
When one goes for the bottom level (squares), which has lower modularity and smaller clusters than the top level 
partition, 
there is a far better agreement with the 
planted partition and the performance gets closer to that of Infomap (circles). The squares follow the performance curves 
used in the comparative analysis by Lancichinetti and Fortunato~\cite{lancichinetti09c}.
Courtesy from Andrea Lancichinetti.}
\label{louvainL}
\end{center}
\end{figure}

We stress that the resolution limit is a feature of modularity itself, not of the specific way adopted to maximise it. Therefore, there is no magic heuristic 
that can circumvent this issue. The Louvain method~\cite{blondel08} has been held as one such magic heuristic. The method performs a greedy optimisation of $Q$ in a hierarchical manner, 
by assigning each vertex to the community of their neighbours yielding the largest $Q$, and creating a smaller weighted super-network whose vertices are the clusters found previously.
Partitions found on this super-network hence consist of clusters including the ones found earlier, and represent a higher hierarchical level of clustering.
The procedure is repeated until one reaches the level with largest modularity. In the comparative analysis of clustering algorithms 
performed by Lancichinetti and Fortunato on the LFR benchmark~\cite{lancichinetti09c},
the Louvain algorithm was the second best-performing method, after Infomap~\cite{rosvall08}. This has given the impression that the peculiar strategy of the method 
solves the resolution problems above, which is not true. The reason why the performance is so good is that Lancichinetti and Fortunato adopted the lowest partition of the hierarchy, 
the one with the smallest clusters~\cite{lancichinetti14}. By using the partition with highest modularity performance degrades considerably
(Fig.~\ref{louvainL}), as expected. As suggested by the developers of the algorithm themselves, using the lowest level helps avoiding unnatural community mergers; as an example,
they showed that the natural partition of the ring of cliques (Fig.~\ref{reslim}) can be recovered this way. 
However, the bottom level has lower modularity than the top level, so 
we face a sort of contradiction, in that users are encouraged to use suboptimal partitions, even though one assumes that the best clustering corresponds to the highest value of the quality function,
which is what the method is supposed to find.
There is no guarantee that the bottom level yields the most meaningful solution. On the other hand, users have the option of choosing among a few partitions and a slightly higher chance to find what they 
search for.

Moreover, the modularity landscape
is characterised by a larger than exponential\footnote{Exponential in the number $n$ of graph vertices.} number of distinct partitions, whose modularity values are very close
to the global maximum~\cite{good10}. This explains why many heuristic methods of modularity maximisation are able to
come very close to the global maximum of $Q$, but it
also implies that the global maximum is basically impossible
to find. In addition, high-$Q$ partitions are
not necessarily similar to each other, despite the proximity
of their modularity scores. The optimal structural partition, 
which may not correspond
to the modularity maximum due to problems exposed above, may however have a large $Q$-value. Therefore
the optimal partition is basically indistinguishable
from a huge number of high-modularity partitions, which
are in general structurally dissimilar from it. The large
structural diversity of high-modularity partitions
implies that one cannot rely on any of them, at
least in principle. Reliable solutions could be singled out when the domain user imposes some constraints on the clustering
of the system, or when she expects it to have specific features. In the absence of additional information or expectations, 
consensus clustering could be used to derive more robust partitions. Indeed, it has been shown that the consensus of many high-modularity partitions, combined with a hierarchical
approach, could help to solve resolution problems and to avoid to find communities in random graphs without groups~\cite{zhang14b}.

As of today, modularity optimisation is still the most used clustering technique in applications.
This may appear odd, given the serious issues of the method and the fact that nowadays more powerful techniques are available, like a posteriori stochastic block modelling (Section~\ref{sec-WMT}).
Indeed Newman has proven that optimising modularity is equivalent to maximising the likelihood that the planted partition model reproduces the network~\cite{newman16b}. 
But the planted partition model is a very specific case of the general stochastic block model, in that the intra-group edge probabilities are all equal to the same value $p_{in}$ and 
the inter-group edge probabilities are all equal to the same value $p_{out}$. There is no reason to limit the inference to this specific case, when one could use the full model.
 
Optimising partition quality functions may lead to resolution problems, just like 
it happens for modularity\footnote{Traag and Van Dooren have shown that one can design additive quality functions such that the best partition of the network 
induces the optimal partition for any subgraph $S$, i. e., the partition found when the 
detection is performed only on $S$~\cite{traag11}. This is possible when the coefficients of the summand corresponding to each community
does not depend on global properties of the graph. Even those functions have their own preferential community scale, though.}. 
Instead, one could try to optimise {\it cluster quality functions}. One starts with some function $q(C)$ expressing how ``community-like" a subgraph is
and with a seed vertex $s$. The goal is to build a cluster $C_s$ including $s$ such that $q(C_s)$ is maximum\footnote{For some functions the optimum corresponds to their minimum, not the maximum. 
This occurs when they
are related to variables that are supposed to be small when communities are good, like the density of edges between clusters.}. This is usually done by exploring the neighbours of the temporary
subgraph $C_s$, starting from the neighbours of $s$ when $C_s$ includes only $s$. The neighbouring vertex whose inclusion yields the largest increase 
of $q$ is added to the subgraph.
When a new vertex is included, the structure of the subgraph is altered and the other vertices can be examined again, as it might be advantageous to knock some of them out. The process stops when the
quality $q(C_s)$ cannot be increased anymore via the inclusion or the exclusion of vertices.

The optimisation of cluster quality functions offers a number of advantages over the optimisation of partition quality functions. First, it is consistent with the idea that communities are local structures,
which are sensitive to what happens in their neighbourhood, but are fairly unaffected by the rest of the network: the structure of a social circle in Europe is hardly influenced by the dynamics of 
social circles in Australia, though they are parts of the same global social network of humans. 
Consequently, if a network undergoes structural changes in one region,
community structure is altered and is to be recovered only in that region, while the clustering of the rest of the network remains the same. By optimising partition quality functions,
instead, any little change may have an effect on every community of the graph.
Second, since cluster quality functions do not embody any global scale, severe resolution problems are usually avoided\footnote{If one
defines the quality of the cluster with respect to the rest of the network, global scales may still slip into the function.}. Moreover, one can
investigate only parts of the network, which is particularly valuable when the graph is large and a global analysis would be out of reach, computationally. 
The local exploration of the graph allows to reach vertices already assigned to clusters, so overlaps can be naturally detected.
In the last years several 
algorithms based on the optimisation of cluster quality functions have been designed~\cite{baumes05,clauset05,lancichinetti09,lancichinetti11,huang11}.

\subsection{Methods based on dynamics}
\label{sec-dynmet}

Communities can also be identified by running dynamical processes on the network, like diffusion~\cite{zhou03,zhou03b,zhou04,vandongen00,pons05,rosvall08,jeub15}, 
spin dynamics~\cite{reichardt06,ronhovde10,traag11,raghavan07}, 
synchronisation~\cite{arenas06,boccaletti07}, etc.. In this section we focus on diffusion and spin dynamics, that inform most approaches.

Random walk dynamics is by far the most exploited in community detection. If communities have high internal edge density and are well-separated
from each other, random walkers would be trapped in each cluster for quite some time, before finding a way out and migrating to another cluster.
We briefly discuss two broad classes of algorithms:  methods based on vertex similarity and methods based on the map equation.

The first class of techniques consists in using random walk dynamics to estimate the similarity between pairs of vertices. For instance, in the popular 
method {\it Walktrap} the similarity between vertices $i$ and $j$ is given by 
the probability that a random walker moves from
$i$ to $j$ in a fixed number of steps $t$~\cite{pons05}. The parameter $t$ has to be large enough, to allow for the exploration of
a significant portion of the graph, but not too big,
as otherwise one would approach the stationary limit
in which transition probabilities trivially depend on
the degrees of the vertices. 
\begin{figure*}
\begin{center}
\includegraphics[width=\textwidth]{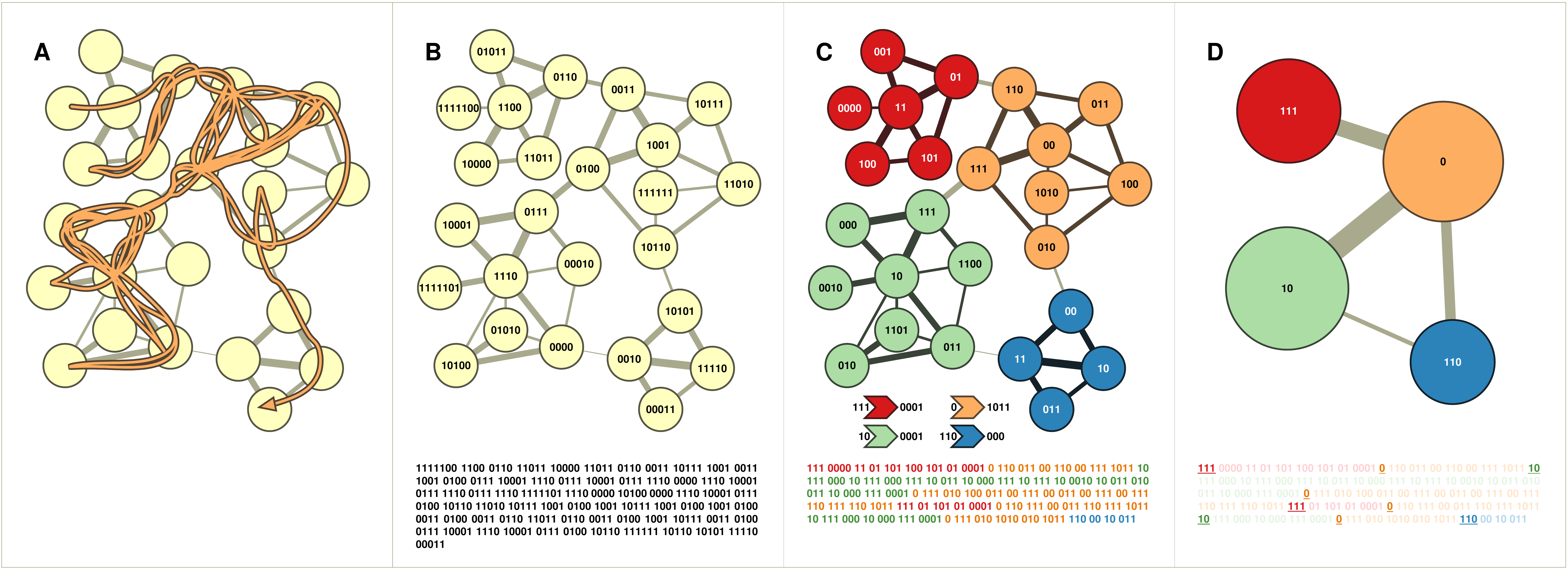}
\caption {Infomap. The random walk in (A) can be described as a sequence of the vertices, each labeled with unique codewords (B), or 
by dividing the graph in regions and using unique codewords only for the vertices of the same region (C). This way the same codeword can be used for multiple vertices,
at the cost of indicating when the random walker leaves a region to enter a new one, as in that case one has to specify the codeword of the new region, to unambiguously locate the walker. 
The network has 
four communities [indicated by the colours in (C)], and in this case the map-like description of (C) is more parsimonious than the one in (B). 
This is shown by looking at the actual code needed in either case (bottom of the figures), which is clearly shorter for (C). In (D) the transitions between the clusters are highlighted.
Reprinted figure with permission from \cite{rosvall08}.
\copyright\,2008, by the National Academy of Sciences, USA.}
\label{figinfomap}
\end{center}
\end{figure*}
If there is a pronounced community structure, pairs of vertices in the same cluster are much more easily reachable by a random walk
than pairs of vertices in different clusters, so the vertex similarity is expected to be considerably higher within groups than between groups\footnote{A related method is the 
{\it Markov Cluster Algorithm} (MCL)~\cite{vandongen00}, which consists of iterating 
two operations: raising to a power the {\it transfer matrix} ${\bf T}$ of the graph, whose element $T_{ij}$ 
equals the probability that a random walker, sitting at $j$, moves to $i$; raising the elements of the resulting matrix to a power, 
such that the larger values are enhanced with respect to the smaller ones, many of which are set to zero to lighten the calculations, while the remaining ones are normalised by dividing them by the sum of 
elements of their column, yielding a new transfer matrix. The process
eventually reaches a stationary state, corresponding to the matrix of
a disconnected graph, whose connected components are the sought
clusters.}.
In that case, clusters can be readily identified via standard hierarchical or partitional clustering techniques~\cite{jain99,xu08}. 
This class of methods have a high computational complexity, higher than quadratic in the number $n$ of 
vertices (on sparse graphs), so they cannot be used on large networks. Besides, they are often parameter-dependent.

The map equation stems from a seminal paper by Rosvall and Bergstrom~\cite{rosvall08}, who asked
what is the most parsimonious way to describe an infinitely long random walk taking place on the graph. The information content of 
any description is given by the total number of bits required to indicate the various stages of the process. The simplest description 
is obtained by listing sequentially all vertices reached by the random walker, each vertex being described by a unique codeword. However, if the network has a community structure,
there may be a more compact description, which follows the principle of geographic maps, where there are multiple cities and streets with the same name across regions.
Vertex codewords could be recycled among different communities, which play the role of regions/states, and vertices with identical name are distinguished by specifying the  
community they belong to. If clusters are well separated from each
other, transitions between clusters
are infrequent, so it is advantageous to use the map,
with the communities as regions, because in the description
of the random walk the codewords of the clusters will
not be repeated many times, while there is a considerable
saving in the description due to the limited length
of the codewords used to denote the vertices (Fig.~\ref{figinfomap}). 
The map equation yields the description length of an infinite random walk consists of two terms, expressing
the Shannon entropy of the walk within and between
clusters. The best partition is the one yielding the
minimum description length.

This method, called Infomap, can be applied to
weighted networks, both undirected and directed. In the
latter case, random walk dynamics is modified by introducing
a teleportation probability, as in the PageRank process~\cite{brin98}, to ensure that a non-trivial stationary state is reached.
It has been successively extended to the detection of hierarchical community structure~\cite{rosvall11} and of overlapping clusters~\cite{esquivel11}.
In classic random walks the probability of reaching a vertex only depends on where the walker stands, not 
on where it is coming from. The map equation has also been extended to random walks whose 
transition probabilities depend on earlier steps too (higher-order Markov dynamics)~\cite{rosvall14,persson16}, retaining memory of the (recent) past.
Applications show that in this way one can recover 
overlapping communities more easily than by using standard first-order random walk dynamics, especially pervasive overlaps, which are usually out of reach for most
clustering algorithms (Section~\ref{ncp}). 

Infomap and its variants usually return different partitions than structure-based methods (e. g., modularity optimisation). This is because 
they are based on flows running across the system, as opposed to structural variables like number of edges, vertex degrees, etc.. The difference is particularly
striking on directed graphs~\cite{rosvall08}, where edge directions heavily constrain the possible flows. 
Structural features obviously play a major role
on the dynamics of processes running on graphs, but dynamics cannot be generally reduced to an interplay of structural elements, at least not simple ones like, e. g.,
vertex degrees. Sometimes structural and dynamic approaches are equivalent, though. 
For instance, Newman-Girvan's modularity is a special case of a general quality function, called
{\it stability}, expressing the persistence of a random walk within communities~\cite{delvenne10,lambiotte08}.

The methods we have discussed so far are global, in that they aim at finding the whole community structure of the system. However, 
random walks along with other dynamical processes can be used
as well to explore the network locally, starting from seed vertices~\cite{jeub15}. Good communities correspond to bottlenecks of the dynamics and
depend on the choice of the seed vertices, the time scale of the dynamics, etc.. Such local perspective enables to identify community overlaps in a natural way, 
due to the possibility of reaching vertices multiple times, even if they are already classified.

Spin dynamics~\cite{baxter07} are also regularly used in network clustering. The first step is to define a spin model on the network, consisting of
a set of spin variables $\{s_i, i=1, 2, \dots, n\}$, assigned to the vertices and a Hamiltonian ${\cal H}(\{s\})$, expressing the energy of the 
spin configuration $\{s\}$. For community detection, spins are usually integers: $s=1, 2, \dots, q$.
Contributions to the energy are usually given by spin-spin interactions. The coupling of a spin-spin interaction can be {\it ferromagnetic} (negative)
or {\it antiferromagnetic} (positive), if the energy is lower when the spins are equal or not, respectively. The goal is to find those spin configurations
that minimise the Hamiltonian ${\cal H}(\{s\})$. If couplings are all ferromagnetic, the minimum energy would be trivially obtained for the configurations where all vertices have identical spin 
values. Instead, one would like to have identical spins for vertices of the same cluster, and different spins for vertices in different clusters, to identify the community structure.
Therefore, Hamiltonians feature both ferromagnetic and antiferromagnetic interactions [{\it spin glass dynamics}~\cite{mezard87}]. A popular model consists in rewarding
edges between vertices in the same cluster, as well as non-edges between
vertices in different clusters, and penalising edges between
vertices of different clusters, along with non-edges between
vertices in the same cluster. This way, if the edge density within communities is appreciably larger than the edge density between communities, as it often happens,
having equal spin values for vertices in the same cluster would considerably lower the energy of the configuration. On the other hand, to bring the energy further down
the spins of vertices in different clusters should be different, as many such vertices would be disjoint from each other, and such non-edges would increase the energy
of the system if the corresponding spin variables were equal. 
A general expression for the Hamiltonian along these lines is~\cite{reichardt06}
\begin{equation}
{\cal H}(\{s\})=-\sum_{ij} [a_{ij}A_{ij}-b_{ij}(1-A_{ij})]\delta(s_i,s_j)\,,
\label{eqRBgen}
\end{equation}
where $A_{ij}$ is the element of the adjacency matrix, $a_{ij}, b_{ij} \geq 0$ are arbitrary coefficients, and 
the Kronecker delta selects only the pairs of vertices with the same spin value.

A popular model is obtained by setting $a_{ij}=1-b_{ij}$ and $b_{ij}=\gamma P_{ij}$, where 
$\gamma$ is a tunable parameter and
$P_{ij}$ a null model term, expressing the expected number of edges running between vertices $i$ and $j$ under a suitable randomisation of the 
graph structure. The resulting Hamiltonian is~\cite{reichardt06}
\begin{equation}
{\cal H}_{RB}(\{s\})=-\sum_{ij} (A_{ij}-\gamma P_{ij})\delta(s_i,s_j)\,.
\label{eqRB}
\end{equation}
If $\gamma=1$ and $P_{ij}=k_ik_j/2m$, $k_i$ ($k_j$) being
the degree of $i$ ($j$) and $m$ the total number of graph edges, 
the Hamiltonian of Eq.~(\ref{eqRB}) coincides with the modularity by Newman and Girvan [Eq.~(\ref{eq:mod})], up to an irrelevant multiplicative constant.
Consequently, modularity can be interpreted as the Hamiltonian of a spin glass as well. 

By setting $a_{ij}=1$ and $b_{ij}=\gamma$ we obtain 
the {\it Absolute Potts Model} (APM)~\cite{ronhovde10}, whose Hamiltonian reads
\begin{equation}
{\cal H}_{APM}(\{s\})=-\sum_{ij} [A_{ij}-\gamma (1-A_{ij})]\delta(s_i,s_j)\,.
\label{eqAPM}
\end{equation}
Here, there is no null model term. The models of Eqs.~(\ref{eqRB}) and (\ref{eqAPM}) can be trivially extended to weighted graphs~\cite{traag11}.
They allow to explore 
the network at different resolutions, by suitably tuning the parameter $\gamma$.
However, there usually is no information about the community sizes,
so it is not possible to decide {\it a priori} the proper value(s) of $\gamma$ for a specific graph. A common heuristic is to 
estimate the {\it stability} of partitions as a function of $\gamma$. 
It is plausible that partitions recovered for a given $\gamma$-value will not be disrupted if $\gamma$ is varied a little. 
So, the whole range of $\gamma$ can be split into intervals, each interval corresponding to the most frequent partition detected in it. Good candidates 
for the unknown community structure of the system could be the partitions found in the widest intervals of $\gamma$, 
as they are likely to be more stable (or robust) than the other partitions\footnote{We stress, however, that the persistence of partitions in intervals is not necessarily related 
to clustering robustness~\cite{lewis10,onnela12}.}. However,
the results of the algorithm do not usually have a linear relationship with $\gamma$, hence the 
width of the intervals is not necessarily correlated with stability, as intervals of the same width but centred at different values of $\gamma$ may have rather different 
importance. 

A good operational definition of stability is based on the stochastic character of optimisation methods, which typically deliver different results for the same system and set of parameters, 
by changing initial conditions and/or random seeds.
If a partition is robust in a given range of $\gamma$-values, most partitions delivered by the algorithm will be very similar. On the other hand, if 
one explores a $\gamma$-region in between two strong partitions, the algorithm will deliver the one or the other partition and the 
individual replicas will be, on average, not so similar to each other. So, by calculating the similarity $S(\gamma)$ of partitions found by the method at a given resolution
parameter $\gamma$ (for different choices of initial conditions and random seeds), stable communities are revealed by peaks of $S(\gamma)$~\cite{ronhovde09}. 
Since clustering in large graphs can be very noisy, peaks may not be well resolved. 
Noise can be reduced by working with consensus partitions of the individual partitions returned by the method for a given $\gamma$
(Section~\ref{sec-consensus}). These manipulations are computationally costly, though. Besides, multi-resolution techniques may miss
relevant cluster sizes, as it happens for multi-resolution modularity~\cite{lancichinetti11} (Section~\ref{sec-modopt}).

\subsection{Dynamic clustering}
\label{sec-dynclus}

Due to the increasing availability of time-stamped network data, there is currently a lot of activity 
on the development of methods to analyse temporal networks~\cite{holme12}. In particular, 
the problem of detecting dynamic communities has received a lot of attention~\cite{fortunato10,spiliopoulou11}. 

Clustering algorithms used for static graphs can be (and often are) used for dynamic networks as well.
What needs to be established is how to handle the evolution. Typically one can describe it as a succession of {\it snapshots} $G_1, G_2, \dots, G_l$, 
where each snapshot $G_t$ corresponds to the configuration of the graph in a given time window\footnote{A graph configuration consists of the sets of vertices and edges that
are active within the given time frame, along with the intensity of their interactions in that frame (weights) and possibly other aspects of the dynamics, 
like burstiness~\cite{barabasi10}, duration of the interactions, etc..}. There are two possible strategies.

The simplest approach is to detect the community structure for each individual snapshot, which is a static graph~\cite{hopcroft04,asur07,palla07}.
Next, pairs of communities of consecutive windows are associated.
A standard procedure is finding the cluster $C_{t}^i$ in window $t$ that is most similar to cluster $C_{t+1}^j$ in window $t+1$, for instance by using Jaccard 
similarity score [Eq.~(\ref{eqt20})]~\cite{palla07}. This way every community has an image in each phase of the network evolution and one can track 
its dynamics. Various scenarios are possible. Communities may disappear at some point and
new communities may appear, following the exclusion or the introduction of vertices and edges, respectively. 
Furthermore, a cluster may fragment into smaller ones or merge with others.
However, since snapshots are handled separately, this strategy
often produces significant variations between partitions
close in time, especially when the
data sets are noisy, as it usually happens in applications. 

It would be preferable to have a unified framework,
in which communities are inferred both from the current
structure of the graph and from the knowledge of
the community structure at previous times.
An interesting implementation of this strategy is
{\it evolutionary clustering}~\cite{chakrabarti06}. The goal of the approach
is finding a partition that is both faithful to the system configuration at snapshot $t$ and close to
the partition derived for the previous snapshot $t-1$.
A cost function is introduced, whose optimisation yields a tradeoff between such two constraints.
There is ample flexibility on how this can be done, in practice.
Many known clustering techniques normally used for static graphs
can be reformulated within this evolutionary framework. Some interesting 
algorithms based on evolutionary clustering have been proposed~\cite{chi07,lin08}.
Mucha et al. have also presented a method that couples the system's configurations of different snapshots, within a modularity-based framework~\cite{mucha10}.
In the resulting quality function ({\it multislice modularity}), all configurations are simultaneously taken into account and the coupling between them 
is expressed by a tunable parameter. The approach can handle 
general multilayer networks~\cite{boccaletti14,kivela14}, where layers are either networks whose vertices are connected by a specific edge type (e. g., friendship, 
work relationships, etc., in social networks), or networks whose vertices have connections (interactions, dependencies) with the vertices of other networks/layers.
On the other hand, since the approach is based on modularity optimisation, it has the drawbacks exposed in Section~\ref{sec-modopt}.

\begin{figure}[h!]
\begin{center}
\includegraphics[width=\columnwidth]{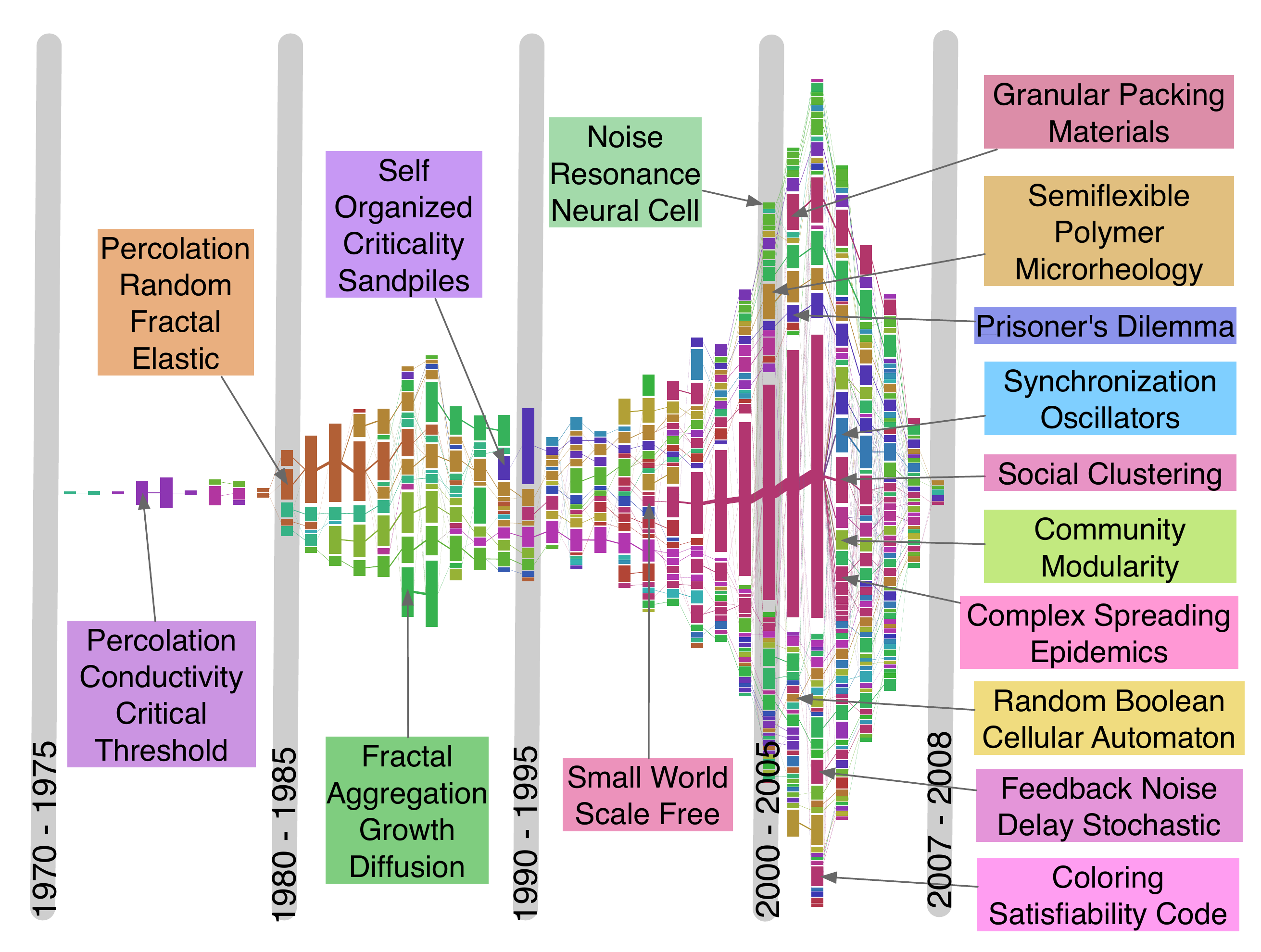}
\caption {Consensus clustering on dynamic networks. Time evolution of clusters of the citation network of 
papers published in journals of the American Physical Society (APS).
The clusters with  
the keyword {\it Network(s)}
among the top 15 most frequent words appearing in the title of the
papers were selected. Communities were detected with Infomap~\cite{rosvall08} on 
snapshots spanning each a window of $5$ years, except at the right end
of each diagram: since there is no data after $2008$, the last windows
must have $2008$ as upper limit, so their size shrinks ($2004-2008$, $2005-2008$,
$2006-2008$, $2007-2008$). Each vertical bar represents a consensus partition combining pairs of consecutive snapshots.
A color uniquely identifies a community, 
the width of the links between clusters is proportional to the number of papers they have in
common. The rapid growth of the field {\it Complex Networks} is clearly visible, as well as its later split into a number of smaller
subtopics, like {\it Community Structure}, {\it Epidemic Spreading}, {\it Robustness}, etc.. Reprinted figure with permission from~\cite{lancichinetti12}. \copyright\,2012, by the Nature Publishing Group.}
\label{dyncons}
\end{center}
\end{figure}

Consensus clustering (Section~\ref{sec-consensus}) is a natural approach to find stable dynamic clusterings by combining
multiple snapshots.  Let us suppose we have a time
range going from $t_0$ to $t_m$, that we want to divide into $w$ windows of size $\Delta t$.
For the sake of stability, one should consider sliding windows, i. e., overlapping time intervals. 
This way consecutive partitions will be based on system configurations sharing a lot of vertices and edges, and change is (typically) smooth.
In order to have exactly $w$ frames, each of them has to be shifted by an interval $\delta t=(t_m-t_0)/w$ with respect to the previous one.
So we obtain the windows $[t_0, t_0+\Delta t]$, $[t_0+\delta t, t_0+\Delta t+\delta t]$,
$[t_0+2\delta t, t_0+\Delta t+2\delta t]$, ...,
$[t_{m}-\Delta t, t_m]$. The community structure of each snapshot can be found via any reliable static clustering technique.
Next, the consensus partition from the clusterings of $r$ consecutive
snapshots, with $r$ suitably chosen, is derived~\cite{lancichinetti12}. 
Again, one could consider sliding windows: for instance, the first window could consist of the first
$r$ snapshots, the second one by those from $2$ to $r+1$, and so on until the 
interval spanned by the last $r$ snapshots. In Fig.~\ref{dyncons} we show an application of this procedure on the citation network of 
papers published in journals of the American Physical Society (APS).

An alternative way to uncover the evolution of communities by accounting for the 
correlation between configurations of neighbouring time intervals is to use probabilistic models~\cite{sarkar05,yang09,peixoto15b,peixoto15c}.

If the system is large and its structure is updated in a stream fashion, instead of working on snapshots one could  
detect the clustering {\it online}, every time the configuration of the system varies due to new information, 
like the addition of a new vertex or edge~\cite{aggarwal05,zanghi08}. 
An advantage of this approach is that change is due to the effect that the 
small variation in the network structure has on the system, and it can be tracked by simply adjusting the partition of the previous configuration, 
which can be usually done rather quickly.

\subsection{Significance}
\label{sec-sign}

Let us suppose that we have identified the communities, somehow. Are we done? Unfortunately,
things are not that simple. 
\begin{figure*}[h!]
\begin{center}
\includegraphics[width=0.9\columnwidth]{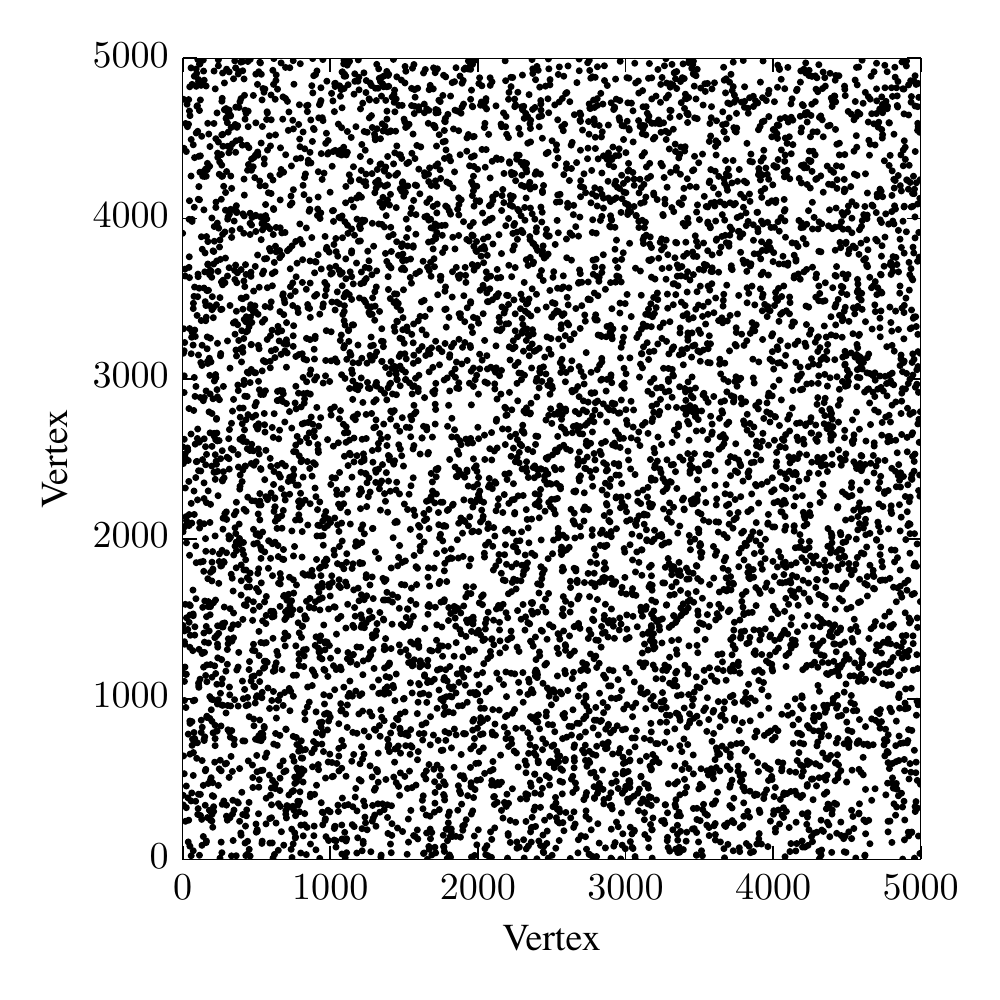}
\includegraphics[width=0.9\columnwidth]{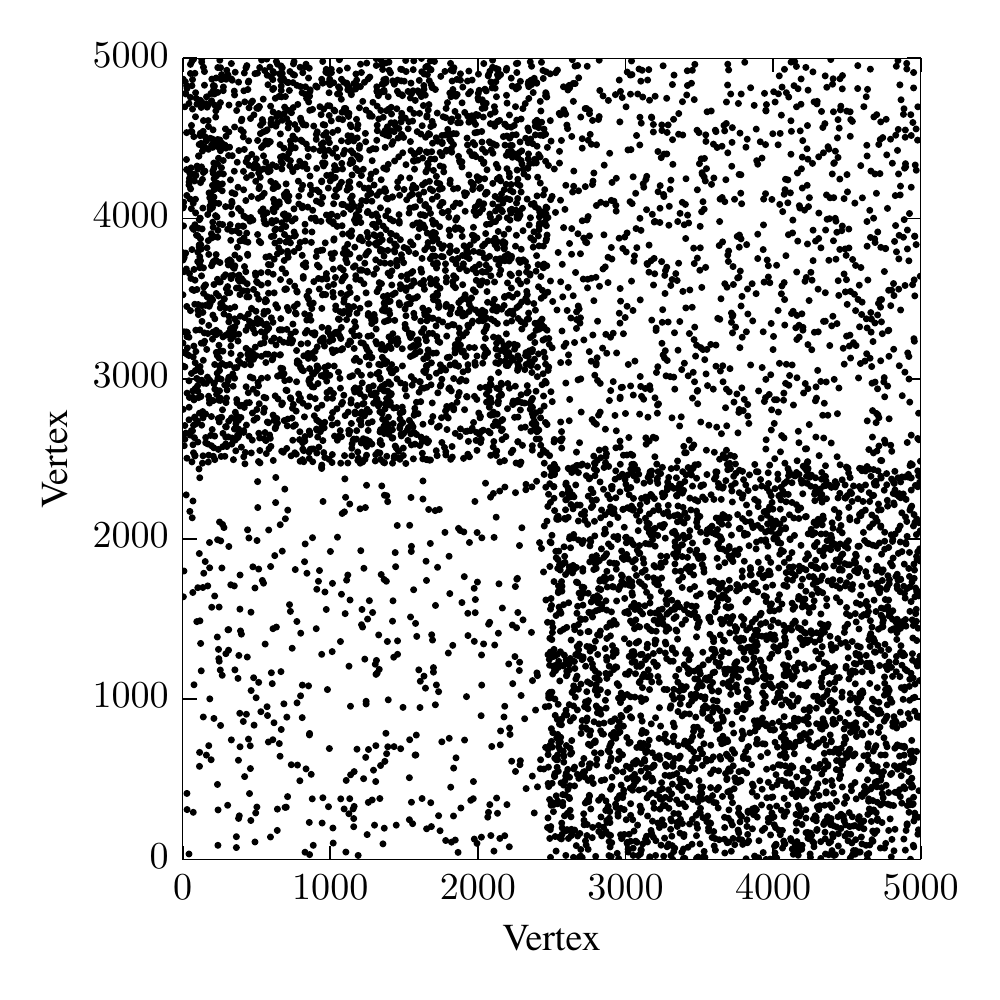}
\includegraphics[width=0.9\columnwidth]{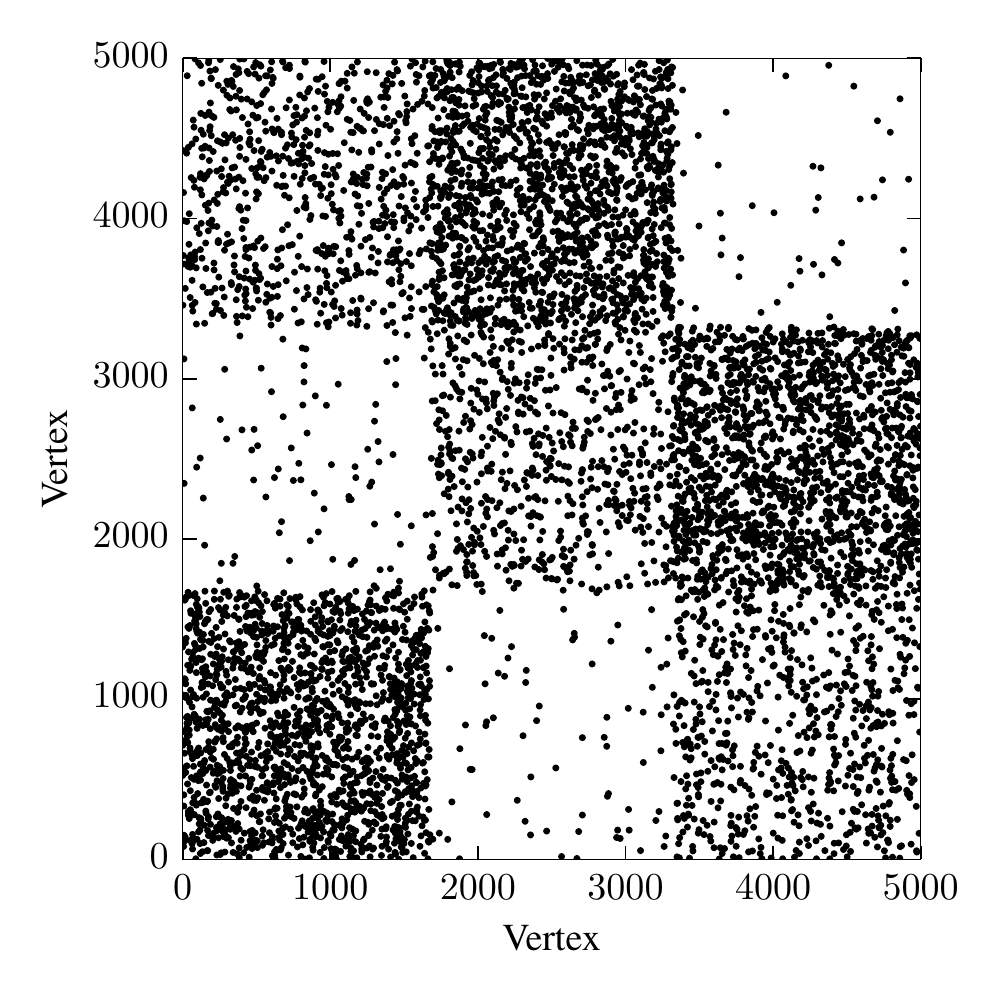}
\includegraphics[width=0.9\columnwidth]{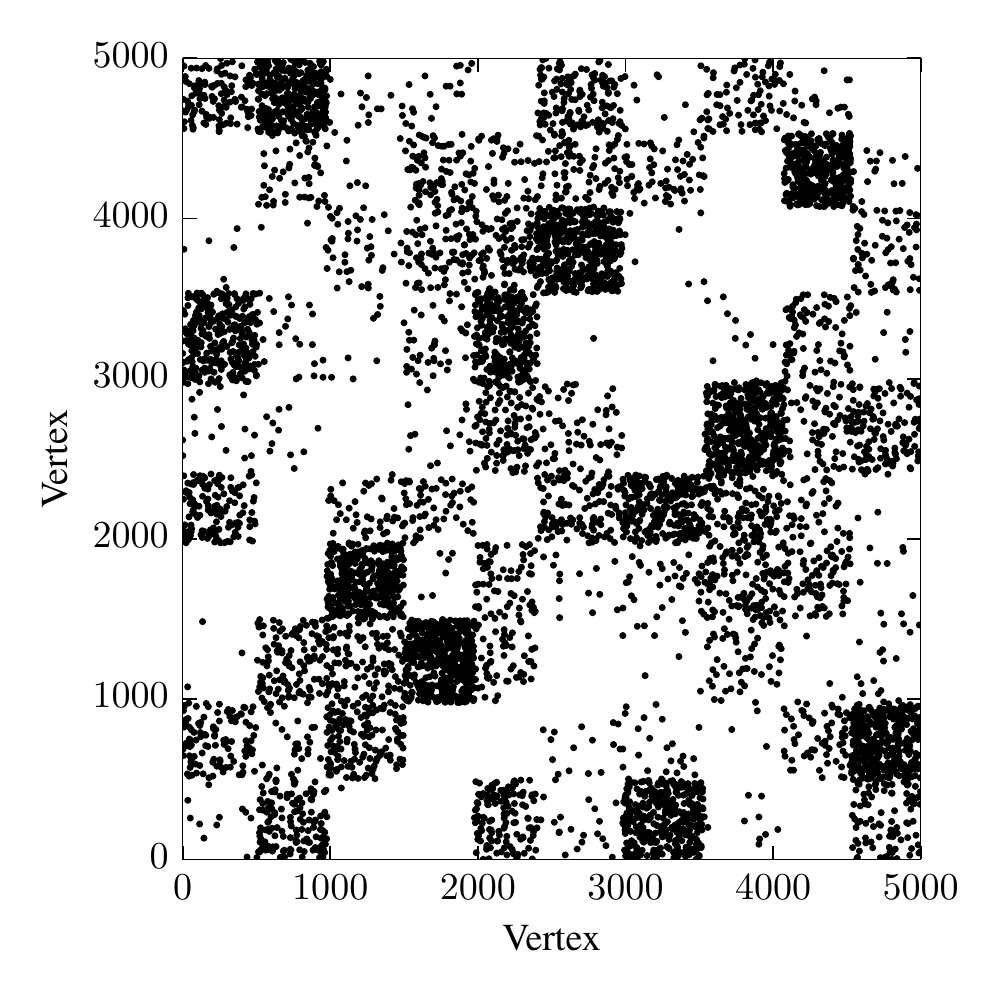}
\caption {Artificial groups in random networks. The elements of the adjacency matrix of an Erd\H{o}s-R\'enyi random graph (top left) can be rearranged, by suitably
reshuffling the list of vertices. This procedure may produce a block structure, with the blocks becoming increasingly more visible the smaller their size.
Such groups are not real, though, but they are generated by random fluctuations in the edge patterns among the vertices.
The construction principle of the network does not give any special role to groups of vertices, since 
all pairs of vertices have identical probability to be joined. Courtesy by Tiago P. Peixoto.}
\label{figrand0}
\end{center}
\end{figure*}

In Fig.~\ref{figrand0} (top left) we show the adjacency matrix 
of the random graph {\'a} la Erd\H{o}s-R\'enyi illustrated in Fig.~\ref{fig:stylized-random}.
The graph has $5\,000$ vertices, so the matrix is $5\,000\times 5\,000$.
Black dots indicate the existence of an edge between the corresponding vertices, while missing edges are represented in white.
By construction, there is no group structure. However, we can rearrange the elements of the matrix, by reordering the vertex labels.
In Fig.~\ref{figrand0} (top right) we see that, by doing that, one can generate a group structure, of the assortative type, 
with two blocks of equal size. If we increase the number of blocks to three Fig.~\ref{figrand0} (bottom left) and ten Fig.~\ref{figrand0} (bottom right)
we can make the matrix look more and more modular. This is why many clustering techniques detect communities in random networks as well, though they should not.
Where do the groups come from? Since they cannot be real by construction, they must be generated
by random fluctuations in the network construction process. Random fluctuations are particularly relevant on sparse graphs (Section~\ref{detectab}).

The lesson we learn from this example is that it is not sufficient to identify groups in the network, but one should also ask {\it how significant}, 
or non-random, they are. Unfortunately, most clustering algorithms are not able to assess the significance of their results. 
If the groups are compatible with random fluctuations, they are not proper groups and should be disregarded. The lower
the chance that they are generated by 
randomness, the more confident we can be that the blocks reflect some actual group structure. Naturally, this can be done only if one
has a reliable {\it null model}, describing how the structure of the network at study can be randomised and allowing us to estimate how likely it is that 
the candidate group structure is generated this way. The configuration model~\cite{bollobas80,molloy95} is a popular null model in the literature.
It generates all possible configurations preserving the number of vertices and edges of the network at study, and the degrees of its vertices.
One may compute some variables of the original network,
and estimate the probability that the model reproduces them, or {\it p-value}, i. e., the fraction of model configurations yielding 
values of the variables compatible with those measured on the original graph. If the p-value is sufficiently low ($5\%$ is a standard threshold), one concludes that the 
property at study cannot be generated by randomness only. For community structure, one can compute various properties of the clusters, e. g., their 
internal density, and compare them with the model values. Some clustering algorithms, like OSLOM~\cite{lancichinetti11} are based on this principle.
Along the same lines, $z$-scores can be used as well [see the example of Eq.~(\ref{eq:zscore})].
Degree-corrected stochastic block models~\cite{karrer11} (Section~\ref{sec-inference}) also  
include the configuration model, which corresponds to the case without group structure\footnote{Actually it would be a variant of the configuration model, as the degree sequence of the vertices
would be preserved only on average, not exactly. It is the same null model used in the standard formulation of modularity (Section~\ref{sec-modopt}).}.
In this case significance can be estimated by doing model 
selection between the versions with and without blocks (Section~\ref{sec-inference}).

A concept very related to significance is that of {\it robustness}. If clusters are significant it means that they are resilient if the network structure is perturbed, to some extent.
One way to quantitatively assess this is introducing into the system a controlled amount of noise, and checking how much noise it takes 
to disrupt the group structure. The greater the required perturbation, the more robust the communities. 
For instance, a perturbation could be rewiring a fraction of randomly chosen edges~\cite{karrer08}. 
After the network is perturbed, the community structure is derived and compared to the one of the original network\footnote{For reliable results multiple configurations have to be 
generated, for a given amount of noise, and the similarity scores have to be averaged.}. The trend of the 
partition similarity shows how the group structure responds to perturbations.

A similar approach consists in sampling network configurations from a population which the original network is supposed to belong to ({\it bootstrapping}), 
and comparing the clusterings found in those configurations, to check how frequently subsets of vertices are clustered together in different samples, which is an index of 
the robustness (significance) of their clusters~\cite{rosvall10}. 

\subsection{Which method then?}
\label{sec-WMT}

At the end of the day, what most people want to know is: which method shall I use on my data?
Since the clustering problem is ill-defined, there is no clear-cut answer to it. 

Popular techniques are based on similar ideas of communities, like the ones we reviewed in Sections~\ref{sec-defs} and \ref{sec-MV}.
What makes the difference is the way clusters are sought. The specific procedure affects the 
reliability of the results (e. g., because of resolution problems) and the time complexity of the calculation, 
determining the scope of the method and constraining its applicability.

Most methods propose a universal recipe, that is supposed to hold on every data set. In so doing, one 
neglects the peculiarities of the network at study, which is valuable information that could orient the 
method towards more reliable solutions. But algorithms are usually not so flexible to account for 
specific network features. For instance, in some cases, there is no straightforward extension capable to handle
high-level features like edge direction or overlapping communities\footnote{Especially extensions of clustering algorithms to the case of directed 
graphs are not straightforward and often impossible. Spectral methods may not work because spectra of directed graphs may be rather involved
(for instance the eigenvalues of the adjacency matrix are typically not real).
Likewise, some processes on directed graphs may not reach a stationary state, like 
simple random walks.}.

Validation of algorithms, like the comparative analysis of~\cite{lancichinetti09c}, have allowed
to identify a set of methods that perform well on artificial benchmarks. There are two important issues, though.
First, we do not know how well real networks are described by currently used benchmark models. Therefore, 
there is no guarantee that methods performing well on benchmarks also give reliable results on real data sets.
Structural analyses like the ones discussed in Section~\ref{ncp} might allow to identify more promising benchmark models.
Second, if we rely so much on current benchmarks, which are versions of the stochastic block model (SBM), we already know what the
best method is: a posteriori block modelling, i. e., fitting a SBM on the data. Indeed, there are several advantages to this approach.
It is more general, it does not only discover communities but several types of group structures, like disassortative groups (Fig.~\ref{fig:stylized-bipartite}) and
core-periphery structure (Fig.~\ref{fig:stylized-coreper}). It can also capture the existence of hierarchies among the clusters.
Moreover, it yields much richer results than standard clustering algorithms, as it delivers the entire set of parameters of the most likely SBM, with which one 
can construct the whole network, instead of just grouping vertices.
SBMs are very versatile as well. They can be extended to a variety of contexts, e. g., directed networks~\cite{peixoto14b}, networks with weighted edges~\cite{aicher14}, 
with overlapping communities~\cite{airoldi08}, with multiple layers~\cite{peixoto15b}, with annotations~\cite{newman16,hric16}.
Besides, the procedure can be applied to any network model with group structure, not necessarily SBMs.
The choice between alternative models can be done via model selection.
A posteriori block modelling is not among the fastest techniques available. 
Networks with millions of vertices and edges could be investigated this way, but very large networks remain out of reach. Fortunately,
many networks of interest can be attacked. The biggest problem of this class of methods, i. e., the determination of the number of clusters, seems to 
be solvable (Section~\ref{sec-inference}). We recommend to exploit the power of this approach in applications.

Algorithms based on the optimisation of cluster quality functions should be considered as well (Section~\ref{sec-modopt}), because they may avoid resolution problems and 
explore the network locally, which is often the only option when the system is too large to be studied as a whole. 

Algorithms based on the optimisation of partition quality functions, like modularity maximisation, are plagued by the problems we discussed in Section~\ref{sec-modopt}.
Nevertheless, if one knows, or discovers, the correct number of clusters $q$,
and the optimisation is constrained on the subset of partitions having $q$ clusters, such algorithms become competitive~\cite{nadakuditi12,darst14}.

We also encourage to use approaches based on dynamics (Section~\ref{sec-dynmet}).
In principle, the resulting clustering depends on the specific dynamics adopted. In practice, there often is a substantial overlap between the clusters found with 
different dynamics. An important question is whether dynamics may uncover groups that are not recoverable from network structure alone. 
Differences in the clusterings found via dynamical versus structural approaches could be due to the fact that dynamical processes are sensitive to 
more complex structural elements than edges (e. g., paths, motifs)~\cite{arenas08,serrour11,benson16} (Section~\ref{sec-MV}). However, even if that were true,
dynamical approaches could be more natural ways to handle such higher-order structures, and to make sense of the resulting community structure.

In general, however, the final word on the reliability of a clustering algorithm is to be given by the user, and any output is to be taken with care. 
Intuition and domain knowledge are indispensable elements to support or disregard solutions.

\section{Software}
\label{sec-soft}

In this section we provide a number of links where one can find the code of clustering algorithms and related techniques and models.
\begin{itemize}
\item {\it Artificial benchmarks}. Code to generate LFR benchmark graphs (Section~\ref{art-bench}) can be found here 
\burl{https://sites.google.com/site/andrealancichinetti/files}. The code for the dynamic benchmark by Granell et al.~\cite{granell15} is available at 
\burl{https://github.com/rkdarst/dynbench}. 

\item{\it Partition similarity measures}. Many partition similarity measures have their own function in R, Python and MatLab and are easy to find.
The variant of the NMI for covers proposed by Lancichinetti et al.~\cite{lancichinetti09} can be found at \burl{https://sites.google.com/site/andrealancichinetti/mutual}, the one by
Esquivel and Rosvall~\cite{esquivel12} at \burl{https://bitbucket.org/dsign/gecmi/wiki/Home}. 

\item {\it Consensus clustering}. The technique proposed by Lancichinetti and Fortunato~\cite{lancichinetti12} 
to derive consensus partitions from multiple outputs of stochastic clustering algorithms can be downloaded from 
\burl{https://sites.google.com/site/andrealancichinetti/software}.

\item {\it Spectral methods}. The spectral clustering method by Krzakala et al.~\cite{krzakala13}, based on the non-backtracking matrix (Sections~\ref{sec-tools} and \ref{sec-spectral}), 
can be downloaded here: \burl{http://lib.itp.ac.cn/html/panzhang/dea/dea.tar.gz}.

\item {\it Edge clustering}. The code for the edge clustering technique by Ahn et al.~\cite{ahn10} can be found here: \burl{http://barabasilab.neu.edu/projects/linkcommunities/}.
The link to the stochastic block model based on edge clustering by Ball et al.~\cite{ball11} is provided below.

\item {\it Methods based on statistical inference}. The code to perform the inference of the  
degree-corrected stochastic block model\footnote{We stress that the method is parametric, in that the number of clusters has to be provided as input. In Section~\ref{sec-inference} we have
pointed to techniques to infer the number of clusters beforehand.} by Karrer and Newman is available at \burl{http://www-personal.umich.edu/~mejn/dcbm/}.
The weighted stochastic block model by Aicher et al.~\cite{aicher14} can be found at \burl{http://tuvalu.santafe.edu/~aaronc/wsbm/}.
The code for the overlapping stochastic block model based on edge clustering by Ball et al.~\cite{ball11} is at \burl{http://www-personal.umich.edu/~mejn/OverlappingLinkCommunities.zip}.
The model combining structure and metadata by Newman and Clauset~\cite{newman16} is coded at \burl{http://www-personal.umich.edu/~mejn/Newman_Clauset_code.zip}.
The program to infer the bipartite stochastic block model by Larremore et al.~\cite{larremore14} can be found at \burl{http://danlarremore.com/bipartiteSBM/}.

The algorithms for the inference of community structure developed by Tiago Peixoto
are implemented within the Python module {\tt graph-tool} and can be found at \burl{https://graph-tool.skewed.de/static/doc/dev/community.html}. They allow us to perform model selection of various
kinds of stochastic block models: degree-corrected~\cite{karrer11}, with overlapping groups~\cite{peixoto15}, and for networks with layers, with valued edges
and evolving in time~\cite{peixoto15b}. The hierarchical block model of~\cite{peixoto14b}, that searches for clusters at high resolution, is also available.
All such variants can be combined at ease by selecting a suitable set of options. 

The algorithms for the inference of overlapping communities via the Community-Affiliation Graph Model (AGM)~\cite{yang12c} and the 
Cluster Affiliation Model for Big Networks (BIGCLAM)~\cite{yang13}
(Section~\ref{ncp}) can be found in the package \burl{http://infolab.stanford.edu/~crucis/code/agm-package.zip}.

\item {\it Methods based on optimisation}. There is a lot of free software for modularity optimisation. In the {\tt igraph} library (\burl{http://igraph.org}) there are several functions, both in the $R$ and in the 
Python package:
{\tt cluster\_fast\_greedy} (R) and {\tt community\_fastgreedy} (Python), implementing the fast greedy optimisation by Clauset et al.~\cite{clauset04}; 
{\tt cluster\_leading\_eigen} (R) and {\tt community\_leading\_eigenvector} (Python) for the optimisation based on the leading
eigenvector of the modularity matrix~\cite{newman06};  {\tt cluster\_louvain} (R) and {\tt community\_multilevel} (Python) 
are the implementations of the Louvain method~\cite{blondel08}; {\tt cluster\_optimal} (R) and {\tt community\_optimal\_modularity} turn the 
task into an integer programming problem~\cite{brandes08b}; {\tt cluster\_spinglass} (R) and {\tt community\_spinglass} (Python)
optimise the multi-resolution modularity proposed by Reichardt and Bornholdt~\cite{reichardt06}. 

Some methods based on the optimisation of cluster quality functions are also available. The code for the optimisation of the local modularity by Clauset~\cite{clauset05} can be found at
\burl{http://tuvalu.santafe.edu/~aaronc/shared/LocalCommunity2005_GPL.zip}. The code for OSLOM is downloadable from the dedicated website \burl{http://www.oslom.org}.

\item {\it Methods based on dynamics}. Infomap~\cite{rosvall08} is currently a very popular algorithm and its code can be found in various places. 
It has a dedicated website \burl{http://www.mapequation.org}, where several extensions can be downloaded, including hierarchical community structure~\cite{rosvall11}, overlapping 
clusters~\cite{esquivel11} and memory~\cite{rosvall14}. Infomap has also its own functions on {\tt igraph}, both in the R and in the Python
package ({\tt cluster\_infomap} and {\tt community\_infomap}, respectively). {\it Walktrap}~\cite{pons05}, another popular method based on random walk dynamics, is available on
{\tt igraph}, via the functions {\tt cluster\_walktrap} (R) and {\tt community\_walktrap} (Python). The local community detection algorithms proposed in~\cite{jeub15} can be downloaded from
\burl{http://people.maths.ox.ac.uk/jeub/code.html}.

\item {\it Dynamic clustering}. The code to optimise the multislice modularity by Mucha et al.~\cite{mucha10} is available at \burl{http://netwiki.amath.unc.edu/GenLouvain/GenLouvain}.
Detection of dynamic communities can be performed as well with consensus clustering (Section~\ref{sec-consensus}) and via stochastic block models (Section~\ref{sec-inference}).
Links to the related code have been provided above.

\end{itemize}

\section{Outlook}
\label{sec-OL}

As long as there will be networks, there will be people looking for communities in them.
So it is of uttermost importance to have a set of reliable concepts and principles guiding 
scholars towards promising solutions for network clustering. We have presented established 
views of the main aspects of the problem, and exposed the strengths as well as the limits of popular 
notions and approaches. 

What's next? We believe that there will be a trend towards the development of domain-dependent algorithms, exploiting as much 
as possible information and peculiarities of network data sets. Generalist methods could still be used to 
get first indications about community structure and orient the investigation in promising directions. 
Some existing approaches are sufficiently flexible to accommodate
various features of networks and community structure (Section~\ref{sec-WMT}).
 
At the same time, we believe that it is necessary to find accurate models of networks with community structure, both for 
the purpose of designing realistic benchmark graphs for validation, and for a more precise inference of the groups and of their features.
Investigations of real networks at the level of subgraphs, along the lines of those discussed in Section~\ref{ncp}, are instrumental to 
the definition of such models.

While benchmark graphs can be improved, there is one test that one can rely on to assess the performance of clustering algorithms: 
applying methods on random graphs without group structure. We know that many popular techniques find groups in such graphs as well, failing the test.
On a related note, it is critical to determine how non-random the clusters detected on real networks are, i. e., to estimate their significance (Section~\ref{sec-sign}).

We stress that this exposition is by no means complete. The emphasis is on the fundamental aspects of network clustering and on main stream approaches. 
We discussed works and listed references which are of more immediate relevance to the topics discussed. A number of topics have not been dealt with.
Still we hope that this work will help practitioners to design more and more reliable methods and domain users to extract 
useful information from their data.

\vskip0.5cm

\begin{acknowledgments}

We thank Alex Arenas, Florian Kimm, Tiago Peixoto, Mason Porter and Martin Rosvall for a careful reading of the manuscript and many valuable comments.
We gratefully acknowledge MULTIPLEX, grant No. 317532 of the
European Commission.

\end{acknowledgments}

\end{sloppypar}

\end{document}